\documentclass{aa}
\usepackage{graphicx}
\usepackage{layouts}
\usepackage{txfonts}
\usepackage{natbib}
\usepackage{color}
\usepackage{amssymb}
\usepackage{amsmath}
\newcommand{\ria}[1]{{\color{blue}{#1}}}

\begin{document}

%   \title{Tuning in on Cepheids III. -- Investigation of the Twin Peak Phenomenon in {\it OGLE }LMC Cepheids}
   %\title{Continuous-time period and light curve shape changes \\ in 53 {\it OGLE }LMC Cepheids}
   \title{Investigating Light Curve Modulation via Kernel Smoothing}
   \subtitle{I. Application to 53 fundamental mode and first-overtone Cepheids in the LMC}
   
%   \titlerunning{Period and light curve shape changes in {\it OGLE }LMC Cepheids}
\titlerunning{Cepheid Period and Light Curve Shape Modulations in Continuous Time}

   \author{Maria S\"uveges\inst{1}\fnmsep\thanks{Present address: Max Planck Institute for Astronomy, K\"onigstuhl 17, 69117 Heidelberg, Germany; } \and Richard I. Anderson\inst{2}\fnmsep\thanks{Swiss National Science Foundation Fellow}}

   \institute{Observatoire de Gen\`eve, Universit\'e de Gen\`eve, Ch.
   d'Ecogia 1290, Versoix, Switzerland \\
    \email{maria.suveges@unige.ch;sueveges@mpia.de} \\
\and
   Department of Physics \& Astronomy, The Johns Hopkins University, 3400 N Charles St, Baltimore, MD 21218, USA \\
    \email{ria@jhu.edu}
    }

   \date{Received 7 May 2016; accepted ???}

\abstract {Recent studies have revealed a hitherto unknown complexity of Cepheid
  pulsations by discovering modulated variability using photometry, radial
  velocities, and interferometry.
  However, a statistically rigorous search and characterization of such
  phenomena has so far been missing.} 
{We have employed a method as yet unused in time series analysis of variable stars for detecting and characterizing modulated variability in continuous time. %This method is capable of handling large data sets originating from present-day surveys to examine the statistical distribution of such phenomena. 
Here we test this new method on 53 classical Cepheids from the {\it OGLE-III} catalog.}
{We implement local kernel regression to search for both period and amplitude modulations simultaneously in continuous time and to investigate their detectability. We determine confidence intervals using parametric and non-parametric bootstrap sampling to estimate significance and investigate multi-periodicity using a modified pre-whitening approach that relies on time-dependent light curve parameters.} 
{We find a wide variety of period and amplitude modulations and confirm that first overtone pulsators are less stable than fundamental mode Cepheids. Significant temporal variations in period are more frequently detected than those in amplitude. We find a range of modulation intensities, suggesting that both amplitude and period modulations are ubiquitous among Cepheids. Over the 12-year baseline offered by {\it OGLE-III}, we find that period changes are often non-linear, sometimes cyclic, suggesting physical origins beyond secular evolution. Our method more efficiently detects modulations (period and amplitude) than conventional methods reliant on pre-whitening with constant light curve parameters and more accurately pre-whitens time series, removing spurious secondary peaks effectively.} 
{Period and amplitude modulations appear to be ubiquitous among Cepheids.  Current detectability is limited by observational cadence and photometric  precision: detection of amplitude modulation below 3 mmag requires space-based   facilities. Recent and ongoing space missions ({\it K2, BRITE, MOST, CoRoT}) as well as upcoming ones ({\it TESS, PLATO}) will significantly improve detectability of fast modulations, such as cycle-to-cycle variations, by providing high-cadence high-precision photometry. High-quality long-term ground-based photometric time series will remain crucial to study longer-term modulations and to disentangle random fluctuations from secular evolution.}

   \keywords{methods:statistical -- stars:oscillations -- stars:variables:Cepheids -- Magellanic Cloud}

   \maketitle

%\printinunitsof{in}\prntlen{\textwidth}
%\printinunitsof{in}\prntlen{\textheight}

% ________________________________________________________________
\section{Introduction}\label{sec:Intro}

%\riacomm{A thought: Is the terminology between frequency and period unambiguous throughout the paper?}

Classical Cepheid variable stars (from hereon: Cepheids) have been the focus of
a great deal of research since their discovery by \citet{1786RSPT...76...48G},
who suggested that their study "may probably lead to some better knowledge of
the fixed stars". Indeed, Cepheids have been of great historical importance for
the understanding of stellar evolution and structure. Although Cepheids are
usually considered to be highly regular variable stars, Cepheid pulsations were
shown very early on to exhibit time-dependencies
\cite[e.g.][]{1919MNRAS..79Q.177E}. %\riacomm{FIND DIFFERENT CITATION?}.

In particular the changing periods of Cepheids have received much attention
\citep[e.g.][]{szabados83,2000ASPC..203..244B,2001AcA....51..247P,2002AcA....52..177P,2006PASP..118..410T},
since they offer an immense opportunity for studying the {\it secular}
evolution of stars on human timescales (decades) and provide important tests of stellar evolution models \citep[e.g.][]{2013AstL...39..746F,2016A&A...591A...8A}. In addition, changing periods
complicate phase-folding of time-series data obtained over long temporal
baselines and often have to be accounted for when determining the orbit of
long-period binary systems containing Cepheids
\cite[e.g.][]{2013MNRAS.434..870S,2015ApJ...804..144A}. In addition, cyclic
variations of pulsation periods exhibited by some Cepheids have been
discussed in terms of the light-time effect due to orbital motion
\citep{1989CoKon..94....1S}, although only few cases have
been confirmed using radial velocities \citep{1991CoKon..96..123S}. Period changes may also be related to the much-discussed linearity of the period-luminosity relation, see \citet{2016arXiv160404814G} and references therein.

 Recent detailed studies of both large samples of Cepheids in the LMC \citep{poleski08} and of individual stars in the Galaxy \citep[e.g.][]{2000NewA....4..625B,2017arXiv170105192K} have revealed intricate, possibly periodic period change patterns that are not necessarily consistent with the classical picture of \textit{secular} evolution.

One of the most notorious and intricate cases of period and amplitude variations
is that of Polaris, the North Star \citep{1983ApJ...274..755A}, which
identifies this Cepheid to be crossing the classical instability strip for the
first time \citep{2006PASP..118..410T}.
Polaris' amplitude seemed to diminish to the point of disappearing, which had been
interpreted as the Cepheid leaving the instability strip. However, more recent
observations have shown that the pulsation amplitude has increased again, and
Polaris remains a puzzle. Another unique well-known
example is that of V473 Lyrae \citep{1980A&A....91..115B,1982A&A...109..258B}.
Combining many years worth of observations, \citet{2013AN....334..980M} were
able to trace this star's amplitude modulation cycles and determined a
modulation period of $1204$\,d. They discussed these modulations in the context
of the \citet{1907AN....175..325B} effect, which is better known in RR Lyrae
stars and has seen a boost in research thanks to the {\it Kepler} mission
\citep[e.g.][]{2010ApJ...713L.198K,2011MNRAS.417..974B,2014ApJS..213...31B}. 

\citet{2014JAVSO..42..267P} presented evidence that some long-period Cepheids
exhibit amplitude changes of up to a few hundredths of a magnitude over
timescales of a few hundred or thousands of days, potentially exhibiting cyclic
behavior (e.g. for U Carinae). Such strong modulations are presumably not very
common among classical Cepheids, or else they would likely be found more
frequently in long-term photometric surveys such as the All Sky Automated Survey 
\citep{2002AcA....52..397P} or in other long-term Cepheid photometry
\citep[e.g.][]{2014AstL...40..125B}.
\citet{soszynskietal08} mention that about $4\%$ of fundamental-mode (FU) and
$28\%$ of first-overtone (FO) Cepheids are ``Bla\v{z}ko Cepheids'', identified
via secondary period peaks near the primary period (from hereon ``twin peaks''),
which are found after pre-whitening the light curve using the primary period.
Additionally, among the entire set of 3374 Cepheids, 8 (all FO) were labeled as
having variable amplitude. \citet{soszynskietal15b} have since
provided additional targets of interest in this regard. While this paper was under review, \citet{smolec17} further reported light curve modulation in 51 Cepheids located in the Small and Large Magellanic Clouds, none of which overlap with the stars discussed in the present work.
 Evidence for non-radial modes in Cepheids has been found in a sample of 138 Small Magellanic Cloud (SMC) FO Cepheids that exhibit light curve modulation \citep{2010AcA....60...17S,2015arXiv151203708D,2016MNRAS.tmp..341S}.
Periodicity of such light curve modulations, if it
can be firmly established, would be strongly indicative of Cepheids pulsating in more than one mode \citep{2009MNRAS.394.1649M}.

Though difficult to detect with ground-based observatories, small amplitude
light curve fluctuations appear to be rather common, if photometry is
sufficiently precise and densely-sampled. \citet{2012MNRAS.425.1312D} first
showed this for the {\it Kepler} (fundamental-mode) Cepheid V1154 Cygni, and
\citet{evansetal15} recently used the {\it MOST} satellite to demonstrate the
different types of irregularities seen in the fundamental-mode Cepheid RT
Aurigae and the first-overtone Cepheid SZ Tauri. Such low-amplitude modulations
and period ``jitter'' may be explained by convection and/or granulation
\citep{2014A&A...563L...4N}. \citet{stothers09} furthermore proposed a model involving activity cycles to explain period and light amplitude changes in short-period Cepheids. However, light curve modulations remain difficult
to detect, even with precise space-based photometry \citep{2015MNRAS.454..849P}.

While most amplitude modulations in Cepheids are found using photometric
measurements, the extreme precision afforded by state-of-the-art planet hunting
instruments has recently enabled the discovery of small amplitude spectral
modulations in Cepheids \citep{2014A&A...566L..10A,2016MNRAS.463.1707A}. Furthermore,
tentative evidence for modulated angular diameter variability in the long-period Cepheid
$\ell$\,Carinae based on long-baseline interferometry has recently been
presented by \citet{2016MNRAS.455.4231A}.

In summary, recent advances in instrumentation have enabled the discovery that Cepheid
variability is not as regular as often assumed. Whether or not
irregularities are detected is dominated by observational precision and
time-sampling. Moreover, there is evidence that not all irregularities of
Cepheid variability share the same origin; for instance, the time-scales
of radial velocity modulation in short- and long-period Cepheids are very
different \citep{2014A&A...566L..10A}.

Given the patchy evidence for irregularities and modulations in Cepheid
variability, it is important to characterize how and how often these phenomena
occur. To this end, we have implemented a method based on local kernel estimation to detect irregularities in Cepheid pulsations. For the first time, this method allows to search for smooth
variations of light curve amplitudes and periods in continuous time and enables
the quantification of the significance of the detected effects. We here
describe our technique, which we apply to a total of 53 FU and FO Cepheids from
the {\it OGLE-III} catalog \citep{soszynskietal08}. In a follow-up paper, we will then
apply this technique to the full sample of {\it OGLE-IV} classical Cepheids
\citep{soszynskietal15a} to investigate limits of detectability, the rate of
occurrence of period and amplitude modulations in Cepheids, and to characterize
them. This will be a crucial step toward a physical understanding of these
phenomena.

This paper is structured as follows. We describe our method for analyzing light
curves in Sec.\,\ref{sec:Method}, which is divided into target selection
(Sec.\,\ref{sec:targetselection}) and a description of the sliding-windows based
light curve modeling (Sec.\,\ref{sec:slidingwindows}). We present the results of
this modeling in Sec.\,\ref{sec:Results}, which we divide into subsections
dedicated to changing periods (Sec.\,\ref{subsubsec:perch}), changing amplitudes
(Sec.\,\ref{subsubsec:ampch}), and a discussion of how light curve shapes change
with time (Sec.\,\ref{subsubsec:lcch}). We present the
implications of the new method on results of a multiperiodicity analysis and on
pre-whitening artefacts (Sec.\,\ref{subsec:resper}), compare the trends and
fluctuations discovered among different groups of Cepheids
(Sec.\,\ref{sec:trends+fluctuations}), investigate their relationships with
physical parameters of the Cepheids (Sec.\,\ref{sec:comparephysical}), and
compare our results with the literature. We summarize and
conclude in Sec.\,\ref{sec:Conclusions}.  We explain the statistical methodology in detail in Appendix \ref{app:detailedmethods}. Using simulations of modulated periodicity with parameters taken from real Cepheids, we 
benchmark the detectability and performance of the kernel method
in Appendices \ref{app:sim} and \ref{app:bias}.  Figures illustrating the results for all 53 Cepheids and tables containing the numerical results from the fitting procedures are given in in Appendices\,\ref{app:figtemporal} and \ref{app:tables}.

\section{Methodology}\label{sec:Method}

\subsection{Data and target selection}\label{sec:targetselection}

\begin{figure*}
\begin{tabular}{lr}
\includegraphics[scale=.34]{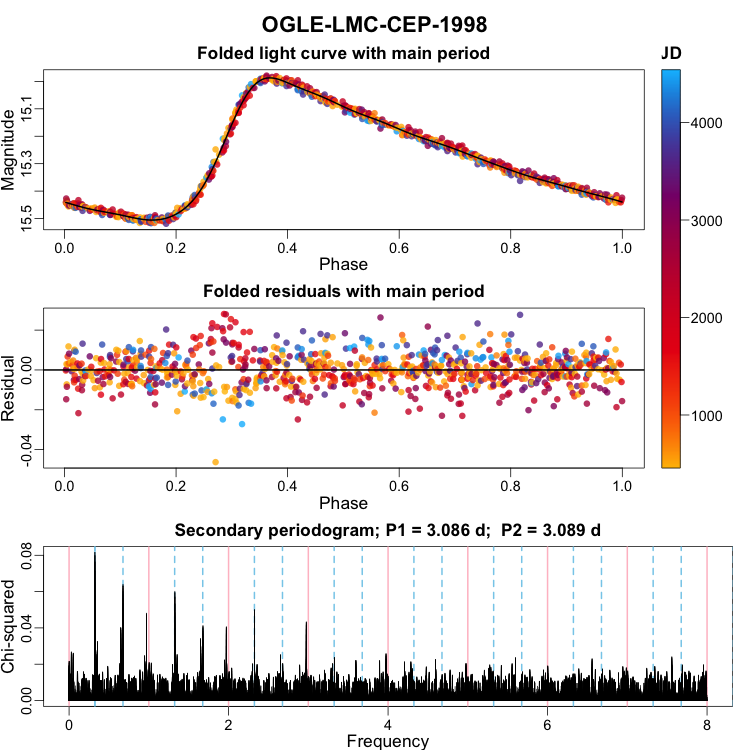}
\includegraphics[scale=.34]{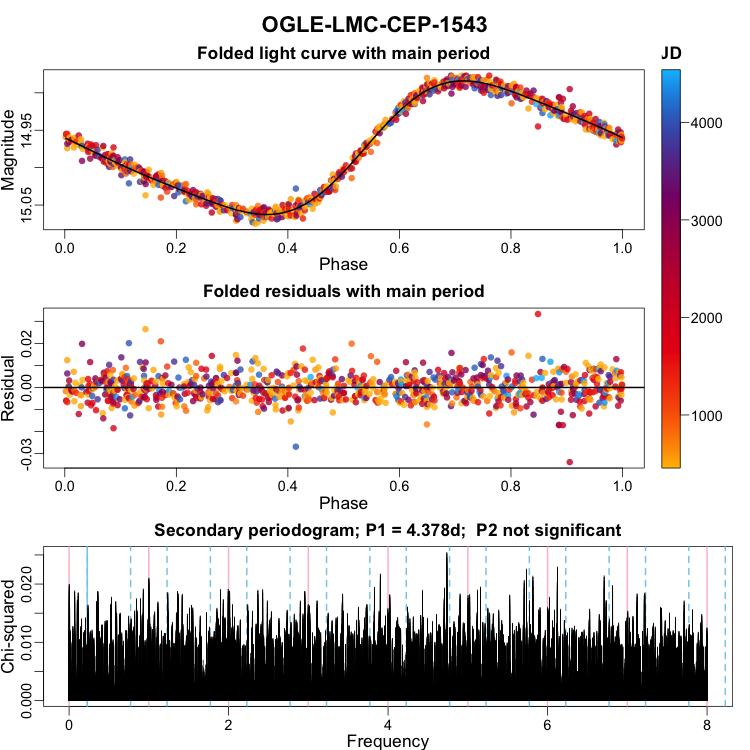}
\end{tabular}
\caption{{\it Left:} CEP-1998 as an example of a twin-peak target, for which
secondary periods are found after pre-whitening the light curve using the
primary period. {\it Right:} CEP-1543 as an example from the control group (no
twin peaks).  Colors trace observation date, increasing from yellow to blue.
{\it Top panels:} Phase-folded (with primary period) {\it OGLE-III} I-band light
curve. {\it Center panels:} Residuals after subtracting Fourier series model.
{\it Bottom panels:} Periodogram of the residuals shown in the center panel.
Pink vertical lines at $0, 1, \ldots d^{-1}$ indicate parasite frequencies due
to residual trends in the time series. The solid light blue vertical line
indicates the {\it OGLE }catalog (primary) frequency. The dashed light blue lines
correspond to its $\pm 1, 2, \ldots $ aliases. }
\label{fig:twinpgrams}
\end{figure*}

We analyze I-band photometric time-series data (light curves) of classical Cepheids in the LMC published by  the second- and third-generation {\it Optical Graviational Lensing Experiment} \citep[{\it OGLE-II} and {\it -III}]{soszynskietal08, udalskietal99}. These data were taken with the 1.3m Warsaw telescope at
Las Campanas, Chile and reduced using difference imaging \citep{alardlupton98,
alard00, wozniak00, udalskietal08}. {\it OGLE } photometry is particularly
well-suited for the study of modulations in Cepheid variability due to its high
quality (precision of up to $0.02$\,mag), long temporal baseline (spanning up to 12 years), large number of observations (up to $\gtrsim 1500$ per target),
excellent homogeneity, and large number of Cepheids available for study.  The time sampling of the two surveys is seasonal, with the longest gap (between the end of {\it OGLE-II} and the start of {\it OGLE-III}) up to 300 $d$ for some stars. The instrumentation changed between the two phases of the survey, around HJD $- 2450000 \approx 2000$, so I-band photometry in the LMC from the third phase was carefully calibrated to seamlessly fit with the photometric data from the second based on more than 620000 stars in 78 overlapping subfields \citep{udalskietal08}. All data were obtained from the {\it OGLE-III} server\footnote{\url{http://ogledb.astrouw.edu.pl/~ogle/CVS/}; \\
\url{ftp://ftp.astrouw.edu.pl/ogle/ogle3/}}.

This work has a dual goal of assessing the performance of our methodology
and of investigating the phenomenology of the modulated variability exhibited by
Cepheids. To this end, we selected a sample of 53 Cepheids consisting both of
ones likely to exhibit modulations and ones likely to be stable
pulsators. Investigations of all Cepheids within the {\it OGLE } catalog of
variable stars will be presented in the future. 

Whether or not a given Cepheid is likely to show the effect is difficult to determine a priori. A sign of modulations may be the presence of peaks in the secondary periodogram at a frequency very close to the primary one (a ``twin" of the primary peak;  $|f_1 -f_2| < 0.001$ except for one Cepheid, CEP-1564, for which this was 0.0176).
The reason is that if the harmonic decomposition of the oscillation or its
period varies with time, then pre-whitening with a constant model will not lead
to a perfect removal of the primary pulsation from the light curve, and the
residual signal would appear in the secondary periodogram as a twin peak. 

We thus selected as \emph{likely irregular candidates} those Cepheids for which a
multiperiodicity analysis reveals a twin peak after pre-whitening. We refer to
these targets here as ``twin-peak Cepheids" rather than adopting the
terminology of \citet{soszynskietal08} who refer to them as ``Bla\v{z}ko
Cepheids'', since it is still unclear whether the origin of the irregularities
of the Cepheid variability is the same as that of the amplitude modulation found
in RR Lyrae stars. Moreover, in order to assess the efficacy of the twin
peaks phenomenon as a diagnostic for identifying pulsation irregularities
% via the twin peaks phenomenon
and to provide a baseline for comparison, we also selected a \emph{likely regular}, or ``control'', sample
consisting of target stars that do not exhibit twin peaks. Since
first-overtone (FO) Cepheids are considered to be more irregular (less stable)
than fundamental-mode (FU) Cepheids, we treat these two groups separately.
To create a basis for target selection, we first modelled all individual light curves of fundamental (FU) and first-overtone (FO) Cepheids that consisted of more 700 observations, and inspected their residual periodograms. For the
purpose of sample selection, we modelled light curves using a non-periodic
fifth-order polynomial trend (to account for a possible temporal evolution
of the instrument zero-point as well as other spurious or true changes in mean
magnitude) and a Fourier series with 10 harmonics using the period from
the {\it OGLE-III} catalog. Secondary (residual) periodograms  were computed for each star using the method of \citet{zechmeisterkurster09}.
Including the polynomial trends proved efficient at removing artefacts from the
residual periodograms, such as high peaks near $0,1,2, \ldots d^{-1}$ that
otherwise dominated. Although a precise light curve modeling could have called
for more or less than 10 harmonics, we found this a sufficient approach for
sample selection. In all later stages of the analysis, in particular during the sliding window analysis, we used a more detailed modeling with
different harmonic orders, which were determined separately for each Cepheid (cf. Section \ref{sec:slidingwindows}  and Appendix \ref{app:detailedmethods}).

Figure\,\ref{fig:twinpgrams} shows examples of each of the twin peaks and control Cepheid groups. The left hand side exemplifies the twin peaks case with
(OGLE-LMC-)CEP-1998, which has a high secondary peak in the residual
periodogram. For this star, the appearance of the twin peak occurs together with
a prominent separation of the folded residual light curves according to
observation times: the early observations (color-coded in yellow) trace a
different line than later ones (red, violet and blue dots). Such a visual
separation is not observed in every twin peaks Cepheid. Instead, various degrees
of this pattern can be found, although its prominence does seem to correlate
with the height of the peak in the residual periodogram. The fact that the
yellow (earliest) and the light blue (latest) observations are close together,
while the reddish dots of mid-survey observations are far from them, indicate a
variation nonlinear in time, and thus excludes a misestimated period from the
possible explanations. 
For comparison, the right hand panel shows CEP-1543, for which the residuals are
flat and no significant secondary peaks are found in the residual periodogram.

Since our method aims to smoothly and continuously trace temporal variations of
pulsation periods and amplitudes, it is necessary for all objects
studied to have a large number of sufficiently densely and uniformly distributed
observations over the survey time span. We thus limited possible targets to
those with at least 700 I-band observations, distributed in a way that
ensured none of the time intervals used in our sliding windows fits contained
less than 90 observations.  

Of all the Cepheids that satisfied these conditions,
we selected a sample of 12 FU and 12 FO Cepheids that exhibit twin peaks. We further included five additional FO Cepheids with clearly visible changing
amplitudes, which we encountered during light curve inspections. These were included despite having a slightly lower limit of 70 observations for each data sub-interval. This was deemed acceptable, since FO Cepheids have nearly sinusoidal light curve shapes that require fewer harmonics for an adequate light curve representation.
All of these amplitude-changing Cepheids exhibit
twin peaks, and therefore will be treated as part of the twin-peak FO group. Similarly, we selected 12 FU and 12 FO Cepheids for which pre-whitening did not reveal twin peaks as the control samples.

\subsection{Detecting period and amplitude modulation using sliding windows}
\label{sec:slidingwindows}
%\pagebreak

\begin{figure}
\includegraphics[scale=.5]{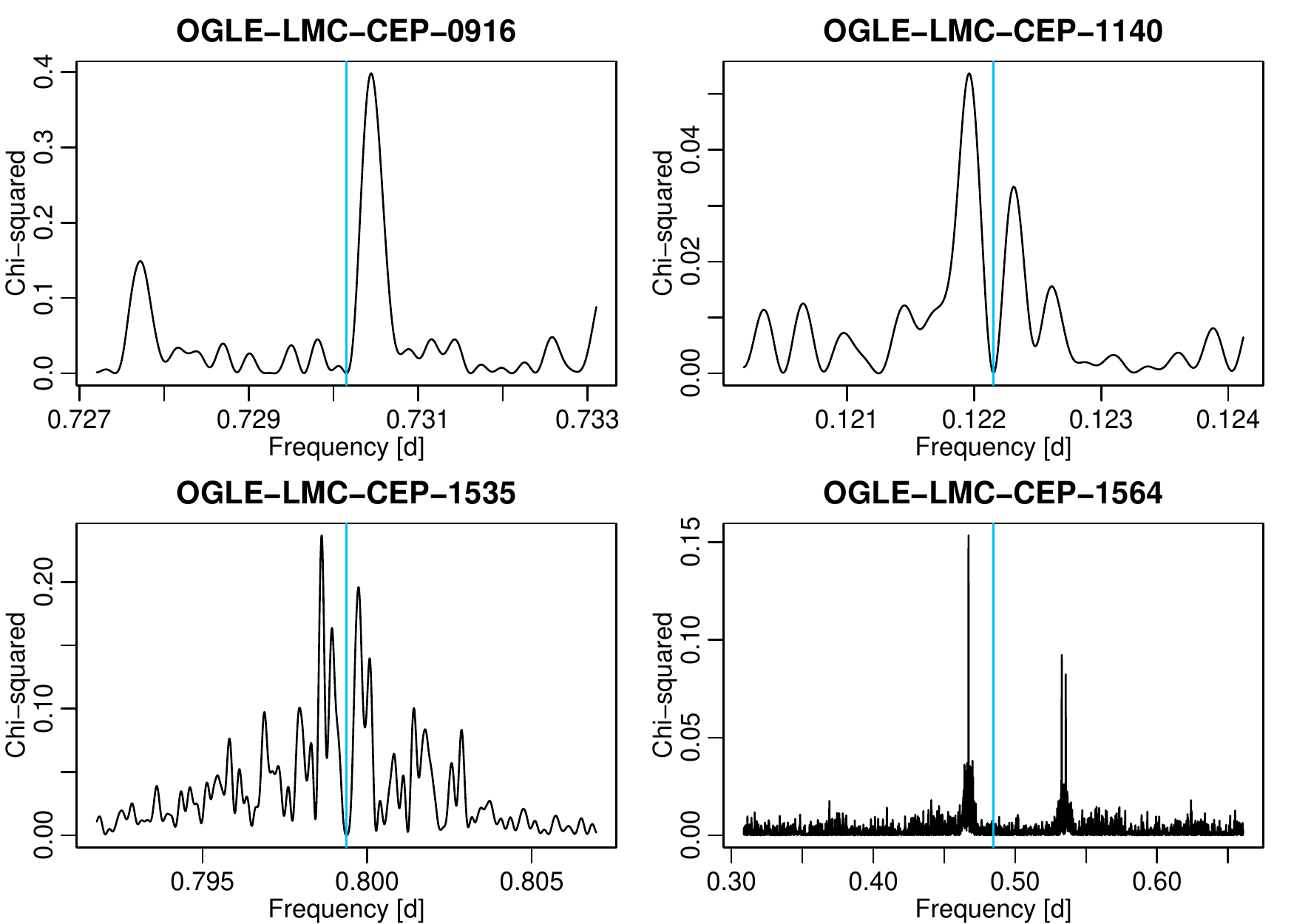}
\caption{Enlarged environment of the removed {\it OGLE } catalog frequency and the secondary peak in the secondary periodogram for four twin-peak Cepheids. The {\it OGLE } catalog frequency is indicated by a vertical blue line.}
\label{fig:twincloseup}
\end{figure}

 Studying time-dependent variability phenomena has been gaining traction for some time. A so-termed \textit{time-dependent Fourier analysis} has been applied to non-linear pulsation models of RR Lyrae stars \citep{1987ApJ...319..247K}, RRc stars observed by the \textit{Kepler} spacecraft \citep{2015MNRAS.447.2348M}, RR Lyrae stars in the Galactic Bulge observed by {\it OGLE }  \citep{2015MNRAS.447.1173N}, and 138 FO Cepheids in the SMC \citep{2016MNRAS.458.3561S}. As an alternative, the \textit{analytic signal processing} method has also been applied to hydrodynamical models \citep{2002A&A...385..932K} and to investigate the period doubling phenomenon in RR Lyrae stars using \textit{Kepler} data \citep{2010MNRAS.409.1244S}. Given the sensitivity of the method to (seasonal) gaps in the ground-based {\it OGLE } data, analytic signal processing is not a suitable choice for the present investigation.

%\ria{While the present work was under review, \citet{smolec17} presented an investigation of periodic light curve modulation in SMC and LMC Cepheids based on a systematic search for double modulation side peaks in periodograms following a standard pre-whitening technique. Such an approach differs significantly from ours, since it assumes periodicity of any detected modulations as well as a stable primary pulsation period. Interestingly, \citet{smolec17} reportedly investigated all Cepheids in the {\it OGLE }Magellanic Cloud collection \citet{2008AcA....58..163S,2010AcA....60...17S,2015AcA....65..297S}. Since not one of the 53 Cepheids discussed in the present work is mentioned in his paper, it appears that there are significant differences in sensitivity of the respective methods. Our future analysis of the full {\it OGLE }Cepheid collection using the kernel method will provide further insights into this matter.}

 Since previous studies of time-dependent Fourier and analytic signal processing techniques adopted fixed oscillation frequencies, any fluctuations in pulsation period were absorbed as phase-shifts in the Fourier phase coefficient \citep{2015MNRAS.447.2348M,2010MNRAS.409.1244S}. 
Here, we seek to go one step further and develop a method capable of efficiently dealing with data gaps that simultaneously determines Fourier coefficients and changes in pulsation period. Moreover, our method makes no assumptions as to the periodicity or repeatability of any detected modulations.

We adopt a highly flexible model % is required to achieve this and to 
to describe the potentially diverse  and presumably small
types of Cepheid light curve variations. Physical causes for changes in
period, for instance, may originate from secular evolution or binarity
(light-time effect). However, amplitude modulations are not so easily explained
and modeled. In our sample, the separation of the residual light curves shown in the left panel of Figure~\ref{fig:twinpgrams} suggests nonlinearity of the changes. %\ria{The diversity of twin-peak structures in secondary periodograms (obtained by pre-whitening ) suggests the presence of potentially complex, i.e., non-repeating, modulation patterns, although contributions from the irregular {\it OGLE }time sampling may also contribute.}
 Moreover, the diversity of twin-peak structures in the examined secondary periodograms, though the irregular {\it OGLE } time sampling and the photometric noise  undoubtedly affect the shapes, also suggests that modulation patterns may not be strictly repetitive. Figure \ref{fig:twincloseup} gives a few examples of this diversity. While the present work was under review, \citet{smolec17} presented an investigation of periodic light curve modulation in SMC and LMC Cepheids based on a systematic search for double modulation side peaks in periodograms following a standard pre-whitening technique. Such an approach assumes periodicity of any detected modulations as well as only mild effects of the uneven time sampling and noise. Our visualizations of the residual light curves, with the observing time colour-coded, support that many kinds of secondary peak structures can be associated with visible modulations of the period and/or the harmonic parameters, and these are not necessarily strictly periodic or linear.  Our goal is therefore to be open to all possibilities, and allow for nonlinear, cyclic, and non-cyclic components in the modelling of the modulations. Local kernel modelling \citep[sliding window technique; e.g.,][]{fangijbels} is a simple option that provides this flexibility.

We define these sliding windows using a grid of times that covers the entire
observational baseline. The first grid point $\tau_1$ is fixed to the time of
the first observation of the star, and every subsequent grid point is defined as
$\tau_i = \tau_1 + (i-1) \times 30$ days. For most of our Cepheids, 137 or 128
grid points were thus defined. One target (CEP-2580)  was limited to 80, since
observations of this target started roughly five years after the others. We wish
to obtain a local estimate of pulsation period and harmonic amplitudes at each
gridpoint $\tau_i$. Therefore, at each $\tau_i$, we select the data within a
3-year window centered at that point, and fit a harmonic model with a
third-order polynomial trend:
\begin{equation}\label{eq:localmodel}
Y_i = \sum_{k = 0}^3 a_k t^k + \sum_{m = 1}^{M} \left( s_m \sin 2 \pi m f t_i + c_m \cos 2 \pi m f t_i \right) + \epsilon_i,
\end{equation}
where $\epsilon_i \sim \mathcal{N} (0, \sigma_i)$ are assumed to be independent Gaussian errors. The harmonic order $M$, individually selected for each Cepheid, is based on the constant models using the whole time span: $M = M_0 + 2$, where $M_0$ is the order of the best constant model for all data by the Bayes Information Criterion \citep{schwarz78}, and the two extra terms are added to allow for variations in the light curve shape. $M$ was subsequently kept fixed at all gridpoints, though all the model parameters were refitted in each window. Thus, we obtain best-fit pulsation
period, harmonic amplitudes, and polynomial coefficients at each gridpoint.  Any changes in mean brightness are thus absorbed by the polynomial term in eq.\,\ref{eq:localmodel} and treated as nuisance parameters.

The parameter estimation is
performed by a weighted nonlinear least squares procedure that optimizes
both the harmonic parameters and the pulsation period separately in each window.
The result is a time series of the best-fit model parameters $\theta(\tau_i)$ with
corresponding point-wise confidence bands, where
$\theta(\tau_i)$ can stand for any of the pulsation period, harmonic parameters,
or peak-to-peak amplitude. We estimate both approximate
theoretical point-wise confidence bands and ones based on a Monte Carlo
experiment; the two were in general very close to each other.

However, the weighting scheme in this estimation procedure is chosen in a special way, to put emphasis on the
process near the central gridpoint. To decrease bias there, we combine the usual
error weighting with the kernel weights of local modelling: inverse variances
(as given by the square of the photometric uncertainty) were multiplied with a
factor derived from a normal density centered at $\tau_i$ with a standard
deviation of 182.5 days  (at the ends of the survey timespan, the definition remains the same; we do not lengthen the series by the addition of artificial points based on some rule of extrapolation from the observed values). Doing so attributes higher influence to observations
made closer to $\tau_i$ (important not to oversmooth, and to trace
irregularities better), while also weighting data according to their
reliability. The result is a weighting scheme that increases the impact of the
most relevant and most reliable observations, while it reduces the high
correlation observed between estimates in neighboring windows when using simple
error weighting.

In addition to its flexibility, the local sliding window model has a few more advantages over complex global models with constant parameters. Since it is local, abrupt shifts in mean magnitude due to calibration effects between {\it OGLE-II} and {\it -III} data affect it only in windows including the time of the shift, and thanks to the weighting scheme, the effect is attenuated in windows centred relatively far from this time. 
Moreover, the third-order polynomial component in model \eqref{eq:localmodel} accounts for any calibration residuals (between {\it OGLE-II} and III). 
%%%% Moreover, since model \eqref{eq:localmodel} includes also a third-order polynomial component, it can partly account for shifts, especially because there are no observations in the gap around the date of the possible shift, and therefore no precise modelling is needed. 
In the absence of information in the sampling gaps, the estimates of course may be biased or have a higher than average variance. 

The finite window size has two main effects on the estimates. First, the detectable timescales of period and amplitude modulations of the Cepheids
are determined by the 12-year total survey timespan and the window size. Periods longer than approximately the full timespan cannot be
distinguished from trends, so $\sim 12$ years is the long-period limit of
modulation cycles we can identify. In the short-period limit, any fluctuations
on timescales shorter than about 2\,years are smoothed out due to our use of sliding windows giving high weight to observations within a 2\,year time interval. Second, the faster the modulation, the more downward biased (underestimated) the estimated modulation amplitude. This bias is due to the local estimate being a kind of average value within the temporal sliding window. It is %%%%  because within the window, the local estimate is a sort of average value. 
proportional to the characteristic frequency of the modulation and can be estimated in a simple way if this latter is known or estimated. We discuss this and provide an empirical bias correction formula in  Appendix \ref{app:bias}.

Bias is also present at the start and end of the observation period, if the fitted parameters are not approximately constant there, since we lack information in the subinterval of the window that stretches over the end. This bias is strongest at the endpoints, then decrease as the window includes more data, and vanishes at 1.5 years from the ends (the half-width of the window). We investigated the effect of the sampling gaps and the finite observation timespan using simulations, and found this to be a minor effect, although sampling gaps can indeed cause some systematic distortions in the estimates of an underlying trendlike and oscillatory pattern. Moreover, data gaps do not lead to false detections in the absence of modulation, cf. App.~\ref{app:sim} for a detailed discussion. 

To assess whether a constant model sufficiently explains the observed
photometric time series, we repeat our sliding windows analysis on a simulated,
perfectly repetitive stable reference model using the best-fit constant
parameters from a global light curve modeling. Confidence bands are added to
this curve by applying a non-parametric Monte Carlo resampling based on the
residuals\footnote{The probability levels of the found residuals $r_i$ at time
$t_i$ are computed with respect to a Gaussian $\mathcal{N}(0,\sigma_i)$, where
$\sigma_i$ is the error at $t_i$. These probability levels are then resampled
with repetition, and corresponding repetition residuals are computed with
respect to the error bar at their new location. Due to the potential overdispersion
originating possibly in various effects as well as in time-dependent variations, these confidence intervals are very conservative, and provide a very careful, strict estimation of
significance.}.
The estimated functions $\theta(\tau_i)$ are then compared to the obtained
confidence bands. In order to assess significance of departures from the
constant parameter values, we employ the multiple hypothesis testing procedure
of \citet{benjaminiyekutieli01} to avoid spurious detections due to random
fluctuations amplified by strong correlation between neighboring windows.

The fitted magnitudes from the sliding window method enable an improved
pre-whitening. We approximate the noise-free magnitude of the star at time $t_i$
by a weighted average of the fitted value from the models at the two closest
window, with the weights based on the differences between $t_i$ and the window
centres. We compute the improved residuals by subtracting these fitted values
from the observed magnitudes. Then, we perform a secondary period search
\citep{zechmeisterkurster09} on these improved residuals, and we check whether
twin peaks are still apparent, or if other, weak secondary modes appear.

In summary, the local kernel modeling method presented here does not impose
assumptions on the type of period changes encountered and simultaneously traces
temporal variations in pulsation period and light curve shape. This is an
improvement over the $O-C$ technique, which does not allow to fully disentangle
fluctuations due to period and Fourier amplitude changes. While the use of
sliding windows represents a limitation on the time resolution, it provides
tools to attribute statistical significance to the time-variation of signals,
and to obtain a coherent, continuous-time picture of the simultaneous variations
of the period and Fourier composition.

A detailed description of the fitted local model, the model selection, the stable reference model, the error analysis and the assessment of significance taking into account the correlation caused by the overlap of the windows are given in Appendix \ref{app:detailedmethods}. We complement this with realistic simulated examples illustrating the detection power of the model and the limitations imposed by the {\it OGLE} time sampling and the kernel size in Appendix \ref{app:sim}, using real {\it OGLE} observation times and light curve profiles, trends and fluctuations similar to those found in our Cepheid sample. Appendix \ref{app:bias} considers the bias of the method and provides a means of  correcting for it.

The analysis presented in
the paper was performed using the statistical computing environment R \citep{R}. 

\section{Results}\label{sec:Results}

%\subsection{Phenomenology of the time variations}
% latex table generated in R 3.2.2 by xtable 1.7-4 package
% Mon Dec 21 20:27:19 2015

\begin{table*}[!ht] \centering  %\resizebox{\linewidth}{!}{
\begin{tabular}{lrrrrrrrrrrr}
  \hline
% \rule{0mm}{4mm} CEP & $I^{\mathrm{(cat)}}[mag]$ & $DI[mag]$ & $P^{\mathrm{(cat)}}[d]$ & $\sigma_{P^{\mathrm{(cat)}}}[d]$ & $DP[d]$ & $T_P[\%]$ & $A^{\mathrm{(cat)}}[mmag]$ & $A_1[mmag]$ & $\sigma_{A_1}[mmag]$ & $DA_1[mmag]$ & $T_{A_1}[\%]$ \\ 

\rule{0mm}{4mm} CEP & $I^{\mathrm{(cat)}}$ & $DI$ & $P^{\mathrm{(cat)}}$ & $\sigma_{P^{\mathrm{(cat)}}}$ & $DP$ & $T_P$ & $A^{\mathrm{(cat)}}$ & $A_1$ & $\sigma_{A_1}$ & $DA_1$ & $T_{A_1}$ \\ 
% \rule{0mm}{4mm} & mag & mag & $d$ & $10^{-7}d$ & $10^{-7}d$ & $\%$ & mmag & mmag & mmag & mmag & \% \\ 

  \hline
\multicolumn{12}{c}{\smallskip{\bf\rule{0mm}{4mm} FU Cepheids with twin peaks}} \\
  1140 & 13.910 & 0.008 & 8.1864915 & 94 & 5920 & 0 & 495 & 220.2 & 0.3 & 7.1 & 22 \\ 
  1418 & 15.518 & 0.008 & 2.8957105 & 27 & 2317 & 1 & 274 & 120.7 & 0.3 & 3.4 & 0 \\ 
  1621 & 15.138 & 0.017 & 3.3588290 & 19 & 5930 & 59 & 544 & 218.3 & 0.5 & 7.1 & 0 \\ 
  1748 & 14.267 & 0.007 & 6.6320016 & 53 & 7411 & 10 & 547 & 224.1 & 0.3 & 9.4 & 76 \\ 
  1833 & 13.091 & 0.033 & 19.1635310 & 587 & 92244 & 20 & 575 & 267.9 & 0.8 & 9.6 & 0 \\ 
  1932 & 15.171 & 0.019 & 2.9467014 & 16 & 5484 & 74 & 532 & 208.3 & 0.6 & 11.3 & 0 \\ 
  1998 & 15.267 & 0.015 & 3.0863817 & 15 & 1504 & 25 & 519 & 208.3 & 0.4 & 4.7 & 0 \\ 
  2132 & 14.634 & 0.011 & 4.6823873 & 74 & 19867 & 68 & 402 & 163.8 & 0.9 & 4.3 & 0 \\ 
  2180 & 15.495 & 0.010 & 2.5877856 & 9 & 2674 & 56 & 563 & 217.4 & 0.4 & 7.5 & 0 \\ 
  2191 & 14.819 & 0.019 & 4.2070310 & 19 & 4997 & 69 & 619 & 235.3 & 0.4 & 9.2 & 0 \\ 
  2289 & 15.155 & 0.011 & 3.7126964 & 17 & 2436 & 19 & 470 & 191.6 & 0.3 & 6.3 & 0 \\ 
  2470 & 15.291 & 0.008 & 3.2211085 & 19 & 2478 & 31 & 555 & 211.8 & 0.5 & 6.5 & 0 \\ 
\multicolumn{12}{c}{\smallskip{\bf FO Cepheids with twin peaks}} \\
  1405 & 14.857 & 0.012 & 2.7807164 & 57 & 8209 & 33 & 217 & 107.3 & 0.6 & 13.1 & 15 \\ 
  1521 & 14.680 & 0.013 & 3.2147769 & 97 & 16927 & 61 & 184 & 89.7 & 0.7 & 8.4 & 0 \\ 
  1527 & 15.830 & 0.019 & 1.4923973 & 52 & 8598 & 52 & 136 & 64.7 & 1.1 & 34.8 & 55 \\ 
  1535* & 15.928 & 0.036 & 1.2509786 & 42 & 6899 & 31 & 139 & 63.7 & 1.4 & 59.4 & 46 \\ 
  1536 & 15.503 & 0.009 & 1.6112568 & 68 & 10112 & 0 & 54 & 25.1 & 0.5 & 27.5 & 73 \\ 
  1561 & 15.813 & 0.028 & 1.5299801 & 12 & 1746 & 27 & 264 & 128.2 & 0.5 & 9.1 & 5 \\ 
  1564 & 15.362 & 0.009 & 2.0629824 & 25 & 1920 & 0 & 152 & 74.6 & 0.4 & 2.1 & 0 \\ 
  1605 & 14.601 & 0.010 & 3.0881609 & 31 & 4351 & 35 & 236 & 114.5 & 0.3 & 5.7 & 14 \\ 
  1693 & 16.141 & 0.017 & 1.6639538 & 18 & 1841 & 0 & 174 & 85.3 & 0.5 & 7.0 & 0 \\ 
  1704 & 15.376 & 0.010 & 1.8820681 & 49 & 7939 & 57 & 180 & 89.2 & 1.0 & 5.3 & 0 \\ 
  2118 & 15.060 & 0.009 & 2.4461347 & 26 & 3158 & 32 & 176 & 87.0 & 0.3 & 4.4 & 0 \\ 
  2217 & 14.895 & 0.008 & 2.3120543 & 187 & 27774 & 35 & 133 & 63.4 & 1.9 & 6.0 & 101 \\ 
\multicolumn{12}{c}{\smallskip{\bf FO Cepheids with changing amplitudes}} \\
  0916 & 16.062 & 0.016 & 1.3695732 & 67 & 6725 & 26 & 88 & 42.5 & 1.1 & 53.3 & 58 \\ 
  1119* & 15.617 & 0.010 & 1.7431985 & 119 & 19536 & 59 & 76 & 35.7 & 1.0 & 51.5 & 68 \\ 
  1275* & 16.417 & 0.009 & 0.9474017 & 33 & 5796 & 48 & 92 & 43.9 & 1.3 & 45.0 & 69 \\ 
  1955 & 15.668 & 0.016 & 1.8879224 & 47 & 5830 & 18 & 109 & 53.0 & 0.6 & 24.8 & 78 \\ 
  2820 & 16.119 & 0.034 & 2.1127656 & 39 & 4134 & 45 & 230 & 110.9 & 0.7 & 16.8 & 34 \\ 
  \hline
\end{tabular}%}
\caption{Basic parameters and their variation for the twin-peak and amplitude-changing Cepheid sample. The columns of the table are the following: CEP: Cepheid identifier (without the common prefix OGLE-LMC-CEP-); $I^{\mathrm{(cat)}}$: mean $I$ magnitude from the {\it OGLE-III} catalog; $DI$: maximal span $\max I_0(t) - \min I_0(t)$ of the variation of the kernel-estimated mean magnitude $I_0(t)$; $P^{\mathrm{(cat)}}$: primary pulsation period from the {\it OGLE-III} catalog (in days); $\sigma_{P^{\mathrm{(cat)}}}$: the error of the primary pulsation period from the {\it OGLE-III} catalog (in $10^{-7}$ days); $DP$: maximal span of the variation of the primary pulsation period $\max P(t) - \min P(t)$ from the kernel fits (in $10^{-7}$ days); $T_P$: length of time intervals exhibiting significant deviation from $P^{\mathrm{(cat)}}$, in percentage of the total time span; $A^{\mathrm{(cat)}}$:  peak-to-peak amplitude from the {\it OGLE-III} catalog (in millimagnitudes); $A_1$: estimated amplitude of the first harmonic term from the average global fit using the fixed period $P^{\mathrm{(cat)}}$ (in millimagnitudes); $\sigma_{A_1}$: standard error of the former (in millimagnitudes); $DA_1$: maximal span of the variation of the first harmonic amplitude $\max A_1(t) - \min A_1(t)$ from the kernel fits (in millimagnitudes); $T_{A_1}$: length of time intervals exhibiting significant deviation from $A_1$, in percentage of the total time span. Asterisk beside the identifier indicates the three stars which are commented in the {\it OGLE-III} catalog as having variable amplitude. }
\label{tab:deviationintervals_T}
\end{table*}

\begin{table*}[!ht] \centering  %\resizebox{\linewidth}{!}{
\begin{tabular}{lrrrrrrrrrrr}
  \hline
% \rule{0mm}{4mm} CEP & $I^{\mathrm{(cat)}}[mag]$ & $DI[mag]$ & $P^{\mathrm{(cat)}}[d]$ & $\sigma_{P^{\mathrm{(cat)}}}[d]$ & $DP[d]$ & $T_P[\%]$ & $A^{\mathrm{(cat)}}[mmag]$ & $A_1[mmag]$ & $\sigma_{A_1}[mmag]$ & $DA_1[mmag]$ & $T_{A_1}[\%]$ \\ 
\rule{0mm}{4mm} CEP & $I^{\mathrm{(cat)}}$ & $DI$ & $P^{\mathrm{(cat)}}$ & $\sigma_{P^{\mathrm{(cat)}}}$ & $DP$ & $T_P$ & $A^{\mathrm{(cat)}}$ & $A_1$ & $\sigma_{A_1}$ & $DA_1$ & $T_{A_1}$ \\ 
% \rule{0mm}{4mm} & mag & mag & $d$ & $10^{-7}d$ & $10^{-7}d$ & $\%$ & mmag & mmag & mmag & mmag & \% \\ 
  \hline
\multicolumn{12}{c}{\smallskip{\bf FU Cepheids without twin peaks}} \\
  0727 & 13.251 & 0.012 & 14.4891397 & 273 & 20508 & 4 & 630 & 240.3 & 0.4 & 3.7 & 0 \\ 
  1514 & 15.442 & 0.008 & 3.0459183 & 11 & 1324 & 0 & 487 & 187.0 & 0.3 & 4.7 & 0 \\ 
  1543 & 14.972 & 0.006 & 4.3781838 & 71 & 5302 & 0 & 178 & 83.2 & 0.3 & 2.9 & 0 \\ 
  1559 & 15.141 & 0.010 & 3.6469543 & 19 & 1351 & 0 & 452 & 181.1 & 0.3 & 4.4 & 0 \\ 
  1711 & 14.677 & 0.012 & 2.9189534 & 47 & 2080 & 0 & 122 & 51.8 & 0.3 & 4.8 & 0 \\ 
  1753 & 15.573 & 0.021 & 2.5745626 & 14 & 970 & 0 & 393 & 158.4 & 0.4 & 5.0 & 0 \\ 
  2215 & 15.354 & 0.009 & 3.2408617 & 10 & 1766 & 6 & 486 & 193.7 & 0.3 & 4.5 & 0 \\ 
  2229 & 14.724 & 0.013 & 4.8745455 & 19 & 1144 & 0 & 545 & 221.8 & 0.2 & 3.8 & 0 \\ 
  2264 & 15.240 & 0.007 & 3.8808875 & 30 & 2677 & 0 & 253 & 116.8 & 0.3 & 5.6 & 0 \\ 
  2500 & 15.226 & 0.010 & 3.2002470 & 15 & 1551 & 0 & 454 & 185.2 & 0.3 & 3.5 & 0 \\ 
  2580 & 15.439 & 0.009 & 2.9569833 & 28 & 1091 & 4 & 440 & 183.3 & 0.4 & 3.3 & 0 \\ 
  2774 & 15.733 & 0.012 & 3.8131712 & 60 & 3133 & 0 & 297 & 127.5 & 0.5 & 6.1 & 0 \\ 
\multicolumn{12}{c}{\smallskip{\bf FO Cepheids without twin peaks}} \\
  1582 & 18.296 & 0.033 & 0.3187867 & 5 & 393 & 0 & 81 & 41.0 & 1.5 & 15.5 & 0 \\ 
  1638 & 15.222 & 0.009 & 2.3630077 & 174 & 17110 & 0 & 21 & 10.5 & 0.3 & 3.7 & 0 \\ 
  1671 & 15.824 & 0.010 & 1.3642921 & 13 & 924 & 0 & 125 & 59.7 & 0.3 & 4.4 & 0 \\ 
  1698 & 15.828 & 0.015 & 1.5955712 & 15 & 1293 & 0 & 175 & 83.8 & 0.4 & 3.7 & 0 \\ 
  1754 & 16.871 & 0.006 & 0.7113208 & 4 & 364 & 0 & 185 & 84.2 & 0.5 & 6.9 & 0 \\ 
  1772 & 16.442 & 0.009 & 0.9471192 & 6 & 382 & 0 & 175 & 85.1 & 0.4 & 5.6 & 0 \\ 
  1838 & 15.648 & 0.008 & 1.6431880 & 16 & 795 & 0 & 124 & 61.2 & 0.3 & 2.1 & 0 \\ 
  1911 & 16.714 & 0.022 & 0.7485379 & 4 & 327 & 0 & 204 & 96.8 & 0.6 & 10.2 & 5 \\ 
  1937 & 16.364 & 0.013 & 1.0593254 & 7 & 375 & 0 & 188 & 90.6 & 0.5 & 3.9 & 0 \\ 
  2103 & 18.051 & 0.030 & 0.3913544 & 8 & 812 & 0 & 76 & 36.5 & 1.3 & 21.3 & 0 \\ 
  2310 & 16.396 & 0.013 & 1.2713288 & 10 & 1034 & 0 & 177 & 86.2 & 0.5 & 9.2 & 0 \\ 
  2977 & 17.476 & 0.018 & 0.4956663 & 8 & 492 & 0 & 92 & 46.1 & 1.0 & 11.5 & 0 \\ 
   \hline
\end{tabular}%}
\caption{Basic parameters and their variation for the control sample. The columns are the same as those of Table \ref{tab:deviationintervals_T}.} 
\label{tab:deviationintervals_C}
\end{table*}

%\subsubsection{Period changes} \label{subsubsec:perch}

\subsection{Period changes} \label{subsubsec:perch}

\begin{figure*}
%\sidecaption[t]
\begin{center}
\includegraphics[width=\textwidth]{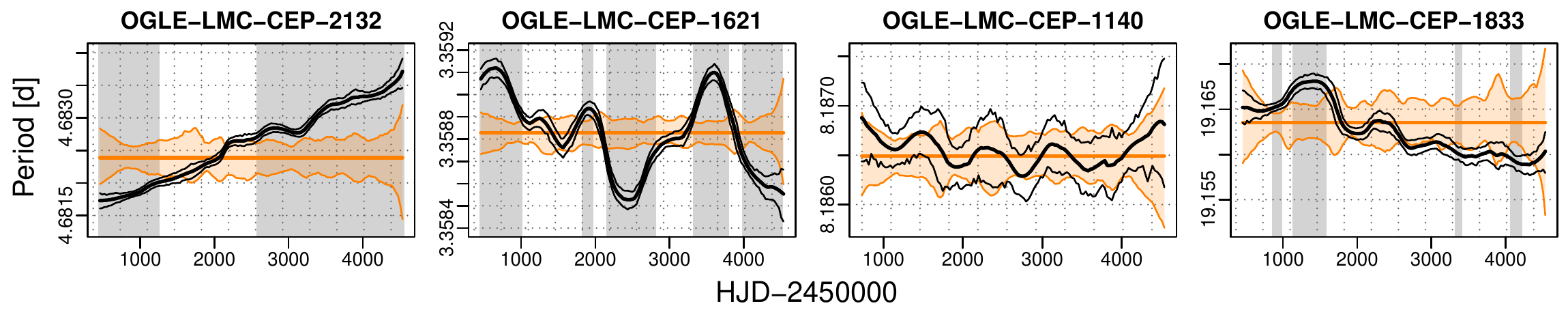}
\caption{Examples for the basic types of period modulation found among the Cepheids.  The plots show the kernel-estimated pulsation period (in days) versus Julian Date (in days), plotted as a solid thick black line, together with its bootstrapped pointwise CI (thin black lines; see Section \ref{sec:slidingwindows}). The heavy orange line denotes the catalog period, which was used as known (non-optimized) value in the fitted stable reference model for the Cepheid. The orange band indicates the nonparametric bootstrap CI around this period, obtained from the procedure described in Section \ref{sec:slidingwindows}. The dotted black horizontal and vertical lines are aids to the eye to estimate the extent and time interval of the changes. The grey background highlights time intervals where the deviation from the stable reference model was found significant by a multiple testing procedure \citep{benjaminiyekutieli01}. Leftmost panel: near-linear trend, middle left: stochastic fluctuations, middle right: quasi-periodic instabilities, rightmost: combination of a trend and weakening quasi-periodic changes.}
\label{fig:sample_period}     
\end{center}
\end{figure*}

Figures~\ref{fig:twins_all_period} and \ref{fig:control_all_period} in the
Appendix give an overview of the variations in pulsation period estimated by the
sliding window technique, as compared to a stable reference Cepheid simulated using the best stable parameter estimates. %The black solid lines indicate the period estimates from the observed magnitudes. The solid orange line indicates the stable reference period (equal to the catalog period), with the orange band being its confidence band (see section \ref{sec:slidingwindows}). The grey shading highlights time intervals of departures of the period from its average value which were found significant by the multiple hypothesis testing \citep[see Sec. \ref{sec:slidingwindows};][]{benjaminiyekutieli01}. 
The sixth column of Tables \ref{tab:deviationintervals_T} and \ref{tab:deviationintervals_C} gives the maximal span of these changes, that is, $\max P(t) - \min P(t)$, while the seventh column
gives the length of these intervals in percentage of the total observation span.    

The figures suggest a continuous broad range of possible variations potentially
extending to many or even all Cepheids, rather than separate subclasses of steady and
non-steady pulsators.  Within the timescale limitations mentioned at the end of
Section \ref{sec:slidingwindows} and discussed in Appendix \ref{app:sim}, the range of the variations may include slow irregular or near-linear trends,
% (examples are the FO CEP-1405 and the FU CEP-2132, respectively) , 
stochastic or multi-scale oscillations,
% (e.g. FO CEP-1561 or FU CEP-1621) , 
quasi-regular changes, 
% (FU CEP-1140) , 
and any combinations of these.
% (e.g. FOs CEP-1704 \& CEP-2118 and FU CEP-1833) .
Figure~\ref{fig:sample_period} presents an example of each (CEP-2132:
near-linear trend; CEP-1621: irregular fluctuations, CEP-1140: quasi-periodic changes;
CEP-1833: combination of damped quasi-periodic changes and near-linear trend). 
 Appendix \ref{app:sim} suggests that though the {\it OGLE } time sampling can cause systematic distortions on the shape of the estimate of a real modulation pattern (see Appendix \ref{app:sim}), it cannot cause the observed phenomena. Neither the trends of CEP-2132 and 1833, nor the large fluctuations of CEP-1621 can be fully explained in this way. Considering the simulation results in Appendix \ref{app:sim} and \ref{app:bias}, the non-significant quasi-periodic modulation in CEP-1140 is much more likely to be attributable to a fast oscillation around the upper frequency detection limit of our window than to noise or effect of sampling gaps (the weak quadratic-like trend may be the consequence of end effects, though these are more likely to cause estimates levelling out than a sharp increase or decrease as here). The exceptionally large scatter of the estimates of the modulation frequency and amplitude presented in Appendix \ref{app:bias} in the case of CEP-1833 warrants prudence in accepting the shape of its modulation, but the existence of a trend seems to be certain and some additional instabilities very likely.

The wide variety of different combinations of these relatively pure
types can be appreciated in Figures~\ref{fig:twins_all_period} and
\ref{fig:control_all_period} among our Cepheid sample. The intervals of deviation from a stable model are much more frequent and last much longer among
twin-peak Cepheids (Figure~\ref{fig:twins_all_period}) than in the control group
(Figure~\ref{fig:control_all_period}), as emphasized by the extent of
grey-shaded areas on the figures, and shown by $T_P$ in Tables
\ref{tab:deviationintervals_T} and \ref{tab:deviationintervals_C}. This is so for both FU and FO Cepheids.

 %We note that period variations are much more common than amplitude variations.
 % For instance, while amplitude modulations are rare among the control group,
 % we do find fast and significant fluctuations of pulsation periods (e.g.
 % CEP-2264). \msucomm{Amplitudes will be mentioned in the next subsection, it's
 % early here.}     

Among the variety of modulation types found (trend-like, stochastic,
oscillation-like or arbitrary combination), visual inspection suggests a
relatively strong trend component for
CEP-1521, CEP-1527, CEP-1704, CEP-2217 (all FOs), and CEP-2132 (FU). Similar
trends are visually less obvious, but potentially present for stars CEP-1405
(FO), CEP-1833 and CEP-2470 (both FUs). For most stars, this trend appears
non-linear, or has added fluctuating components (stochastic or
oscillation-like); almost pure near-linear patterns are shown only by CEP-2132
and CEP-1521, the latter nevertheless with some increasing oscillations towards
the end of the observation period (where end effects may also be present). The strongest trend is observed for CEP-1833: its frequency change within the decade-long {\it OGLE-III} timespan is about 0.008 c/d. Given the
several different types of period changes seen here, however, the observational
baseline  may not be sufficient to ascertain that these period changes are truly
caused by secular evolution \citep[cf. also][]{soszynskietal08}: trends observed
on such short timescales may actually prove to be a portion of fluctuations with
a long characteristic timescale. The frequent presence of relatively strong
fluctuations on various timescales further strengthens this impression.

For four twin peak Cepheids (FU CEP-1140, FO CEP-1536, CEP-1564 and CEP-1693,
the multiple testing procedure did not find significant period changes ($T_P =
0$ in Table \ref{tab:deviationintervals_T}, and thus, instability of the pulsation
period is not  confirmed in these
stars. CEP-1418 also has only a very short interval of period deviation.  A strong instability on a shorter timescale than the sliding window length could cause remaining periodicity after removal of the (average) primary frequency. Although the frequency separation between the primary and the secondary periodogram peaks, which is commonly used as an indicator of the modulation's typical timescale, does not suggest a high-frequency periodic modulation, the sliding window estimates in Figure~\ref{fig:twins_all_period} suggest indeed fast (low-level) variations for these stars. The strength of such fast variations are strongly underestimated by our 3-year window (see Appendix \ref{app:sim}), so it is possible that these stars do have a high-amplitude, fast period oscillation producing perceivable traces in the secondary periodogram.
In addition, CEP-1140 and CEP-1536 have long intervals when their amplitude deviates
significantly from the mean value, which offers an alternative explanation.
A look at the period changes of CEP-1693 and CEP-1418 in Figure
\ref{fig:twins_all_period} reveals a  slow, weak nonlinear trend (together with some
comparatively strong oscillations) in their pulsation period. Although this is
not significant with the small number of observations used per window and with
our very conservative error assessment procedure, it may be sufficient to give
rise to a twin peak when trying to pre-whiten by a constant model fitted to all
data. For the fifth star, CEP-1564, we find no significant deviations from
stable pulsations using our multiple hypothesis testing procedure. % We discuss this star further in section \ref{subsec:resper}.

Within the control group, our results suggest identical variation types except
for trends discernible by eye,  and on average faster quasi-periodic or stochastic changes. The variations appear on average milder than those of the twin-peak group, as shown by the values of $DP$ in tables \ref{tab:deviationintervals_T} and \ref{tab:deviationintervals_C}. The residual-based significance assessment finds these statistically non-significant, so we cannot exclude a noise origin of these modulations. However, as mentioned above, %% and illustrated in Appendix \ref{app:sim},  
variations on timescales comparable to or shorter than the window length are generally under-estimated by the kernel method. %The inspection of the $P(t)$ curves of the control stars in Figure \ref{fig:control_all_period} suggests relatively fast quasi-periodic or stochastic changes with high amplitude, without any noticeable slow trend contribution. 
Fast and strong cyclic variations can cause over-dispersion in the residuals at all phases in the secondary light curve. If these are fast enough with respect to the limitations of the kernel method, they might blur the pattern in the residual light curves seen in the center panels of Fig \ref{fig:twinpgrams}. As a consequence of this strong general over-dispersion, our conservative residual-based significance assessment procedure (cf. Section \ref{sec:slidingwindows} and Appendix \ref{app:detailedmethods}) yields a very broad confidence band around the $P_{\mathrm{cat}}$ value of the constant model, and thus, we find the modulation to be non-significant. It follows that the twin-peak phenomenon, though it seems to be a good indicator of changes that are trend-like on the survey timespan, can fail to indicate even strong oscillatory modulations in the pulsation period, and thus miss many potentially scientifically interesting cases. Two such cases merit to be mentioned, that of CEP-0727 (FU) and CEP-1638 (FO). Their secondary periodogram does not show a twin peak, however, their $DP$ values in Table \ref{tab:deviationintervals_C} are 0.002 $d$ and 0.0017 $d$, respectively, both among the highest in all our sample.

%\subsubsection{Amplitude changes} \label{subsubsec:ampch}

\subsection{Amplitude changes} \label{subsubsec:ampch}

We find much fewer stars to exhibit significant variations in the amplitude of
the leading harmonic term, $A_1 = \left(s_1^2 + c_1^2\right) ^{1/2}$  (cf. equation \eqref{eq:localmodel}, than in
period. Figures \ref{fig:twins_all_A1} and \ref{fig:control_all_A1} present the estimated $A_1(t)$ curves for all stars. Tables \ref{tab:deviationintervals_T} and \ref{tab:deviationintervals_C} summarise the maximum span $\max A_1(t) - \min A_1(t)$ of the first harmonic amplitude $A_1(t)$ from the kernel fits, and the extent of the time intervals of
significant deviations from the average $A_1$ in its last columns $DA_1$ and $T_{A_1}$. It appears indeed that the variations in the
pulsation period are easier to detect, and for this reason they have been
discussed in the literature much more extensively. Nevertheless, we find
significant, more or less cyclic, amplitude variations for the twin peak FO
Cepheids CEP-1527, CEP-1535 and CEP-1536, in addition to those five Cepheids  included because of their clearly visible strong amplitude changes. Several other
FO Cepheids (CEP-1405, -1561 and -1605) exhibit shorter, erratic excursions from otherwise
fairly stable pulsation amplitudes. With the exception of CEP-1536, these stars
also show changes in pulsation period. There are two FU stars as well with
significant amplitude changes, CEP-1748 and CEP-1140, but contrary to the
majority of the FO sample with amplitude modulations, these stars do not undergo
significant period changes, as discussed in Section \ref{subsubsec:perch}. Our method
therefore successfully recovers all Cepheids identified as having variable
amplitudes in the {\it OGLE } catalog and even increases this number.

Long-term, trend-like changes (within the detectability limits of the 12-year
survey) seem to be very rare. Two stars that show a pattern compatible with it
are the FO Cepheids CEP-1561 and -1693. However, even for those, slow
fluctuations around a mean amplitude which is stable on the long-term cannot be
excluded. One FO Cepheid, CEP-2217, has a significantly higher first harmonic
amplitude with the sliding window estimation than with the constant reference
model. This is due to the fact that its strong trend-like period variation causes
strong phase shifts of temporally distant observations, and smears the light curve folded with a constant period.

Our results suggest that small amplitude variations may be present for many more
Cepheids, although these amplitude fluctuations tend to stay below the $95\%$
confidence level with {\it OGLE-III} time cadences and our conservative significance assessment procedure. Taking
this into account, we interpret our results as an indication that amplitude
fluctuations on millimagnitude level and below are a common, possibly ubiquitous, phenomenon whose detectability
is currently limited by the availability of sufficient time resolution and photometric precision.

%\subsubsection{Changes in light curve shape}  \label{subsubsec:lcch}

\subsection{Changes in light curve shape}  \label{subsubsec:lcch}

The morphology of light curves is most commonly described using
relative amplitudes and phases, $\Phi_{j1}$ and $R_{j1}$ for
the leading few harmonics, most importantly for $j = 2,3$ \citep{1981ApJ...248..291S}. However, the continuous-time changes of these parameters are too weak to consider using these quantities, since the variations in the second and third harmonic amplitudes are below the detection limit.

Nevertheless, it is possible to visually compare light curve shapes at different epochs, e.g. where the peak-to-peak amplitudes of the Cepheids are 
very different. Figures \ref{fig:twins_fu_lc}--\ref{fig:control_fo_lc} show the
light curves of the Cepheids in our sample at two such epochs selected
individually for each star, such that furthermore the windows do not overlap. We
plot the folded light curves so that maximum brightness occurs at phase 0.6 to
facilitate visual comparison. Confidence bands are added based on bootstrap
repetitions generated with resampled errors superposed to the reconstructed
estimated local light curve.

The fundamental-mode Cepheids in our sample, both twin-peak and control,
exhibit little scatter in their light curves.
In quite a few cases, there are discernible but tiny-looking differences in the
minimum or maximum brightness or in the pattern of the brightening branch (in a
few cases, these can be due to outliers, e.g. CEP-2215 in Figure
\ref{fig:control_fu_lc}, or to unfortunate phase gaps in the data such as for
CEP-1932 in Figure\,\ref{fig:twins_fu_lc}). There are a few unusual cases, like that of CEP-1833, which has many
downward scattered observations in its more recent window, but none in its
earlier window; or those of CEP-1418 (Figure\,\ref{fig:twins_fu_lc}),
CEP-1543, CEP-1711, and CEP-2774  (Figure\,\ref{fig:control_fu_lc}), which seem
to have stronger overdispersion than other fundamental-mode Cepheids.
These stars have smaller amplitude ($< 0.3$ mag) than those with less noisy light curve (0.4 mag or above), so this can also be due to the relatively smaller magnitude span of the plots, or can indicate possible further variations, not adequately modeled by the 3-year sliding window estimate.

The variations exhibit a larger range for the overtone stars. Beside a
few cases that show light curve variations comparable to the fundamental-mode sample (often those with relatively high average amplitude), and CEP-1536 for which our procedure selected apparently an unnecessarily high harmonic order resulting in a wiggly fit (cf. section \ref{sec:slidingwindows}), there are many that show high dispersion of observed magnitudes together with strong light curve shape
variations. The dispersion of the observations affects the quality of the
estimated light curve shapes, as is shown by the broad confidence bands around
the estimates. In many cases, the observations with high residuals are far off
from the centre of the window in real observation time (light-shaded points are
more distant from the window centre in real observing time than dark-shaded
points). This suggests that relatively strong changes might occur on a shorter
timescale than the kernel length.

\subsection{Residual periodograms} \label{subsec:resper}

Our study presented so far led to the conclusion that for the Cepheids in our sample, the presence of a ``twin peak" in the secondary periodogram after pre-whitening is related to period and/or light curve shape modulations. Pre-whitening a light curve that intrinsically contains such modulations with a model of constant period and harmonic amplitudes will be obviously only approximate. Using instead the local estimates of period and Fourier amplitudes yields a more precise magnitude estimate at every observation time, and hence helps to remove the remnants of the main oscillation (the twin peak) from the secondary periodogram. Moreover, a secondary period search on the residuals of a local pre-whitening can be expected to detect weak secondary modes more efficiently and more precisely than a constant model. We thus constructed residual periodograms for all Cepheids in the sample by subtracting the time-dependent best-fit models.             

\begin{figure*}
\begin{center}
\includegraphics[scale=.21]{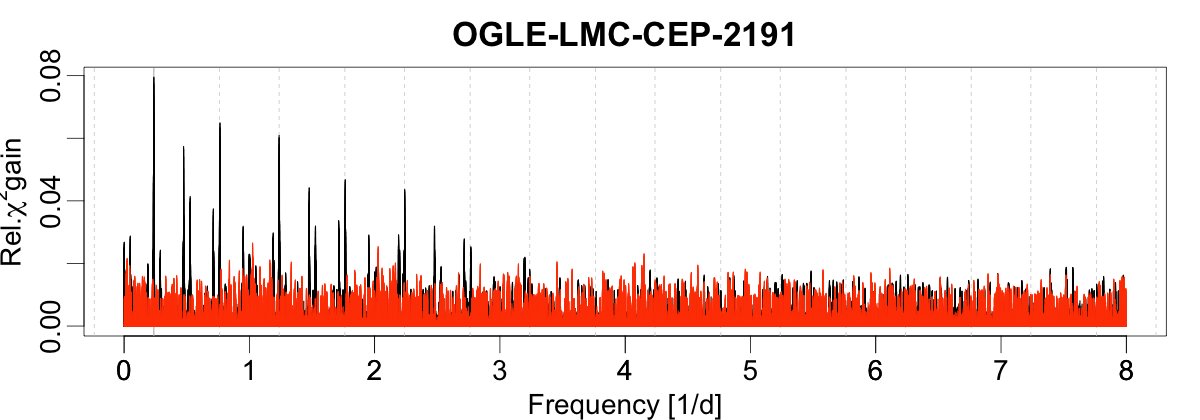}
\includegraphics[scale=.21]{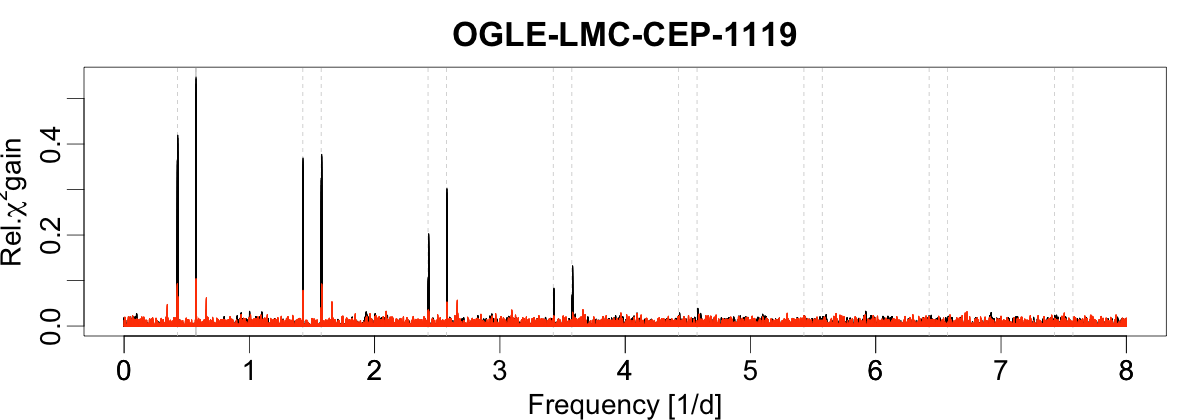}
\includegraphics[scale=.21]{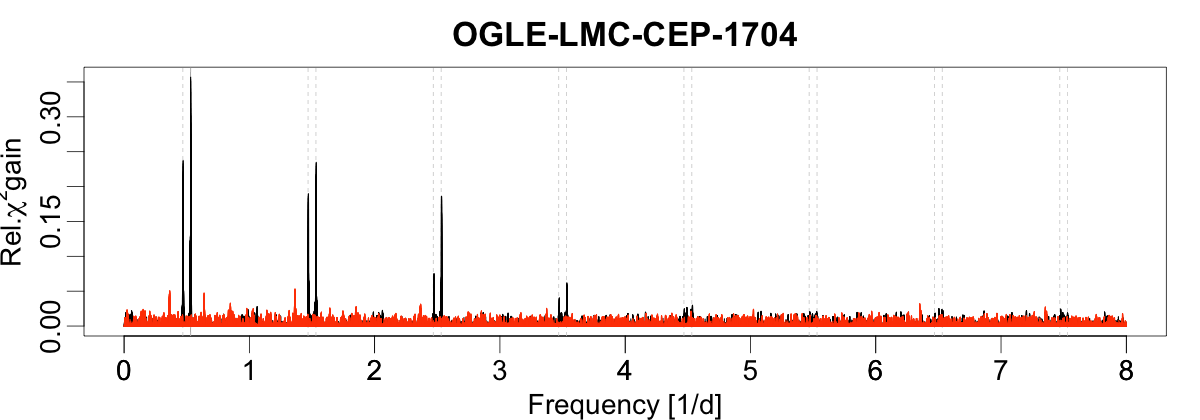}
\includegraphics[scale=.21]{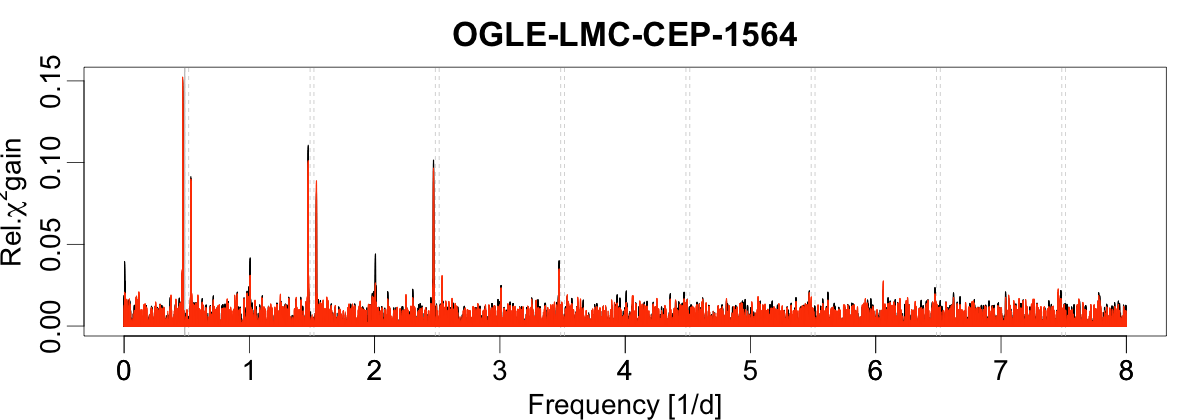}
\caption{Four examples of the effect of the sliding window smoothing on residual periodograms. The GLS periodogram of the residuals from the stable reference model is plotted in black. Twin peaks appear very close to the grey background vertical line that indicates the primary pulsation frequency. Daily aliases of the primary frequency are shown as dashed vertical grey lines. We superposed the periodogram of the residuals from the sliding window fit in red.}
\label{fig:residpgrams}     
\end{center}
\end{figure*}

\begin{figure}
\begin{center}
\includegraphics[scale=.57]{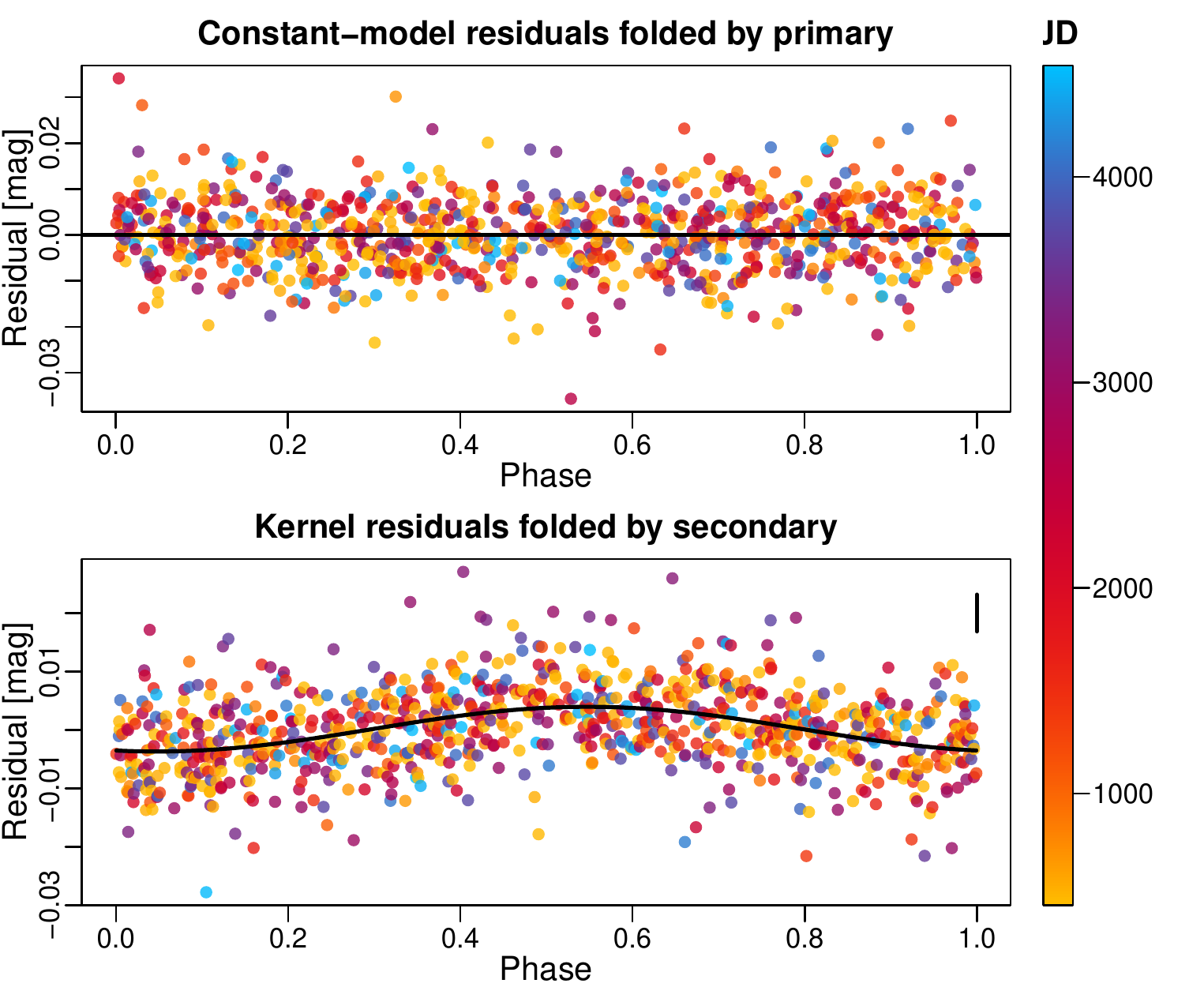}
\caption{Top panel: residuals of CEP-1564 from the constant model, phase-folded with the average primary period. Bottom panel: residuals of CEP-1564 from the kernel fits, phase-folded with the secondary period. The typical error bar is shown in the top right corner.} \label{fig:foldedres_1564}       % Give a unique label
\end{center}
\end{figure}

We compared the residual periodograms computed based on the stable reference fit with constant parameters over the full {\it OGLE-III} timespan to residual periodograms computed using the sliding window fit. In all but two of the 29 twin-peak cases, the sliding window fit removed perfectly the twin peak from the residual periodogram. Moreover, the method did not add any new spurious peaks to the residual periodograms of the 24 control Cepheids, and thus we can safely state that the method does not introduce any artefacts into the results. 

As expected, our more complex pre-whitening method yields more stable residual periodograms than pre-whitening with constant period and harmonic amplitudes. In all but 2 of the 29 Cepheids belonging to the twin peaks group (12 FU, 12 FO, and 5 additional FOs), we find no twin peaks after pre-whitening. Moreover, no new, spurious secondary periods are found in the residual periodograms of the 24 Cepheids belonging to control groups (12 FU, 12 FO). We present four typical examples of Cepheids in the twin peaks samples in Fig.\,\ref{fig:residpgrams}. 

Residual periodograms obtained by pre-whitening with time-dependent light curve parameters are generally uniform and flat for twin-peak FU, control FU, and control FO Cepheids. The top left panel of Figure\,\ref{fig:residpgrams} shows the secondary periodograms after constant pre-whitening (black) and after pre-whitening with local estimates (red) in such a case, for the FU twin-peak Cepheid CEP-2191. In some cases, a slight, albeit most likely non-significant maximum appears at approximately $0, 1, 2, \ldots c/d$. It cannot be decided whether these indicate a parasite frequency of terrestrial origin or a remaining mean trend which is detected by frequency analysis as a low-frequency signal. 

Twin-peak FO Cepheids exhibit more complex patterns in their residual periodograms. Only five of 17 show flat residual periodogram following pre-whitening with our time-dependent best-fit models. Four of these belong to the group of five FO Cepheids that we selected for their obviously changing amplitudes (CEP-0916, CEP-1275, CEP-1955, and CEP-2820), while CEP-1536 was an FO twin-peak group member. CEP-1119, the 5th Cepheid selected for obvious amplitude changes, still exhibits a twin peak after pre-whitening, cf. top right panel in Fig.\,\ref{fig:residpgrams}. This might be caused by imperfect fits in some windows: the sliding window period estimates in the bottom row of Figure \ref{fig:twins_all_period} suggest very sharp period changes that were not precisely fitted. Three other overtone Cepheids exhibit mostly flat residuals similar to those of FU Cepheids with (likely spurious) peaks near $0, 1, 2, \ldots c/d$. For two of them, this peak is weak, for the third (CEP-1535, one of the newly discovered amplitude-changing Cepheids), it is strong.

CEP-1564, shown in the bottom right panel of Figure\,\ref{fig:residpgrams}, represents an exception. This is the only FO twin-peak Cepheid in our analysis for which we find no significant variability of pulsation period ($P_1 \sim 2.063\,$d) and $A_1$ ($A_1 \sim 0.074\,$mag), despite a very prominent secondary peak at $P_{\mathrm{twin-peak}} = 2.141\,$d after pre-whitening with a constant model (see also the bottom right panel Figure \ref{fig:twincloseup}). Pre-whitening with a time-dependent model yields a secondary periodogram and a secondary frequency almost identical to the former one. Folding the residuals with the secondary period yields a scattered, diffuse sinusoidal light curve with a peak-to-peak amplitude of about 7.5 millimagnitudes shown in the bottom panel of Figure\,\ref{fig:foldedres_1564}. Since the period corresponding to the twin peak is relatively far off from the primary pulsation period (the relative difference is larger by a factor of ten than the largest of all other twin-peak stars), extremely large period variations could be expected for this star. However,  the residuals folded by the primary pulsation frequency lack any trace of the pattern which is typical of other twin-peak Cepheids (compare the upper panel of Figure\,\ref{fig:foldedres_1564} to the middle panel of Figure\,\ref{fig:twinpgrams}). Moreover, the kernel fits using a broad search interval for the local pulsation frequency failed to find any strong modulation, either trend-like or fluctuating one. The fits shown in Fig.~\ref{fig:twins_all_period} remained stable for a wide range of search intervals (including also the frequency of the twin peak). Possible explanations for the persistence of the secondary peak can be that it results from fast fluctuations around the detectability limit of our kernel, or that both frequencies are physical, i.e., the star is blended or is a physical binary with another variable star, or that this star is indeed pulsating in two close-by modes.

The residual periodograms of seven remaining FO twin-peak Cepheids exhibit weak secondary peaks similar to those presented in Figure\,\ref{fig:residpgrams}, in the bottom left panel. Their strength ranges from probably non-significant to probably significant. All but one of them are at lower frequencies (longer periods) than the primary pulsation frequency, and the ratio of the lower to the higher frequency ranges from $\sim 0.6$-$0.9$. A few cases of such secondary modes were found after pre-whitening using a stable model by \citet{soszynskietal08}. \citet{soszynskietal15b} reported 206 FO stars exhibiting secondary frequencies with ratios around 0.6 among a total of 3530 FO ($5.8\%$) Cepheids in both Magellanic clouds (82 in the LMC). Our study finds seven out of our  29 FO Cepheids ($24\%$) to show some indication of secondary peaks and spanning a broader interval of frequency ratios, which suggests that these frequencies, though weak, may be even more common than previously thought in overtone pulsators.

\section{Discussion}\label{sec:Discussion}

\subsection{Separation of trends and oscillatory terms}\label{sec:trends+fluctuations}

\subsubsection{Modeling time-dependent light curve parameters}

\begin{figure*}
%\sidecaption[t]
\begin{center}
\includegraphics[scale=.7]{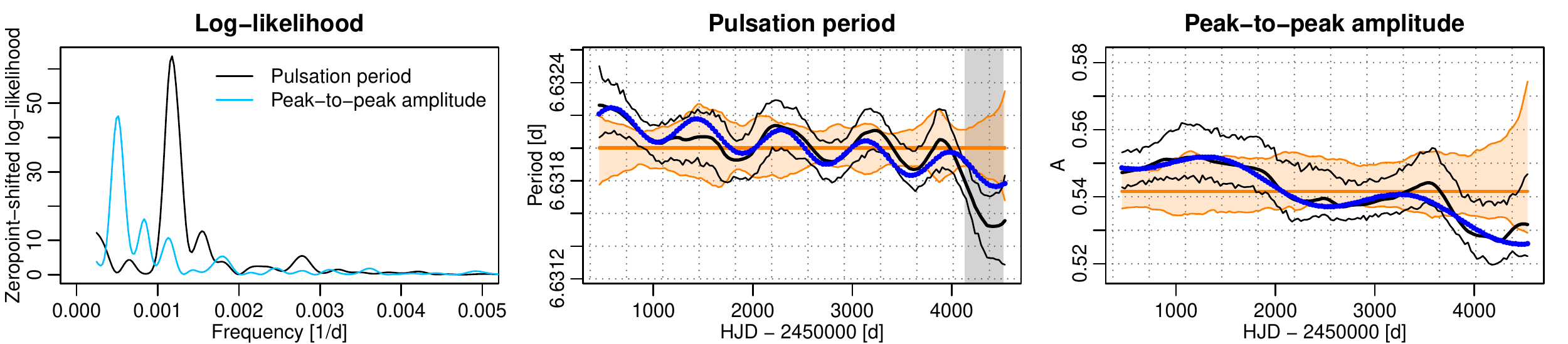}
\caption{CEP-1748 shown as example of the heuristic model
\eqref{eq:armodel} of the changes of the pulsation parameters. Left panel: the
log-likelihoods for the pulsation period (in black) and the peak-to-peak
amplitude (in blue; the log-likelihood functions were shifted to have common
minimum at zero). The other two panels show the fits using the frequency
yielding the highest likelihood (blue dots) against the sliding window estimates
(thick black line), their 95\% confidence interval (thin black lines), the
stable value from the reference fit (thick orange line) and its confidence band
(orange band). The middle panel presents the variations of the pulsation period,
the right panel those of the peak-to-peak amplitude.}
\label{fig:frvar_param1748}     
\end{center}
\end{figure*}

The complex variations seen in the figures of Appendix \ref{app:figtemporal} do
not suggest a simple statistical model.
Aiming in this paper only at a rough separation and quantification of trend-like
and fluctuation-like components, we modeled the time dependence of the pulsation
period, amplitude of the first harmonic term and the peak-to-peak amplitude with
a linear model consisting of a trend and an oscillatory component. This is
purely heuristic and intentionally avoids interpretation of period changes
as evidence for secular evolution (applies only to linear trends) or the
light-time effect (cyclic changes). This is also warranted
since changes in amplitude have no clear theoretical explanation.
The benefit of using this model is its simplicity and
ability to roughly capture the characteristic size of long-term trends and short-term
fluctuations. A physical interpretation of the observed trends can later be
based on the variations described heuristically.

We write our model as
\begin{equation}\label{eq:armodel}
\begin{aligned}
\theta(t_i) = \;&\alpha_{\theta} + \beta_{\theta} t_i  + \gamma_{\theta} \, \cos \, 2 \pi f_{\theta} t_i + \delta_{\theta} \sin 2 \pi f_{\theta} t_i + \eta_{\theta,i}, \\
& \eta_{\theta,i} \sim \mathcal{N} (0, \sigma^2_{\theta,i}),  \quad  \mathrm{Corr} (\eta_{\theta,i}, \eta_{\theta,i+1}) = \rho,
\end{aligned}
\end{equation}
where $\theta(t_i)$ can represent time-dependent pulsation periods $P$, first
harmonic amplitudes $A_1$, or total amplitudes $A$ at time $t_i$. $f_{\theta}$
is the frequency of the oscillatory term, and the error $\eta_{\theta,i}$ is
assumed to follow a normal distribution with the locally estimated error
$\sigma_{\theta,i}$ on the parameter estimate $\theta(t_i)$. The strong
correlations introduced by the overlapping windows make it necessary to include
an autoregressive structure between the consecutive estimates, represented by
the correlation coefficient $\rho$. The model is fitted for a given frequency
$f_{\theta}$ by generalised least squares.         

Searching for the best approximation, we fit this model at a series of test
frequencies $f_{\theta}$ in the range between the minimal and maximal reasonable
frequencies (as constrained by the width of the sliding windows and the full
timespan of the observations). Figure\,\ref{fig:frvar_param1748} shows an
%% RIA (2017-03-26) a {\bf \msu{(particularly well-fitting)}} 
example of these fits. The left panel shows log-likelihoods of the fitted models for pulsation period and
peak-to-peak amplitude as a function of frequency $f_{\theta}$ for the FU
twin-peak CEP-1748. The best fits corresponding to the highest peak of the
log-likelihood profiles are shown in the center and the right panels. The
obtained fits appear to capture the trend and, in most cases, the dominant quasi-regular oscillatory variations, although the model is clearly only a rough approximation.        

In order to quantitatively characterize the period and amplitude modulations in
Cepheids, we extract the key parameters of the model \eqref{eq:armodel}: (a) the slope of the trends $\beta_P$ and $\beta_{A_1}$, cf. Eq.\,\ref{eq:armodel}; (b) the amplitudes of the best-fit oscillatory components $\Delta P$ and $\Delta A_1$, which are defined as $(\gamma_P^2 + \delta_P^2)^{1/2}$ and $(\gamma_{A_1}^2 +
\delta_{A_1}^2)^{1/2}$, respectively; and (c) the frequencies $f_{P}$ and $f_{A_1}$ of the dominant oscillatory components. The estimated parameters for $A_1$ and $P$  are given (with estimated standard errors) in Appendix \ref{app:tables}.  However, there are several important facts to keep in mind. As explained in Appendix \ref{app:bias}, the estimated frequencies can be ``aliased'', that is, frequencies above the kernel method's upper detection limit can be perceived as lower frequencies. Additionally, %%As well, 
the amplitudes $\Delta P$ and $\Delta A_1$ of fast modulations are biased downwards. This bias depends on the modulation frequency, and thus can be corrected through an empirical estimated relationship between the two, if the modulation frequency is well known (see Appendix \ref{app:bias}). However, since it cannot be decided based on our data whether the estimated frequency is correct, the correction may be flawed, particularly for the period modulations in our control sample, for which Figure \ref{fig:control_all_period} suggests the possibility of relatively fast modulations.

We explore the distribution of the estimated fluctuation parameters among the different Cepheid groups and with respect to the average period in the next two subsections (with the above caveats kept in mind).   For amplitudes of the modulations, we give both bias-corrected and uncorrected versions; however, our conclusions do not differ in the two cases.  

\subsubsection{Trends and fluctuations of pulsation periods} \label{sec:per_groups+trends}

\begin{figure*}
%\sidecaption[t]
\begin{center}
\includegraphics[scale=.7]{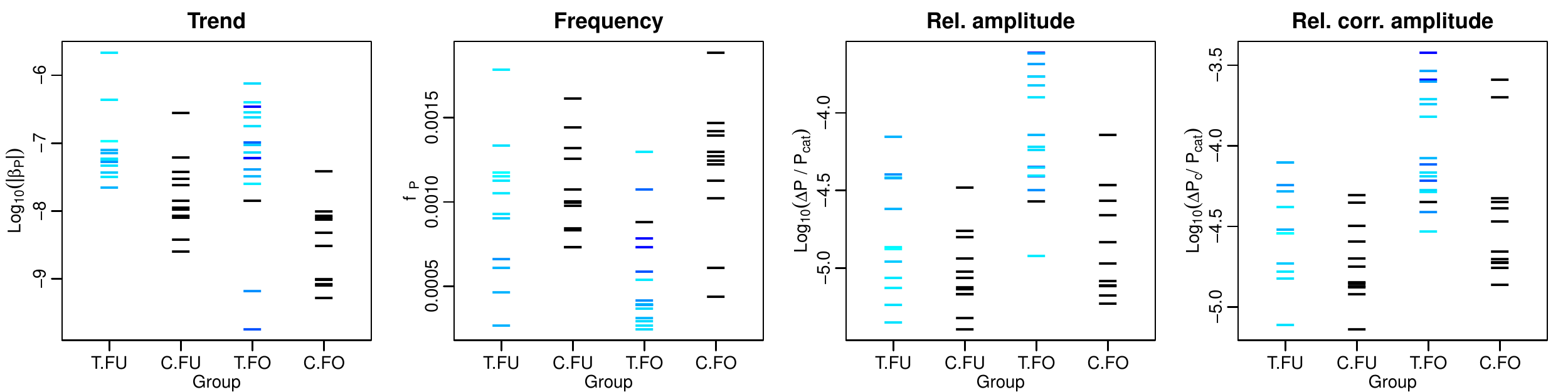}
\caption{Distribution of
the logarithm of the absolute trend $\beta_P$ (left panel), the frequency of the oscillatory term $f_P$ (middle panel) and the
logarithm of the uncorrected and corrected relative amplitude of the modulation, that is, the amplitudes $\Delta P$ and $\Delta P_c$ of the oscillatory term divided by the catalog period $P_{\mathrm{cat}}$ (third and rightmost panels) in model \eqref{eq:armodel}, for the changes in the pulsation period of all Cepheids in our sample. The colours in the twin-peak groups indicate the absolute value of the separation of the twin peaks, ranging from light blue for small separations to dark blue for large ones. The label T.FU and C.FU corresponds to fundamental-mode twin-peak and control Cepheids, respectively; similarly, the labels T.FO and C.FO indicate first overtone twin-peak and control cepheids, respectively. The T.FO group includes the 5 cepheids selected because of their strong amplitude changes.} 
\label{fig:frvar_per_distr}
\end{center}
\end{figure*}

Figure\,\ref{fig:frvar_per_distr} shows the parameters of the trend + oscillation
model fitted to the variations of the pulsation period in the four Cepheid
groups. As a color-coding of twin peak frequency separations shows, there does not appear to be a clear relation between modulation frequency and twin peak frequency separations, despite this being commonly assumed. Any tentative indications of such a relation are strongest for the T.FO group.

%%%%% {\bf\msu{For the twin-peak groups, we added the separation of the twin peaks as a colour code, as a relationship with the estimated modulation parameters could provide an easy-to-access information about these. However, the separation seems to be (weakly) related only to the found frequency of the modulation (corresponding to what is commonly assumed), and even that, only stochastically and only in the T.FO group. %The most obvious outlier, with the highest fitted frequency among the T.FO group but a with a closely-separated twin peak, is CEP-1693.
%%}}

The approximate trend component (leftmost panel) is on average higher in the
twin-peak groups than in the control groups, both for fundamental-mode and
first-overtone Cepheids. Inspecting the two twin-peak overtone stars with the
lowest trend values (CEP-1536 and CEP-1561), the time series of periods from the sliding window fits confirms the absence of an overall trend, and the reason for the appearance of twin peaks in their secondary periodogram may be due to the comparatively strong fluctuating component in their pulsation period, and to some very significant changes in $A_1$ for CEP-1536.  Their primary and secondary peaks had relatively large separation in the stable model analysis (though not exceptionally, as their colour in Figure \ref{fig:frvar_per_distr} indicates), which supports the assumption of the presence of high-frequency modulations.

The \textit{relative} amplitude $\Delta P / P_{\rm{cat}}$ of the
oscillatory component in the right panel of Figure\,\ref{fig:frvar_per_distr} is higher in the overtone groups than in the
fundamental mode groups. The shift between different modes is  larger than the
difference between the control and the twin-peak groups both within the
fundamental and the overtone Cepheid sample. This supports that twin peaks are
more likely to be the result of long-term, trend-like period instability rather
than of fluctuations. The fact that twin-peak groups have, on average, lower
characteristic frequencies of period fluctuations than  control groups,
regardless of pulsation mode, agrees with this: slower semi-regular variations
are more likely than fast oscillating ones to induce twin peaks in residual
periodograms when pre-whitening is carried out with constant light curve
parameters.               

%\riacomm{cite: Turner et al. 2006}.
 
\subsubsection{Trends and fluctuations in brightness amplitude}\label{sec:amp_groups+trends}
% {Comparison of trends and fluctuations of amplitude in different Cepheid groups}

\begin{figure*}
%\sidecaption[t]
\begin{center}
\includegraphics[scale=.7]{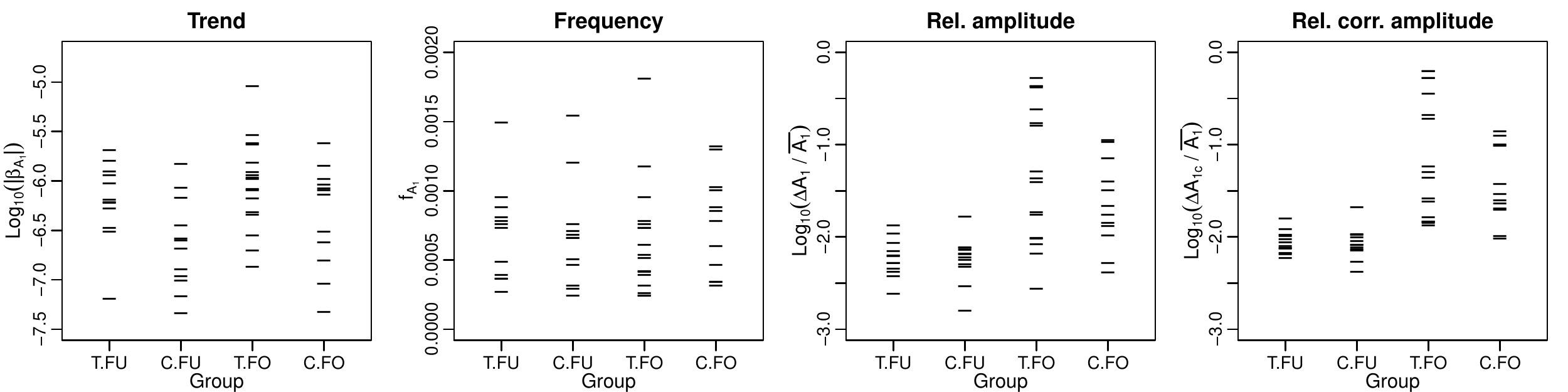}
\caption{ Distribution of
the logarithm of the absolute trend $\beta_{A_1}$ (leftmost panel), the frequency of the oscillatory term $f_{A_1}$ (second panel) and the logarithm of its relative size, that is, the (uncorrected and corrected) size of the changes $\Delta A_1$ and $\Delta A_{1c}$  divided by $\bar A_1$, the average stable-model $A_1$ (third and rightmost panels, respectively), estimated from model \eqref{eq:armodel}. Groups and their labels are the same as in Fig. \ref{fig:frvar_per_distr}.} \label{fig:A1var_distr}       % Give a unique label
\end{center}
\end{figure*}

Figure\,\ref{fig:A1var_distr} shows the temporal variations of the first
harmonic amplitude $A_1$ in the four Cepheid
groups. The trends in $A_1$ (leftmost panel) show a pattern similar to the
trends in the pulsation period: they tend to be on average higher in the
twin-peak groups than in the control groups, though the difference is less
prominent than with the pulsation period. The behaviour of frequencies $f_{A_1}$ of the oscillatory term in the first harmonic amplitude shows no difference across the groups. The relative amplitudes of the oscillatory term in $A_1$ are on average higher among first overtone Cepheids than among fundamental mode objects, especially when we consider the bias-corrected amplitudes.            

In summary, our results suggest that the twin-peak phenomenon is
indicative of trends as well as slow, relatively high-amplitude
fluctuations of primarily the pulsation period. Particularly strong changes
in the amplitude may also be identified by twin peaks, although this ability is
limited to only the strongest or most trend-like cases in {\it OGLE } data.  \citet{soszynskietal08} found 28\% of the FO and 4\% of the FU sample to show the twin-peak phenomenon. However, our study suggests also that potentially interesting cases where the modulations have oscillatory or stochastic character on significantly shorter timescales than the  window length can be missed, as was found in the case of CEP-0727 and CEP-1638 (see Sec. \ref{subsubsec:perch}). Using the twin-peaks phenomenon only to identify cases of modulated pulsation might therefore miss a scientifically interesting subclass, and can give biased estimates about the occurrence and typology of the modulations. 
% \ria{On the other hand, the OGLE} survey characteristics do \ria{enable a}
% straightforward \ria{and clear} detection of period variations in quite a
% large fraction of the {\it OGLE-III} sample (\citealt{soszynskietal08} found 28\% of
% the FO and 4\% of the FU sample to show the phenomenon).

\subsection{Light curve variations versus physical parameters}\label{sec:comparephysical}
% \msucomm{I changed the capitals of each word to small letters, to be consistent with the other titles. If the commonly accepted format is all capitals, please change them.}

\begin{figure}
%\sidecaption[t]
\begin{center}
\includegraphics[scale=.57]{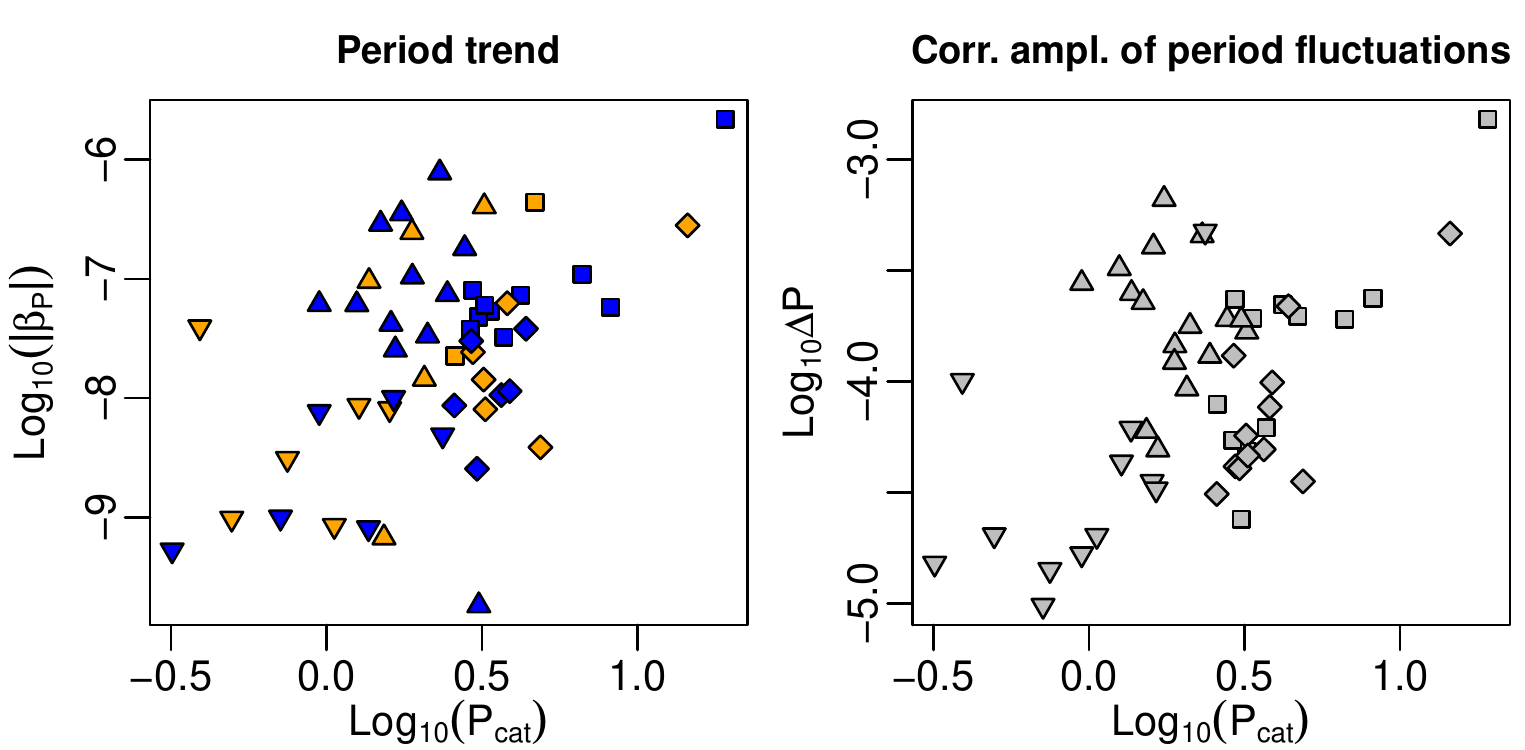}
\caption{Period trend $\mid \! \beta_P
\! \! \mid$ (left panel) and size of period fluctuations $\Delta P$ (right
panel) as estimated from model \eqref{eq:armodel} for the 53 Cepheids. Downward pointing triangles: control overtone Cepheids, upward pointing triangles: overtone twin-peak Cepheids, diamonds: fundamental-mode control Cepheids, squares: fundamental-mode twin-peak stars. In the left plot, negative trends are indicated in blue, positive ones in red. }
\label{fig:pervarpars_vs_catalogP}       % Give a unique label
\end{center}
\end{figure}

\begin{figure*}
\centering
\includegraphics[]{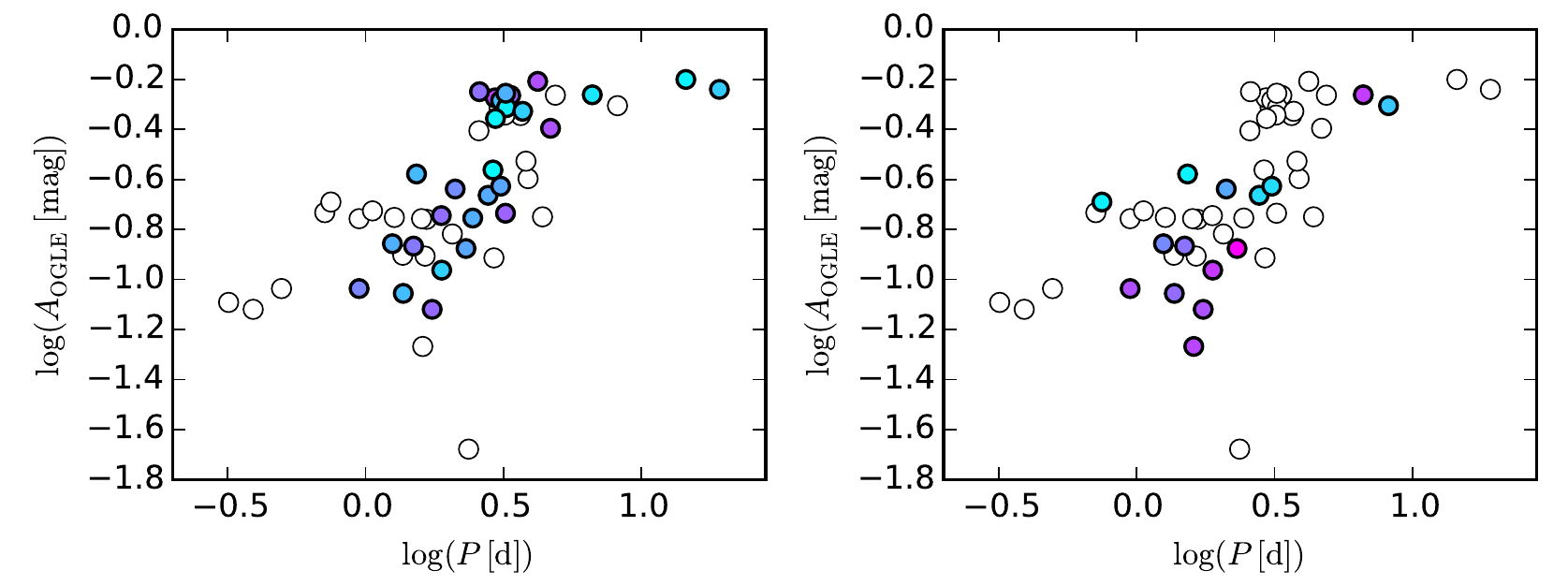}
\caption{Double-logarithmic Period-Amplitude diagram for program Cepheids using
the original {\it OGLE-III} periods and peak-to-peak amplitudes. Constant objects are
shown as open circles, variable objects as filled circles. {\it Left panel:}
Changing periods. The color corresponds to column $T_P$ of Tables \ref{tab:deviationintervals_T} and \ref{tab:deviationintervals_C}, from cyan (0\%)  to magenta (100\%). {\it Right panel:} Variable amplitudes. The color corresponds to column $T_{A_1}$ of Tables \ref{tab:deviationintervals_T} and \ref{tab:deviationintervals_C}, from cyan (0\%)  to magenta (100\%). 
}
\label{fig:logP-logA}
\end{figure*}

In Figure\,\ref{fig:pervarpars_vs_catalogP} we plot the (10-base) logarithm of
absolute values of  period trends and bias-corrected amplitude of period fluctuations, $\log{\Delta
P}$, against average logarithmic pulsation period $\log{P}$. The left panel
shows the well-known monotonically increasing relationship between $\log{P}$ and $\log{\vert \beta_P \vert}$ with  large scatter. This is in agreement with the trends observed for period changes by means of O-C diagrams
\citep[e.g.][]{szabados83,2001AcA....51..247P,2006PASP..118..410T}
and is a consequence of evolutionary timescales, which are shorter
for the higher-mass long-period Cepheids \citep[e.g.][]{2000ApJ...543..955B,2013AstL...39..746F,2016A&A...591A...8A}. The large scatter can be in part due to the 12-year timespan of {\it OGLE}: this may be too short to ascertain whether a slow, trend-like change is indeed a portion of an evolutionary long-term period change, or only a fluctuation on timescales longer than the {\it OGLE } timespan.

Our selection procedure, namely random choice from all Cepheids with twin peaks
or with flat secondary periodograms based only on the presence or absence of the
twin peak, implied that our control FO sample have on average shorter periods
than stars in the twin-peak FO group, as is discernible from Figure
\ref{fig:pervarpars_vs_catalogP}. A similar, though smaller and less clean
separation is also apparent for FU Cepheids. 

The right panel of Figure\,\ref{fig:pervarpars_vs_catalogP} suggests that period fluctuations become stronger with increasing average period. Such a trend is not consistent with an interpretation in terms of the light-time effect. We further find no candidates for binarity based on the light-time effect within this sample: the values of $\Delta P \propto a_1\sin{i}$ determined are too low. Oscillatory period fluctuations have previously been reported for individual long-period Cepheids \citep[e.g. RS Puppis, see][]{2009AstL...35..406B}. 
%%{\bf\msu{and recently for 51 fundamental-mode {\it OGLE }Cepheids \citep{smolec17}}}. 
However, to the best of our knowledge no such relationship has yet been firmly established. The application of our method on the full sample of {\it OGLE } Cepheids will soon enable a more detailed investigation of this relation. 

As a consequence of the period-luminosity relation of Cepheids, similar
near-linear relationships exist between these parameters and mean $I$ or
$V$ magnitude. However, no other relation with colour or position on the
colour-magnitude diagram (CMD) was found.
We find no relationships between the model parameters and
colour, mean magnitude, or period.
Figure\,\ref{fig:logP-logA} shows that Cepheids found to be
variable in period or amplitude occupy all regions of the $\log{P} - \log{A}$
parameter space covered by our sample. This further corroborates the ubiquity of
light curve modulation among Cepheids.

We are currently extending this study using 
the full sample of {\it OGLE-III} and {\it -IV} Cepheids in the Magellanic System \citep{2015AcA....65..297S}. Despite the different observational cadences over the LMC and SMC through {\it OGLE-III} and {\it -IV}, this will provide a more comprehensive picture of the distribution of modulation parameters using a much larger sample in part with a longer timespan, and is aimed at enabling a more in-depth physical interpretation of these phenomena. \citet{smolec17} investigated all Cepheids in the {\it OGLE } Magellanic Cloud collection \citet{2008AcA....58..163S,2010AcA....60...17S,2015AcA....65..297S}. However, in that study, detection was based on the presence of a particular symmetric doublet structure in the residual periodogram after a standard constant-model prewhitening, which does not capture all the possible manifestations of a not necessarily strictly repetitive modulation. As a consequence, none of the 53 Cepheids discussed in the present work is mentioned in \citet{smolec17}. Our ongoing analysis of the full {\it OGLE } Cepheid collection using the kernel method will enable further insights and a more complete comparison. Within its range of detectable modulation frequencies, the sliding window method is much more sensitive to pick out a variety of complex modulation patterns than classical pre-whitening analyses that implicitly assume periodicity of modulations.

\section{Conclusions}\label{sec:Conclusions}

We have applied local kernel modeling, a well-known method of statistics to
investigate  periodic and non-periodic temporal variations of Cepheid light curve parameters. We apply this
method to 53 classical Cepheids from the {\it OGLE-III} catalog of variable stars
\citep{soszynskietal08}, selected according to the presence or absence of a
secondary oscillation frequency close to the primary (referred to as ``twin
peak''), or because of very strong, visually obvious amplitude changes. We
compare on the one hand the behaviour of fundamental-mode with first-overtone
Cepheids, and on the other, the behaviour of Cepheids that exhibit twin
peaks in secondary periodograms with those that do not. Our method yields
estimates of  the parameters of the modulations in the pulsation period and light curve parameters as smooth functions of time; we estimate the significance of the
found  deviations from the best-fit constant parameters by bootstrapping
residuals (Monte Carlo methods) and by multiple hypothesis testing procedures
with respect to the best-fit stable model.             

We find period modulations to be probably a very frequent, possibly ubiquitous
phenomenon among Cepheids. Changes in light curve amplitudes also seem to occur frequently, although they are harder to identify reliably.
Our results suggest that twin peaks are related to
instabilities on longer timescales in the period or in the light curve
parameters. The characteristic size of these instabilities for Cepheids are such
that with a given time sampling pattern and photometric precision, period
changes are easier to detect than variations in the harmonic content of the
light curve. %that ``twin peaks'' identify well Cepheids with period changes on
% longer timescales. In addition, we find that ``twin peaks'' can also be
% related to particularly strong temporal variations in  light curve shapes.            

Over the sample of stars considered, we find a wide range of degrees to which
amplitudes and periods can change, and timescales of the variations ranging from
the shortest to the longest detectable. This suggests an extension of the range
of both amplitude and period changes occurring in Cepheids to beyond our
detection limits, which are imposed by the time sampling and precision of the
photometry used. Applying our method to more densely sampled and more precise
photometry from space ({\it K2, CoRoT}) should allow the detection of smaller
amplitude changes on complementary timescales. 
 
We detect several different types of behaviour (near-linear, oscillatory, and
stochastic) among period and amplitude variations. Specifically, we find that
the majority of Cepheids exhibit period changes beyond or different from
ones expected from secular evolution \citep[see also][]{poleski08}. These may rather
be related to other time-dependent phenomena such as convection, granulation,
rotation (spots), (episodic?) mass-loss, or other forms of modulation such as
the Bla\v{z}ko effect seen in RR\,Lyrae stars.

For Cepheids exhibiting ``twin peaks'', we find that pre-whitening with our
time-dependent best-fit light curve estimates generally removes the ``twin peak"
from the secondary periodogram. After a time-dependent pre-whitening, several
overtone twin-peak stars exhibited weak secondary peaks with frequency ratios of
0.65 to 0.87 to the primary frequency, extending across the
range of period ratios occupied by beat Cepheids in the LMC
\citep{1995AJ....109.1653A,2009A&A...495..249M}.
No such secondary periods were detected in fundamental-mode stars. 

As a next step, we will apply this method to a much larger sample of stars,
such as the {\it OGLE-IV} catalog of Cepheid variable stars, and datasets with
higher photometric precision.
In doing so, we will explore the overall characteristics and occurrence rates of period
and amplitude variations as well as their dependence on the stellar properties. Revealing the time-dependence of Cepheid light curves will thus yield new
constraints on the structure of the outer envelopes of Cepheids and serve to
understand the interaction between the pulsations and the medium they propagate
inside of. Upcoming space-missions such as {\it TESS} and {\it PLATO} will
provide high-quality photometry of many Cepheids, allowing to explore a larger
range of amplitude variations and improve population statistics.        

\begin{acknowledgements}
We thank the anonymous referee for valuable comments that have helped to improve the quality of this manuscript. RIA acknowledges financial support from the Swiss National Science Foundation. The research made use of the public {\it OGLE-III} database (\url{
http://ogledb.astrouw.edu.pl/~ogle/CVS/}) and the NASA's ADS bibliographic services.
% \msucomm{Since R prefers to be acknowledged by a citation, I would like to leave it at the end of the methodology section.}
\end{acknowledgements}

% \listofobjects
%-------------------------------------------------------------------

\bibliographystyle{aa} 
\bibliography{Bib_modulation,Bib_StellarEvolution,Bib_Cepheids,ThesisBibTexRefs,bibfileOrig,bibfileAstro}

\appendix

\section{Detailed statistical methodology}\label{app:detailedmethods}

\subsection{Local estimation}\label{subsec:localest}

\begin{figure}
%\sidecaption[t]
\begin{center}
\includegraphics[scale=.68]{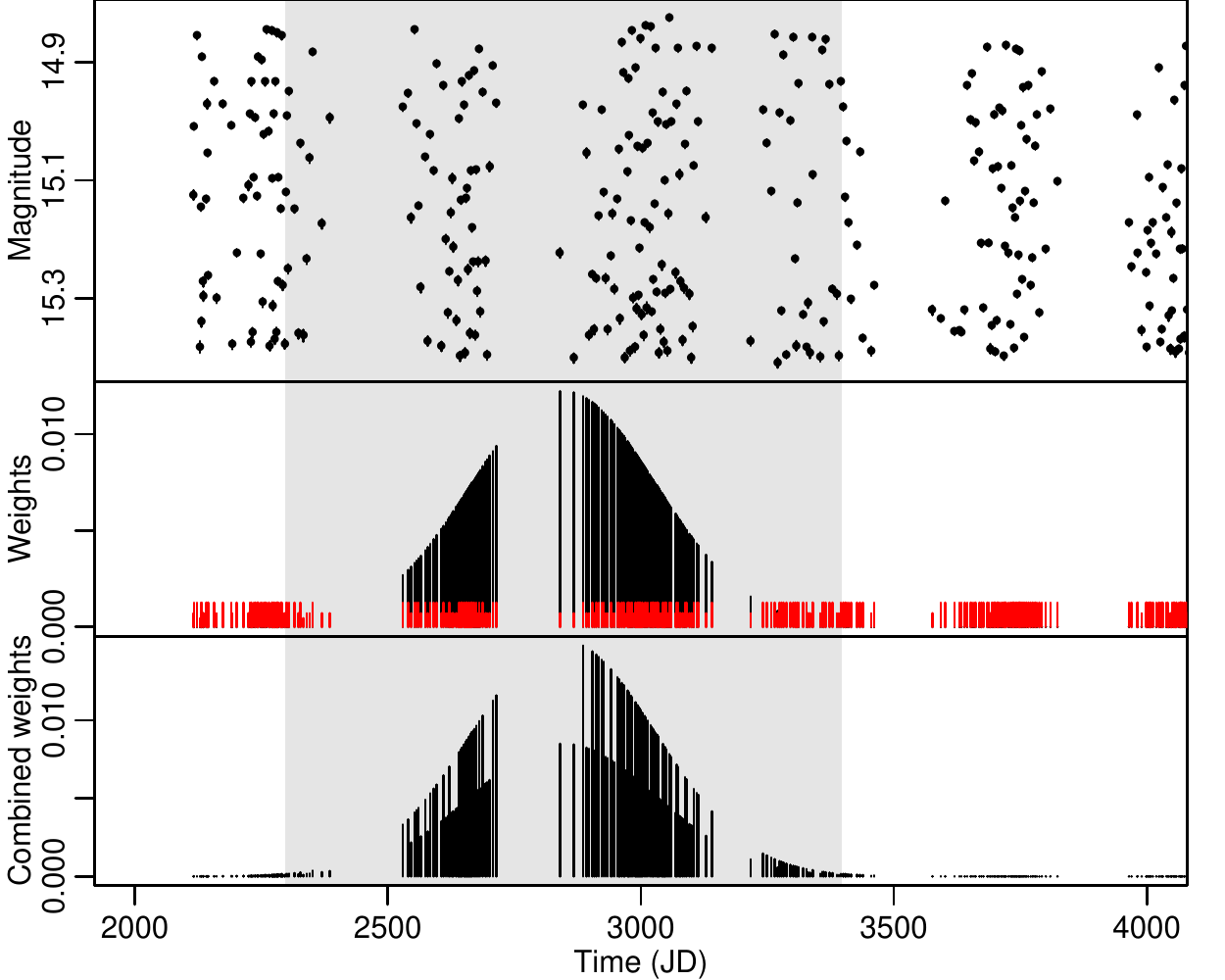}
\caption{Weighting scheme.}
\label{fig:w}       % Give a unique label
\end{center}
\end{figure}

\begin{description}
\item[\it{Windows.}] Corresponding to the aim of the study, we perform nonlinear harmonic series fitting, optimising also over period, in 3-year windows of the time series. We fix the centers  of the windows at a time grid with 30 day separation, which allows to follow the period, amplitude and shape variations with a reasonable temporal resolution, thus fitting about 130 windows for each of our targets.

\item[\it{Weights.}] In order to obtain a smooth picture of the local changes in the light curve shape with most emphasis on the observations close to the centre of the window, we  used a particular weighting scheme, which combined kernel weighting with the usual inverse squared error weights \citep{fangijbels}. Suppose that we have $N$ observations $Y_i$ with errors $\sigma_i$ at times  $t_1,\ldots ,  t_N$. For a window centred at time $\tau_k$, we defined the first component of the weight of an observation at $t_i$ as 
\[
w_i^{(1)} = \left\{ 
   \begin{array}{lr}
      h K\left\{(t_i - \tau_k) / h\right\} &\mathrm{if \;} | t_i - \tau_k | \leq 3h \\
      0 &\mathrm{if \;} | t_i - \tau_k | > 3h
   \end{array}
\right.
\] 

with a Gaussian kernel $K(z) = \frac{1}{\sqrt{2\pi}} \exp \left( - z^2 / 2 \right)$, and with bandwidth $h = 182.5$. This component ensured that the observations close to the window centre  contribute more to the fit than observations farther off, and cut the effect of observations outside a 3-year interval. As an example, Figure \ref{fig:w} shows a window centred at JD $=2850$ as a grey-shaded area of a part of the time series of OGLE-LMC-CEP-1621 (top panel). The Gaussian kernel, computed at the times of the observations, is shown in the middle panel in black.  We used the inverse squared errors 
\[
w_i^{(2)} = 1/\sigma_i^2
\] 
as the other component of the weights, presented in the middle panel of  Figure \ref{fig:w} as red spikes. The final weights, shown in the bottom panel, were determined by
\[
w_i = \frac{1}{W} w_i^{(1)} w_i^{(2)},  \quad W = \sum_{i = 1}^N w_i^{(1)} w_i^{(2)}.
\] 
This scheme ensured both that we obtain an estimate based on the most relevant observations, and that data with comparatively large errors have a weaker influence on the estimate than data with smaller errors. 

\item[\it{Model formula.}] Within each window, and using the above described weighting procedure, we fitted a harmonic + third-order polynomial model of the form
\begin{equation} \label{localmodel}
Y_i = \sum_{k = 0}^3 a_k t_i^k + \sum_{m = 1}^{M} \left( s_m \sin 2 \pi m f t_i + c_m \cos 2 \pi m f t_i \right) + \epsilon_i,
\end{equation}
where $\epsilon_i \sim \mathcal{N} (0, \sigma_i)$ are assumed to be independent Gaussian errors. 
\item[\it{Polynomial order.}] 
Since a neglected nonlinear trend can cause bias in the frequency estimate, we included a third-order polynomial trend into the local model. Inspection of the fits for the Cepheids suggests that the mean magnitude can occasionally vary rapidly, and in such periods, the effect  influences visibly the frequency estimate, as can be seen in the top panel of Figure~\ref{fig:cubicneeded}.  Mean magnitude variations within the window are thus accounted for by the polynomial term in the local model.
\begin{figure}
%\sidecaption[t]
\begin{center}
\includegraphics[scale=.75]{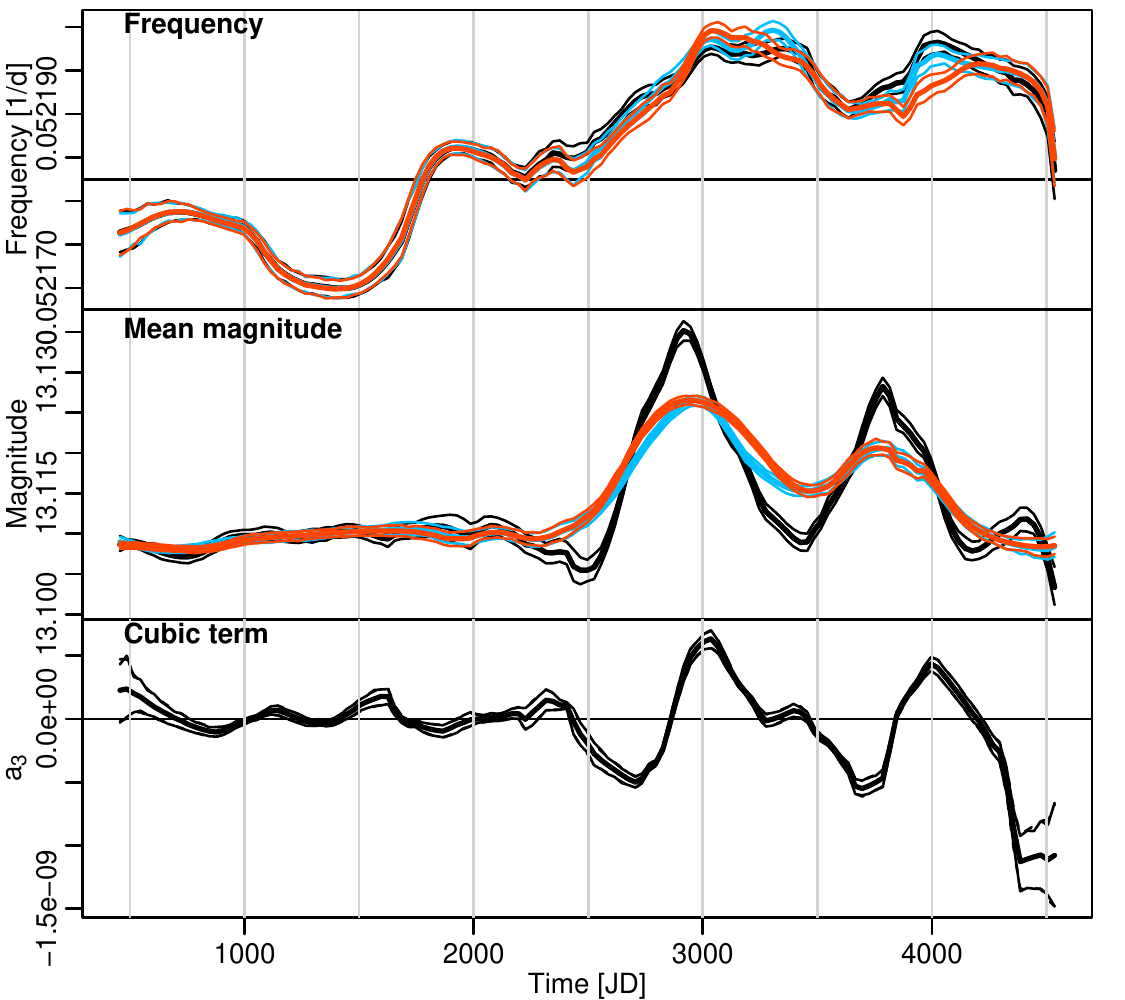}
\caption{Difference of fits with locally constant (red), locally linear (blue) and locally cubic (black) polynomial terms on the example of OGLE-LMC-CEP-1833. The top panel shows the estimates of frequencies as a function of time. The middle panel shows the temporal dependence of the estimated mean magnitude. The bottom panel presents the coefficient of the cubic term in the locally cubic model versus time.}
\label{fig:cubicneeded}       % Give a unique label
\end{center}
\end{figure}
\item[\it{Harmonic order $M$.}] For a sliding window fit, the choice of $M$ is crucial. It must be fixed and the same for all windows, otherwise the appearance of new significant terms or the dropout of formerly significant ones may result in unreasonable jumps of model parameter estimates. As the required $M$ can be very high for some stars such as bump Cepheids and low for others with sine-like light curve, we have to choose $M$ individually. The choice should be such that it  allows for some coefficients becoming temporarily significant and falling out again. It is also desirable to allow for a somewhat higher order than that of a stable model for the complete observation span, since if there are in truth temporal variations, a stable model sees a sort of average, and we can get a too simple model. However, a too high $M$ adds many non-significant parameters to the model, which in turn can change the values estimated for the significant terms. Thus, the individual values were determined by fitting a stable model with $M_{\mathrm{init}}=15$, performing a statistical model selection procedure based on the Bayesian Information Criterion \citep{schwarz78}. After finding a maximal order $M_{\mathrm{final}}$, we finally fixed the harmonic order at $M = M_{\mathrm{final}} + 1$.  
\item[\it{Model fitting.}] According to the aim of investigating the temporal behaviour of the frequency and the light curve parameters, we treated the problem as a nonlinear model, optimising over the linear parameters $a_0, \ldots, a_3, s_1, c_1, \ldots, s_M, c_M$ as well as over the frequency $f$. Optimisation was initialised with the stable model parameters, and was restricted to an adapted frequency interval around the catalog frequency.
\end{description}

\subsection{Error analysis} \label{subsec:erroranal}

We conducted a careful error analysis to assess the significance of the presumably small temporal variations in variability parameters. We therefore employed several different methods of determining confidence intervals for the estimated time-varying parameters, and to compare the time-varying model to the temporally stable alternative reference model for each Cepheid.

%%%% To assess significance of such tiny variations in the parameters as we expect for the Cepheids, a careful error analysis is necessary. We thus used several different methods to obtain confidence intervals of the estimated time-varying parameters and to compare the time-varying model to the alternative of a pulsation stable over time.

\begin{figure*}[!h]
\centering
%\sidecaption[t]
%\begin{center}
\includegraphics[scale=.69]{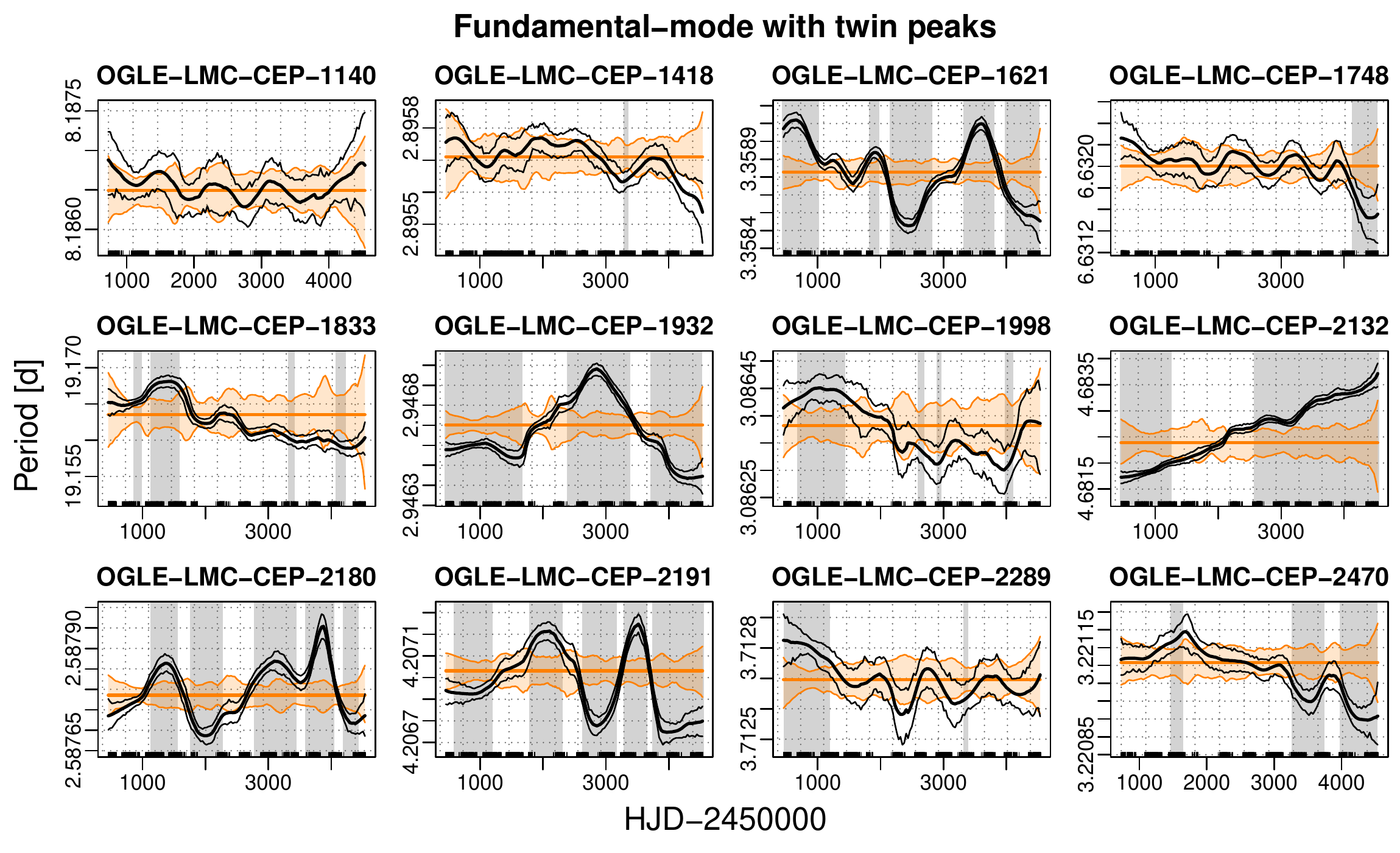}
\includegraphics[scale=.69]{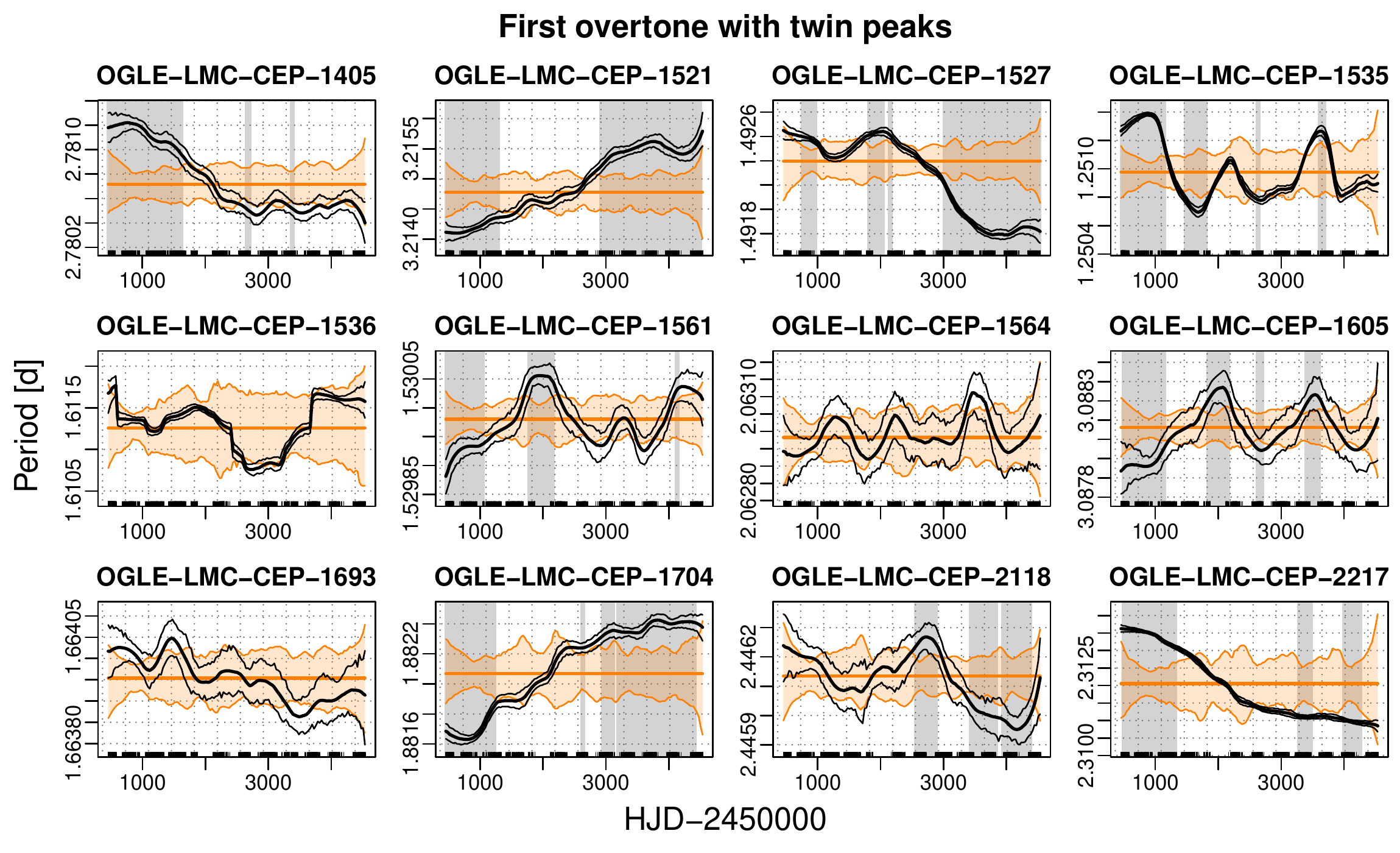}
\includegraphics[scale=.78]{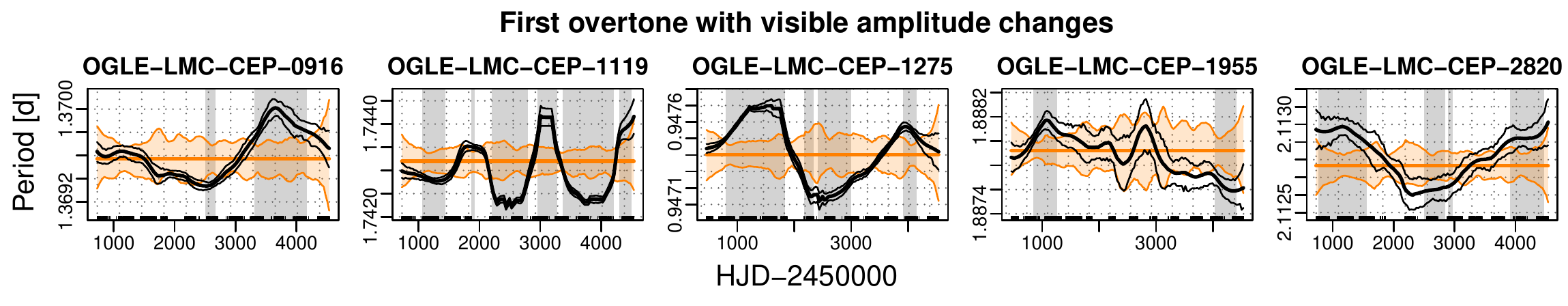}
\caption{Temporal variation of the pulsation period of the twin-peak Cepheids. For the notation, see the beginning of this section. }
\label{fig:twins_all_period}       % Give a unique label
%\end{center}
\end{figure*}

\begin{figure*}[!h]
%\sidecaption[t]
\begin{center}
\includegraphics[scale=.78]{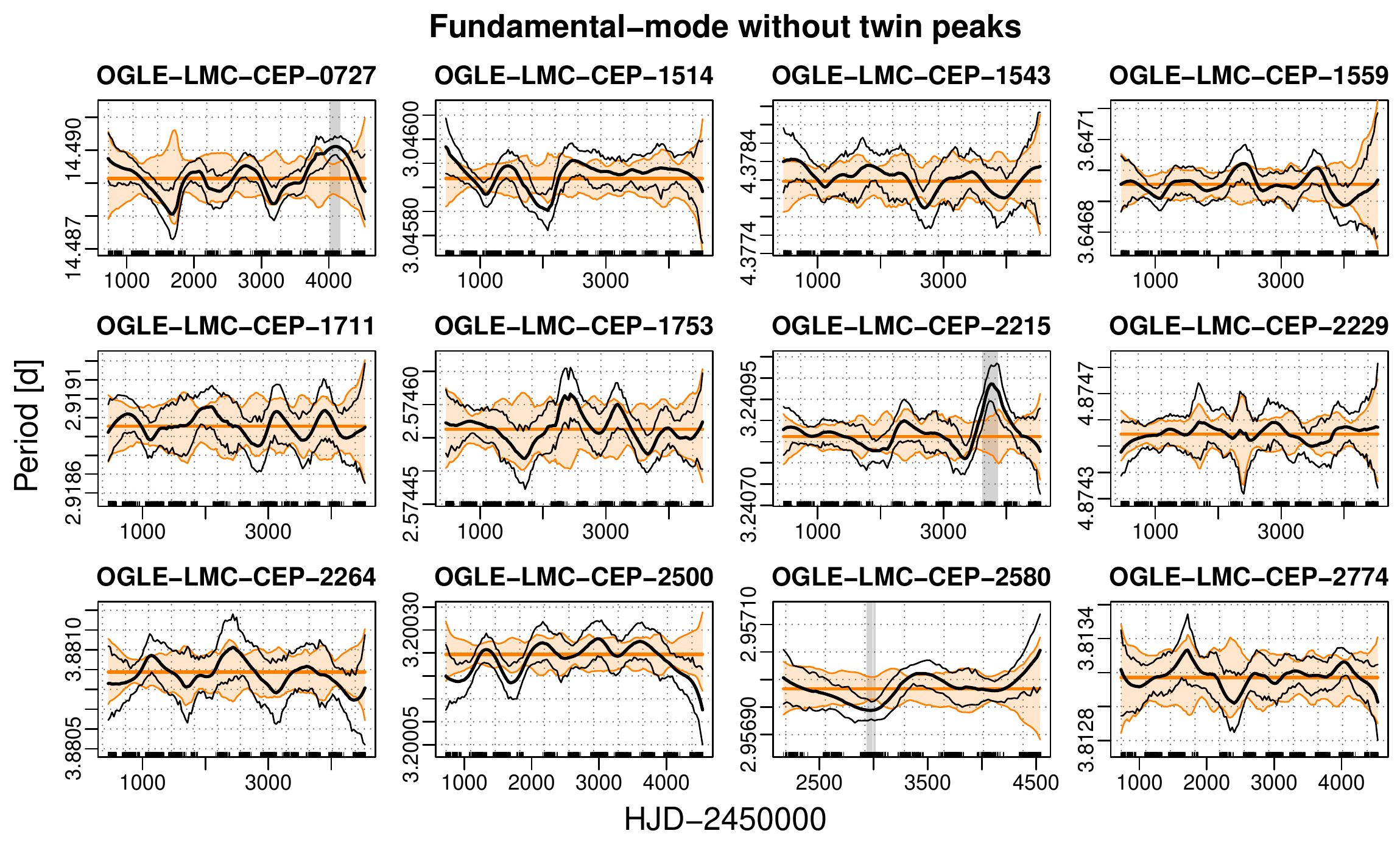}
\includegraphics[scale=.78]{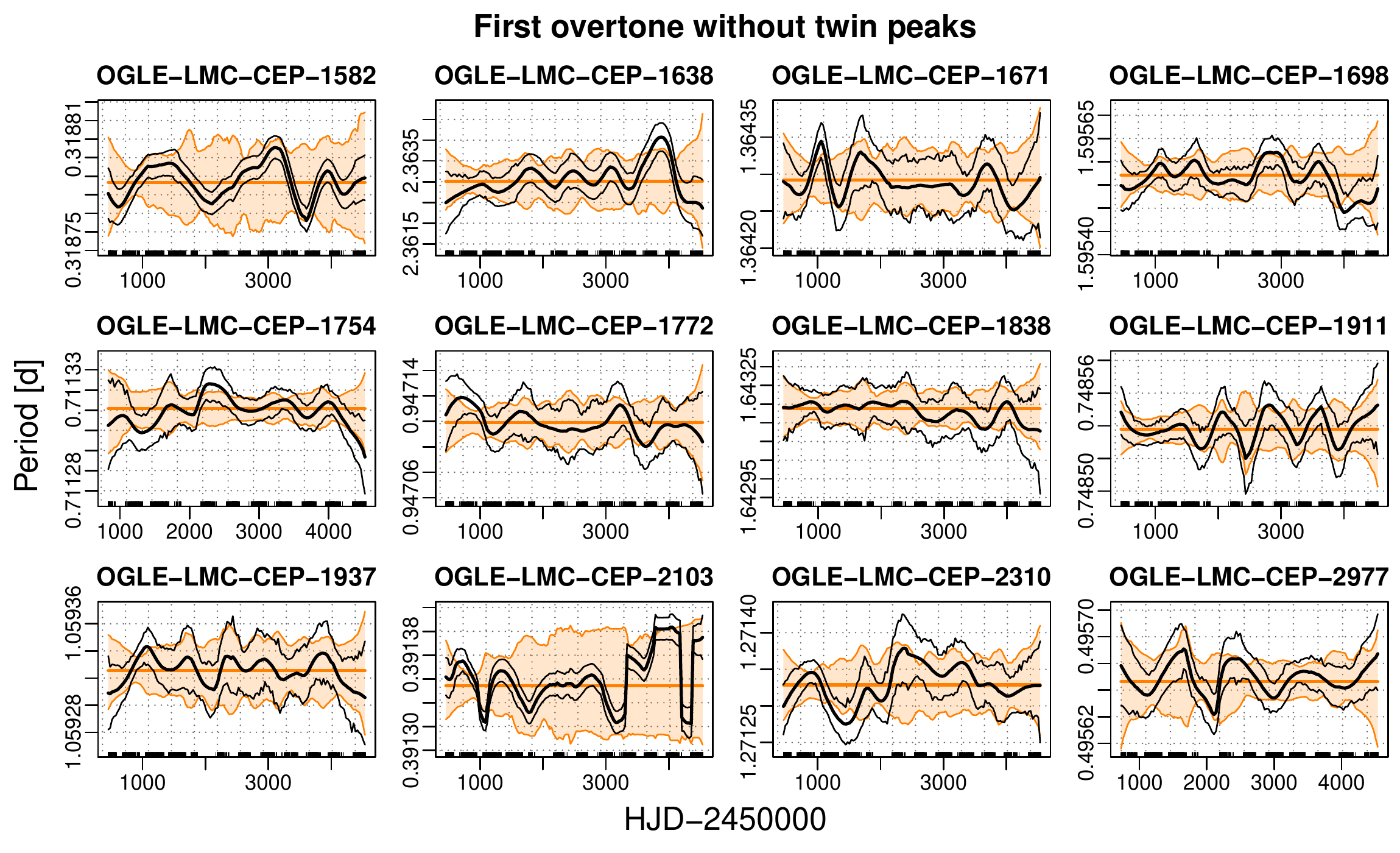}
\caption{Temporal variation of the pulsation period of the control Cepheids. For the notation, see the beginning of this section.}
\label{fig:control_all_period}       % Give a unique label
\end{center}
\end{figure*}

\begin{figure*}[!h]
%\sidecaption[t]
\begin{center}
\includegraphics[scale=.69]{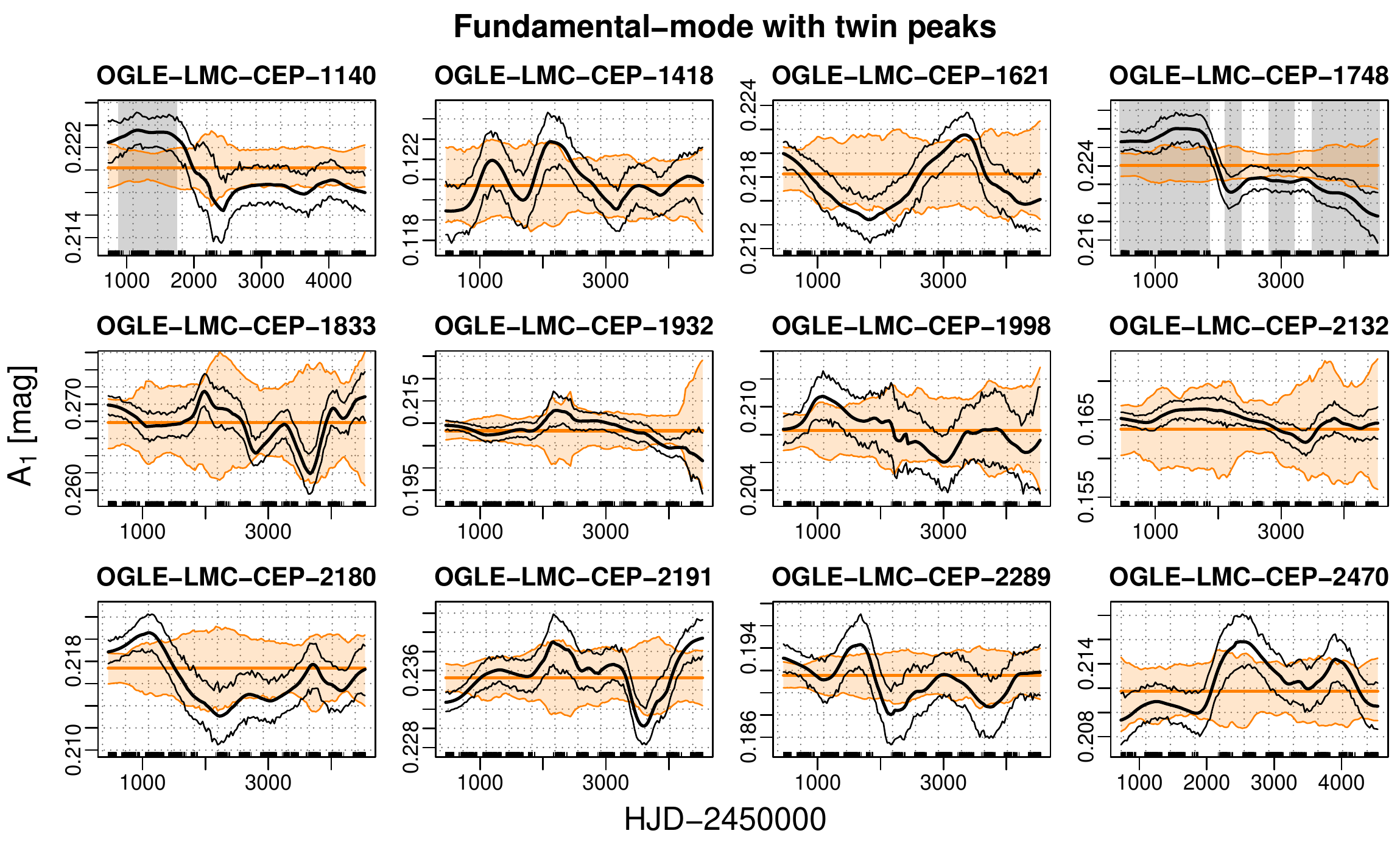}
\includegraphics[scale=.69]{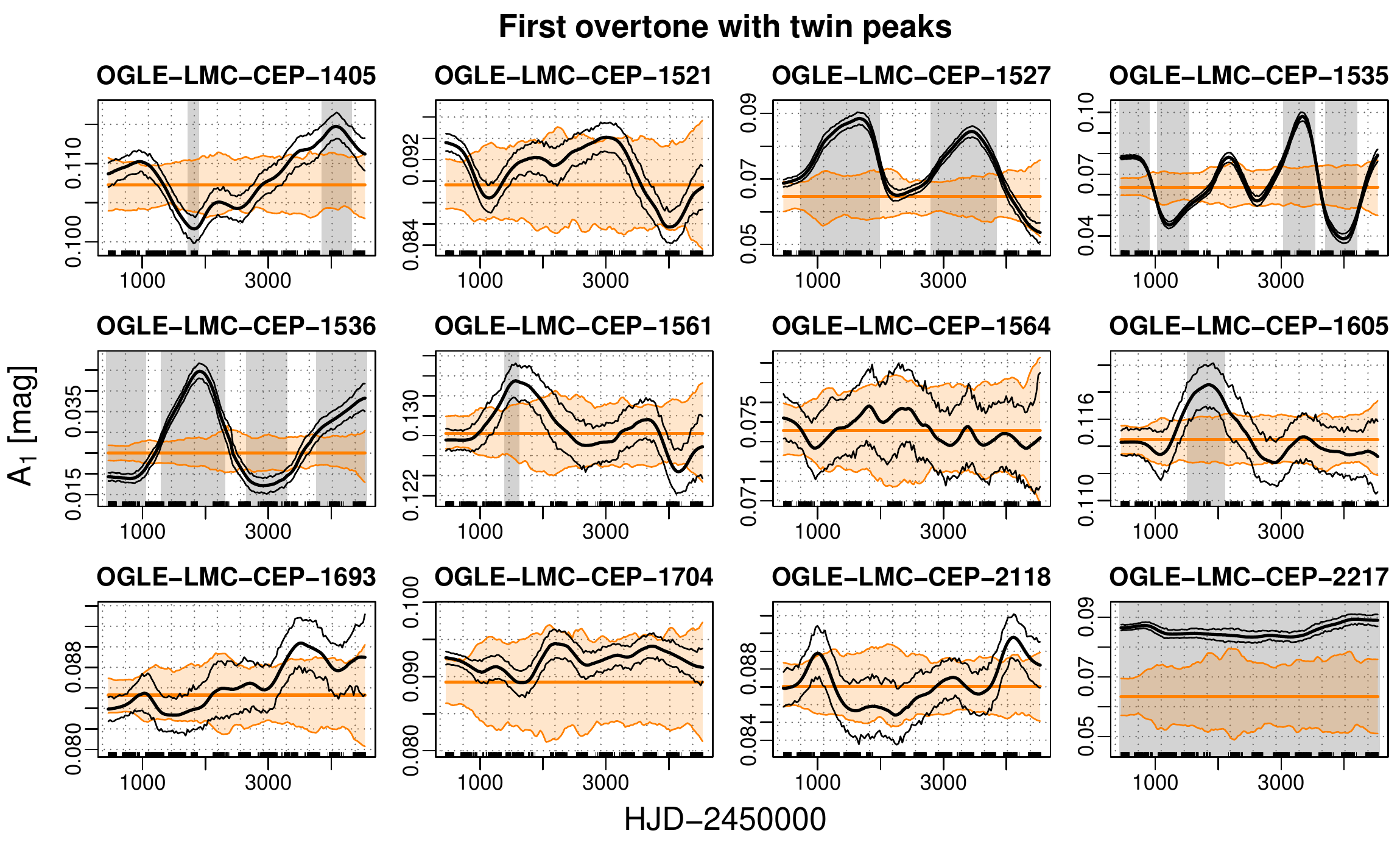}
\includegraphics[scale=.78]{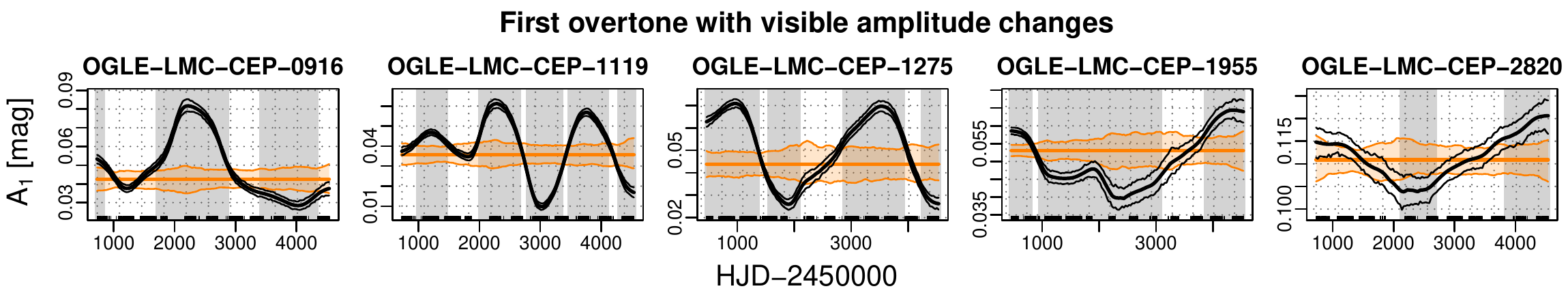}
\caption{Temporal variation of the leading harmonic amplitude $A_1$ of the twin-peak Cepheids. For the notation, see the beginning of this section. }
\label{fig:twins_all_A1}       % Give a unique label
\end{center}
\end{figure*}

\begin{figure*}[!h]
%\sidecaption[t]
\begin{center}
\includegraphics[scale=.78]{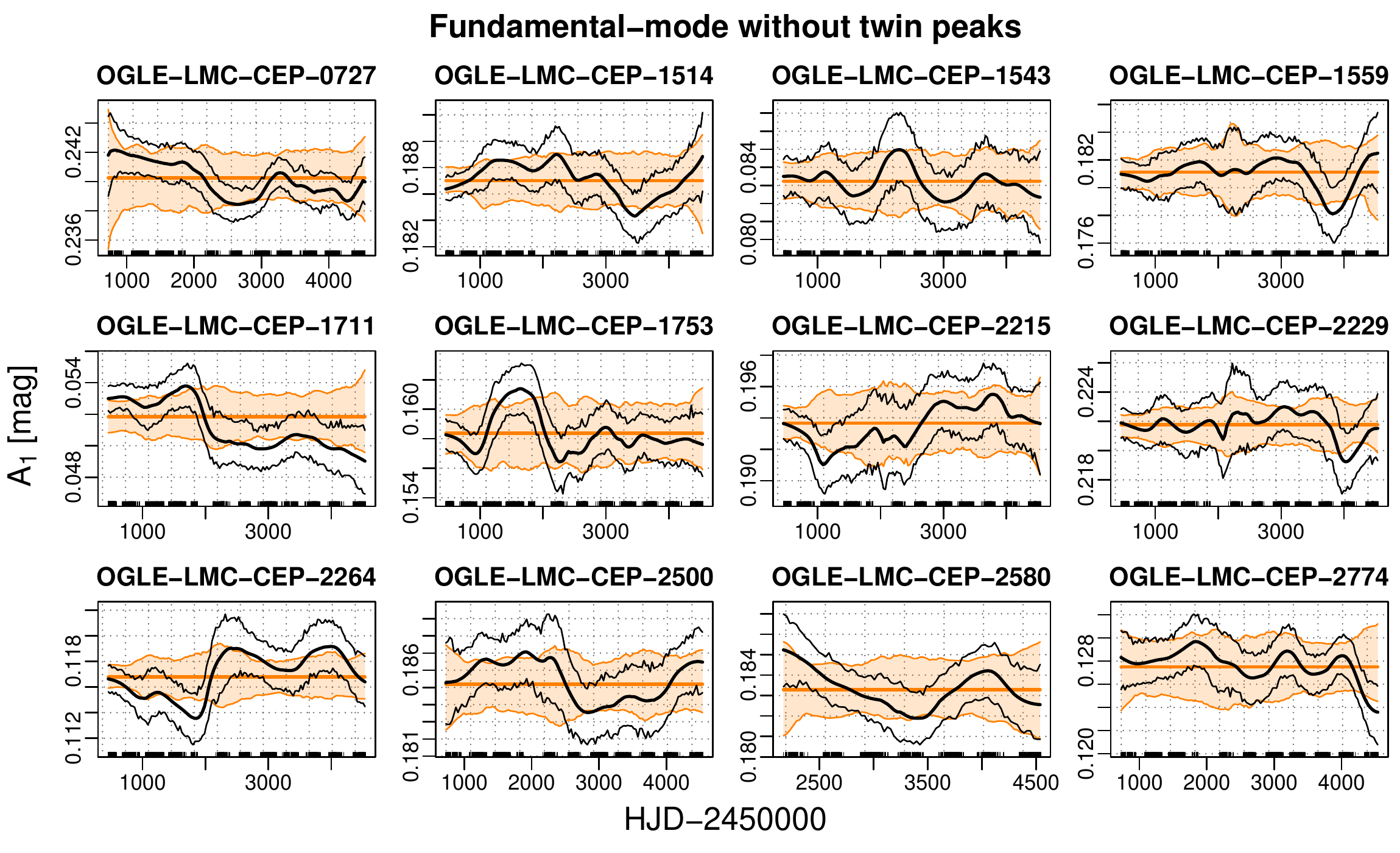}
\includegraphics[scale=.78]{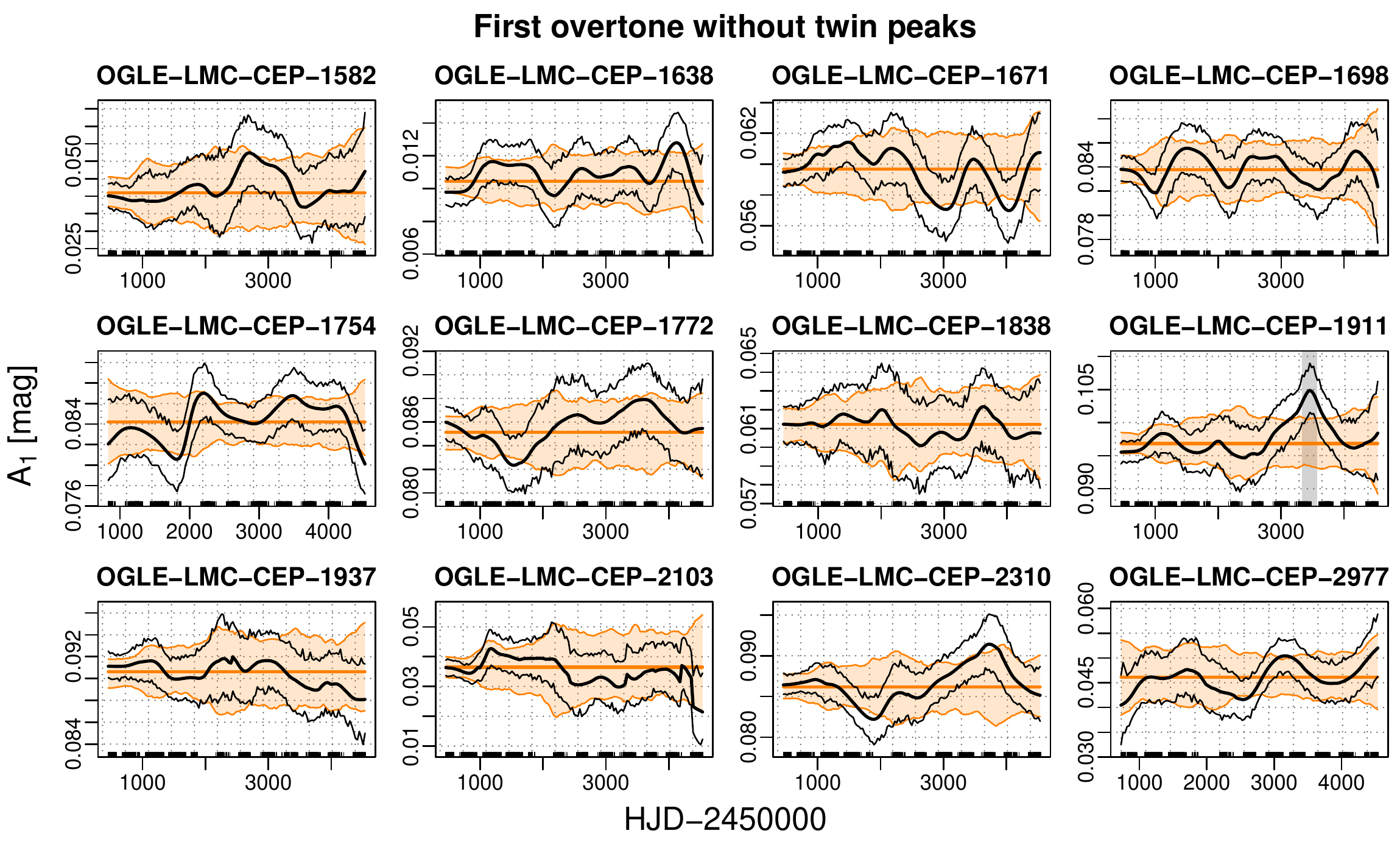}
\caption{Temporal variation of the leading harmonic amplitude $A_1$ of the control Cepheids. For the notation, see the beginning of this section.}
\label{fig:control_all_A1}       % Give a unique label
\end{center}
\end{figure*}

\vspace{-8pt}
\begin{description}
\item[\it{Uncertainty of the estimates.}] For local linear regression models with normally distributed errors, uncertainty of the estimated parameters can be obtained by closed formulae \citep[see equation 3.6 in][]{fangijbels}, if the errors are known. In that case, the error on the estimates follows a multivariate normal distribution. However, we treated the pulsation period of the Cepheid as a parameter to be fitted. Hence, our model becomes a nonlinear regression model, so this formula and the multivariate normality of the estimates are only approximate (and asymptotic). We therefore used a double procedure to obtain uncertainty of the parameters: 
\begin{enumerate}
\item The above cited formula from \citet{fangijbels}, extended for the nonlinear model.   
\item Parametric bootstrap, using the given observational error bars as the standard deviation of the generating normal distribution at each epoch, assuming independent errors, and using the fitted non-stable models interpolated to each observing time as the error-free light curve. The sliding window estimation was then performed on all repetitions. Taking the quantiles 0.025 and 0.975 of the obtained estimates yielded the 95 \% bootstrap confidence intervals (CIs), plotted around the estimates in Figures \ref{fig:sample_period}, \ref{fig:frvar_param1748} and \ref{fig:twins_all_period}-\ref{fig:control_all_A1}.
\end{enumerate}
In all windows of all Cepheids, these two estimates were very close together, so in our nonlinear model, nonlinearity does not cause large deviations in the statistical distribution of the parameter estimates from approximations based on theory.

\vspace{2pt}
\item[\it{Stable reference model.}] We would like to know whether a stable model, with parameters constant over time, can yield a plausible explanation for the estimates from the local sliding window model. For this, we simulated the best-fit stable model with added noise and repeated the sliding window estimation 250 times for each Cepheid. 

However, in the case of the stable model, the added noise cannot be simulated from the error bars using a Gaussian model, since the residuals from the stable model are strongly over-dispersed with respect to the given error bars. We must allow for an underestimation of the errors, if we want to have a plausible zero hypothesis. Therefore, we applied a nonparametric bootstrap of the standardized residuals. We scaled all the residuals $r_1, r_2, \ldots$ at times $t_1, t_2, \ldots$ with the error bars $\sigma_1,\sigma_2, \ldots$ at that time (the mean of the residuals is zero by construction, so we don't need to subtract it). If the error bars are underestimated by a common factor with a good approximation, and the error distribution is a location-scale model (not necessarily Gaussian, including also heavy-tailed distributions such as the Lorentzian/Cauchy), this procedure obtains a homoschedastic error sample  $s_1 = r_1 / \sigma_1, s_2 = r_2 / \sigma_2, \ldots$. We resampled these with repetition, obtaining a (scaled) sample $s^*_1, s^*_2, \ldots$ (each of the starred residuals is equal to an arbitrary one of the original scaled residuals). At each time $t_1, t_2, \ldots$, we computed a simulated noise value by scaling back the scaled residual with the local standard error on the observation:  $r^*_1 = s^*_1  \sigma_1, r^*_2 = s^*_2  \sigma_2, \ldots$. This simulated noise value was added to the fitted magnitudes of the stable reference model, obtaining simulated ``observed" magnitude values $y^*_1, y^*_2, \ldots$ of a noisy Cepheid with stable parameters and errors consistent with the assumption of stability. 

For the detection of such possibly very fine effects as short-term period and amplitude changes in Cepheids, we need also to account for the effect of the precision of times, although this may be a minor effect. The {\it OGLE-III} database gives the HJD dates in days with a precision of 5 digits. In order to assess the effect of rounding, we replaced the times $t_1, t_2, \ldots$ by values $t^*_1, t^*_2, \ldots$ jittered in their 6th digit.

Finally, the time series $\{(t^*_1, y^*_1), (t^*_1, y^*_2), \ldots \}$ was fitted by the local kernel model, obtaining an estimate for every parameter value at every window centre $\tau_i$ ($P^*(\tau_i), a_0^*(\tau_i), \ldots, a_3^*(\tau_i), s_1^*(\tau_i), \ldots, c_M^*(\tau_i)$, denoted generally by $\theta^*(\tau_i)$). This procedure was repeated $R=250$ times, and at each window centre, the  quantiles 0.025 and 0.975 of the obtained estimates $\{\theta^*_1(\tau_i),\ldots, \theta^*_{R}(\tau_i)\}$ were taken to get the 95\% pointwise bootstrap confidence intervals on our hypothetic stable Cepheid. We performed the bootstrap procedure also with unchanged times. We found that the jittering has a negligible effect on the resulting confidence bands, and gives visibly larger CIs only rarely in short time intervals. Thus, our results are robust against the finite precision of the times. We indicate the CIs obtained with jittered times as orange band around the estimated stable bands in Figures \ref{fig:sample_period}, \ref{fig:frvar_param1748} and \ref{fig:twins_all_period}-\ref{fig:control_all_A1}.

The procedure is approximate for many reasons (the fitted Cepheid is itself an estimation, the residuals follow a correlated joint distribution with variances somewhat different from $\sigma_i^2$, the procedure is based on the assumption of homogeneous under-estimation of errors, and although the observational error distribution is not restricted to Gaussian, it must nevertheless be a location-scale distribution), but it accounts for the two dominant effects, namely, the overdispersion of the residuals and the implications of the irregular sparse time sampling on the sliding window estimates.

\end{description}

\subsection{Attribution of significance: multiple hypothesis testing} \label{subsec:mht}
 
The zero hypothesis to be tested for each model parameter (pulsation period, amplitudes) at each window centre $\tau_i$ is that the true local value of $\theta$ is in fact equal to the stable value $\bar\theta$. We compute a probability of seeing a discrepancy between the local and the stable value equal to or higher than the one determined under this zero hypothesis, then compare this probability to a pre-defined confidence level $\alpha$ ($0.05$ in our study): if the probability of the found discrepancy is higher than this level, we cannot reject the zero hypothesis of the local estimate being equal to the stable one.

To compute the probability, we suppose that the discrepancy $\theta(\tau_i)-\bar\theta$ follows a Gaussian distribution with mean 0 and variance computed from the empirical distribution of the repetitions obtained with the bootstrap procedure described above. This distribution represents the null distribution, that is, when the pulsation parameters of the Cepheid are in fact stable. We visually confirmed that the repetitions $\theta^*_1(\tau_i)-\bar\theta, \ldots, \theta^*_R(\tau_i)-\bar\theta$ do follow a Gaussian distribution using quantile-quantile plots \citep[for a short description, see][]{suveges14}. Using this null distribution, we can then compute a $p$-value, namely, the probability that such a discrepancy is given by the kernel method at window centre $\tau_i$ when the Cepheid is a steadily pulsating one (with the best-fit stable reference parameters). 

%\ria{We then calculated $p$-values} for all kernel estimates $\theta(\tau_i)$ \ria{to quantify the probability of estimating a given value via by the kernel method at window centre $\tau_i$ despite a stable light variations (with the best-fit stable reference parameters)}.

%%%% To compute the probability, we suppose that the discrepancy $\theta(\tau_i)-\bar\theta$ follows a Gaussian distribution with mean 0 and variance computed from the empirical distribution of the repetitions obtained with the bootstrap procedure described above. That the repetitions $\theta^*_1(\tau_i), \ldots, \theta^*_R(\tau_i)$ do follow a Gaussian distribution when $\theta$ represents pulsation period or polynomial or harmonic coefficients, we confirmed visually by quantile-quantile plots \citep[for a short description, see][]{suveges14}. We can then obtain a $p$-value for all kernel estimates $\theta(\tau_i)$: this is the probability that such a value is produced by the kernel method at window centre $\tau_i$ when the Cepheid is a steadily pulsating one (with the best-fit stable reference parameters).

The above $p$-values are point-wise and have been computed for each window centre (about 130 over the total time span of {\it OGLE-II} and {\it III}). Their number itself raises a problem. Generally, if we have a true zero hypothesis, and we assess significance of an alternative at level $\alpha$ separately $K$ times using completely independent data, even in the case of a true zero hypothesis we can expect approximately $K\alpha$ significant values (that is, false positives), simply by randomness. Thus, finding a few significant {\it pointwise} $p$-values should not necessarily be considered as having found a significant {\it global} discrepancy from stability. Multiple hypothesis testing methods are developed to deal with this situation.

Moreover, $p$-values computed for neighboring windows are strongly correlated, due to the overlap in the data used for each local estimate. If by sheer randomness, at $\tau_i$ there was a $p$-value smaller than our confidence limit $\alpha$, there is an increased probability that we will have a similarly small $p$-value at the next or next few window centres, creating a false effect of a longer period where the pulsation parameters were significantly different from the stable one. Thus, particular versions of multiple hypothesis testing procedure must be used, which in addition takes into account the dependence between neighbouring $p$-values. Such procedures have been developed (among others) by \citet{hommel88, hochberg88, benjaminihochberg95,benjaminiyekutieli01} and \citet{meskaldjietal13}. Here, we applied the procedure by \citet{benjaminiyekutieli01} that imposes only few restrictions on the dependence structure and is one of the most powerful among the alternatives. The time intervals during which this method indicates significance are highlighted in Figures \ref{fig:twins_all_period}, \ref{fig:control_all_period},  \ref{fig:twins_all_A1} and \ref{fig:control_all_A1} via a grey background.

\section{Simulated instability} \label{app:sim}

In the following, we investigate how the detectability of the minute effects reported in this work are affected by data structure. Specifically, we seek to clarify the ability of the kernel smoothing method to trace modulations of the pulsations using relatively sparse and unevenly sampled data and to quantify detection limits. To this end, we have simulated trend-like, as well as combined trend- and oscillation-like variations of pulsation periods and amplitudes, for the range of parameters representative of our sample.

Noise-free light curves were generated as follows. First, we determine instantaneous phase via (numerical) integration of the time-varying frequency:
\begin{equation} \label{eq:phase}
\phi(t) = \phi_0 + \int_0^t f(s)ds,
\end{equation}
where we replaced $\phi_0 = 0$ without loss of generality. We then substituted $f(s)$ by  combinations of a linear trend and periodic fluctuations in pulsation {\it period}:
\begin{equation} \label{eq:frchange}
f(s) = \left( \alpha_{P} + \beta_{P} s  + \Delta P \, \cos \, 2 \pi f_{P} s  \right)^{-1}.
\end{equation}
%%%Varying first harmonic amplitudes were modeled as 
Varying first harmonic amplitudes were produced by 
\begin{eqnarray}
s_1(t) &=& \alpha_{s_1} \left( 1 + \frac{\beta_{A_1}}{\alpha_{A_1}} t + \frac{\Delta A_1}{\alpha_{A_1}} \sin  2 \pi f_{A_1} t \right), \\
c_1(t) &=& \alpha_{c_1} \left( 1 + \frac{\beta_{A_1}}{\alpha_{A_1}} t + \frac{\Delta A_1}{\alpha_{A_1}} \sin  2 \pi f_{A_1} t \right),
\end{eqnarray}
where the notation agrees with that of eq. \eqref{eq:armodel}, and $\alpha_{s_1}$ and $\alpha_{c_1}$ were taken from similar linear+oscillatory model fits to the time-varying harmonic coefficients, such that $\alpha_{s_1}^2 + \alpha_{c_1}^2 = \alpha_{A_1}^2$. 

Next, $\phi(t)$, $s_1(t)$ and $c_1(t)$ were inserted into the model formula \eqref{localmodel} to obtain the pure noise-free light curves at times $t_i$:
\begin{eqnarray}
Y(t_i) &=& a_0 + s_1(t_i) \sin \phi(t_i) + c_1(t_i) \cos \phi(t_i) \nonumber \\
& &+ \sum_{m = 2}^{M} \left[ s_m \sin m \phi(t_i) + c_m \cos m\phi(t_i) \right], \nonumber
\end{eqnarray}
where $a_0$ and the $m \geq 2$ harmonic coefficients were kept constant. 
%In the simulations, 
We did not add long-term polynomial components $\sum_{k=1}^{3}a_k t_i^k$ in these simulations. 

To create realistic simulations, we adopted real time samplings and our estimates of fitted trend+oscillation model parameters for three real Cepheids: CEP-1405 (FO twin peak featuring oscillations in $A_1$, as well as a trend and oscillation in $P$), CEP-1748 (FU twin peak with significant variations only in $A_1$) and CEP-2470 (FU twin peak with significant variations only in $P$).
We tested the performance of the kernel method in several different modulation types from simple to complex: (1) using period trends of increasing slope up to the trend found in OGLE-LMC-CEP-1405, (2) superposing oscillations of varying frequency and amplitude to the trend of OGLE-LMC-CEP-2470 and (3) simulate the full estimated model of OGLE-LMC-CEP-1748 and OGLE-LMC-CEP-1405.
To these noise-free light curves, we added independent Gaussian error terms using the {\it OGLE }magnitude error estimates as the standard deviation of the Gaussian, generating 250 independent repetitions of noisy light curves for each investigated pure light curve. Finally, we repeated the sliding window estimation on each of the repetitions, using the same tuning parameters as in the investigation of the observed Cepheids.

\subsection{Trends in the pulsation period} \label{subsec:trendpersim}

\begin{figure}
%\sidecaption[t]
\begin{center}
\includegraphics[scale=.67]{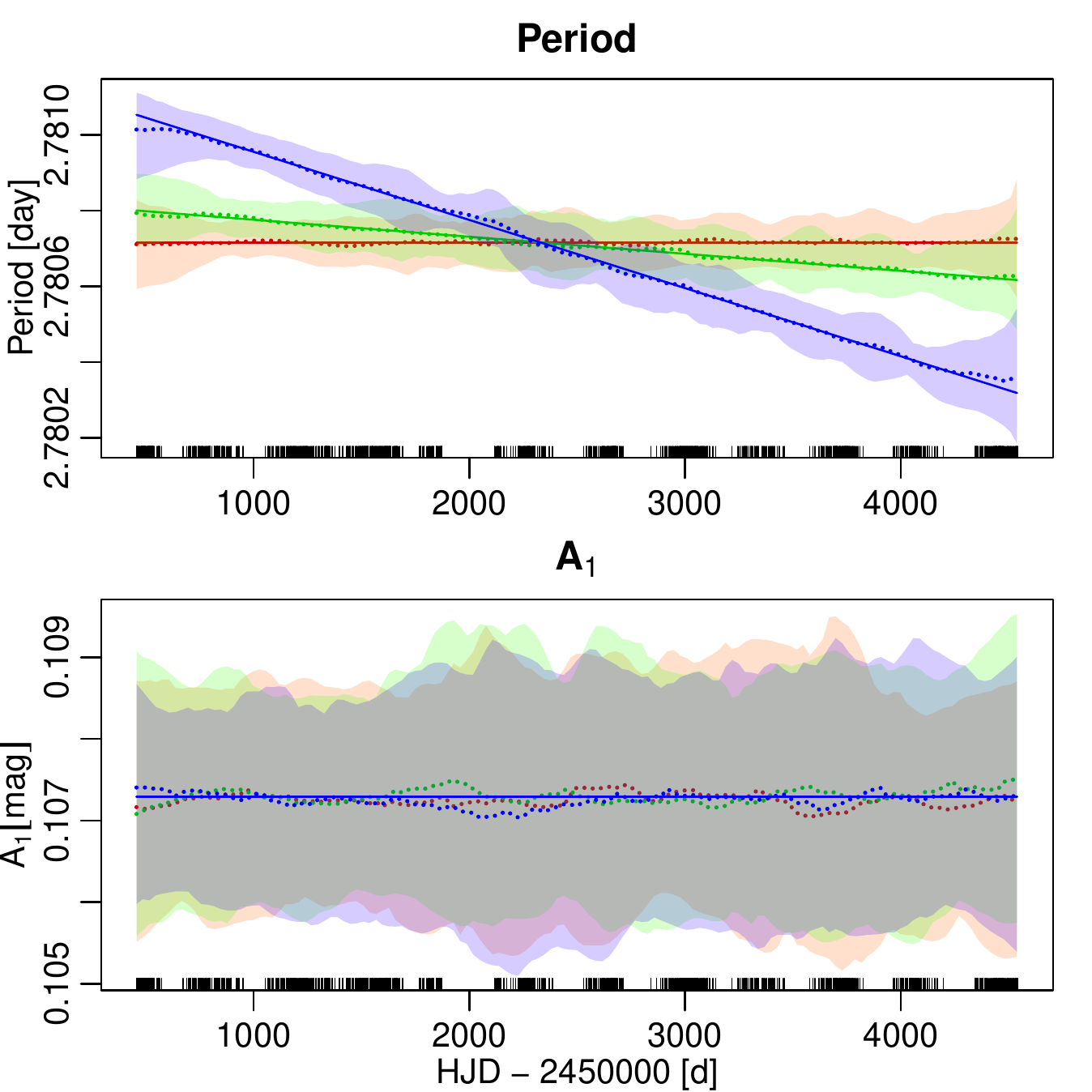}
\caption{Simulated trends in pulsation period (top panel) combined with a constant amplitude (bottom panel) as estimated by the local kernel method. Thin red, green and blue solid lines in the top panel: the simulated values using slopes 0, 0.25 and 1 times that of OGLE-LMC-CEP-1405 (such that the blue line coincides with the trend component found for this Cepheid). The true amplitude, which had the same constant value in all simulations, is represented by a blue line in the bottom panel. Thick dotted red, green and blue lines: the pointwise median of the kernel estimates on the 250 repetitions. The colours show a 95\% symmetric confidence band around the median.}
\label{fig:trendsim}       % Give a unique label
\end{center}
\end{figure}
To investigate the method's power of detecting trends in pulsation period, we simulated a stable light curve with parameters identical to those of CEP-1405, adding linear trends in pulsation period as $\beta_P = a \beta_{P,1405}$, where $\beta_{P,1405} = -1.8 \times 10^{-7} d/d$ (see Table \ref{tab:periodresults}), and $a \in \{0, 0.1, 0.25, 0.5, 1, 2\}$. Figure~\ref{fig:trendsim} shows the median estimates and the 0.025- and 0.975-quantiles for $a = 0, 0.25 \mathrm{\; and \;} 1$. The local kernel estimation is unbiased apart from end effects at the beginning and end of the observation span, both for the period trend and the constant amplitude. The characteristic {\it OGLE }time sampling does not leave a strong systematic imprint on the estimates. The simulation with $a = 0$ (i.e., a perfectly stable light curve) reveals no tendency to incorrectly suggest a non-existing trend or fluctuation. 

%% The latter procedure, bootstrap repetitions and estimation using a simulation of an appropriate stable reference Cepheid, was performed for all Cepheids as described in Section \ref{subsec:erroranal}, with an important difference there. 
%\ria{\riacomm{Make sure this is correct, the original text was a bit unclear:} We carried out this procedure of bootstrap repetitions and estimation for all Cepheids using a simulated stable reference Cepheid with one important change relative to the description in Sec.\,\ref{subsec:erroranal}.}

%The procedure presented in Section \ref{subsec:erroranal}, used to assess the significance of discrepancies from a stable model, is equivalent to this procedure. Since we found strong discrepancies from a Gaussian distribution \riacomm{unclear} and overdispersion of residuals, we used a quantile-based procedure for the bootstrap that preserved the ``extremeness'' of every residual, keeping at least approximately track of the overdispersion, and allowing for a potential under-estimation of the \ria{(photometric)} errors. The 0.025- and 0.975-quantiles of these simulations are plotted as the orange band around each stable reference Cepheid in the Figures of Appendices \ref{subsec:figperiods} and \ref{app:figa1}, and formed the basis of the multiple hypothesis testing to assess the significance of discrepancies from stability.

Inserting non-zero trends into simulations of CEP-1405, we find that 
our method would have detected trends as small as a quarter of the observed trend, cf. the green line in Fig.\,\ref{fig:trendsim}.
%with the time sampling and errors of CEP-1405, even a trend one-fourth of that of CEP-1405 could have been detected, as the green line shows in Figure~\ref{fig:trendsim}.  
Both the trend in the period and the stability of the amplitude is well estimated, which suggests that the model is able to correctly disentangle effects in pulsation period and amplitude.  

\subsection{Combined oscillation and trend in the pulsation period} \label{subsec:combpersim}

\begin{figure}
%\sidecaption[t]
\begin{center}
\includegraphics[scale=.67]{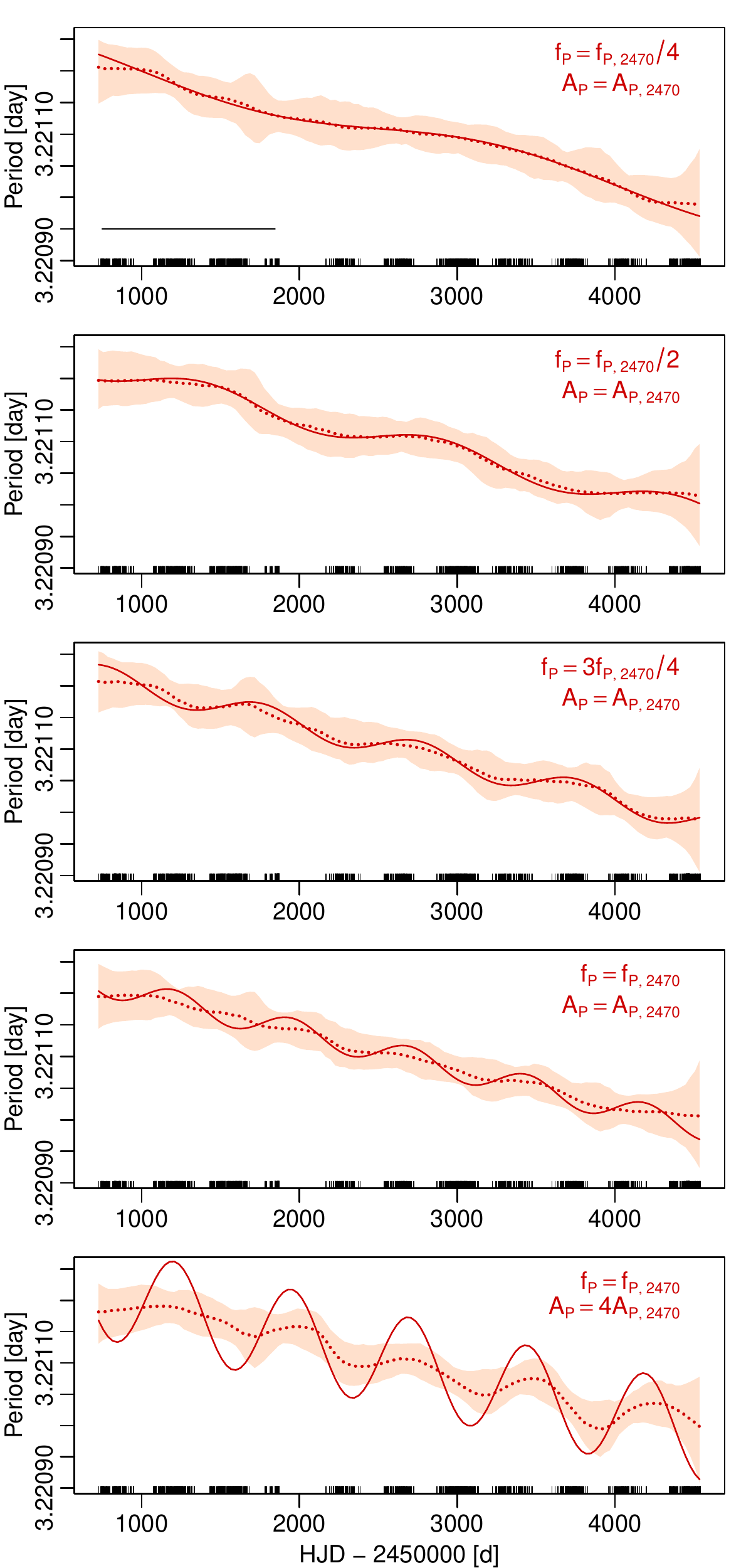}
\caption{Simulated oscillations of the period, superposed to the period trend estimated for CEP-2470. The parameters of the simulated modulations are shown in the upper right corner of the panels. The true simulated modulations of the period are shown with solid red lines. The 0.025, 0.975 quantiles and the median of the kernel estimates are indicated by the coloured band and the dotted thick line, similarly to Figure~\ref{fig:trendsim}. The thin black horizontal line in the top panel shows the full width of the sliding window, to ease comparison with the period of the fluctuations. The rugplot added to the x-axis indicates the observation times.}
\label{fig:trendplusfluctsim}       % Give a unique label
\end{center}
\end{figure}

Trends are smooth long time-scale variations and are usually well-estimated by local models. However, modulations on shorter time-scales can also occur in the Cepheids, and the length of the temporal (sliding) window crucially affects what time-scales can be detected. 

To simulate the case of short time-scale modulations, we adopted the parameters for CEP-2470, whose period and amplitude fluctuations are near the detectability limit of our 3-year-wide sliding window.
We superposed five different oscillations of the pulsation period to a period trend of  $\beta_P = \beta_{P,2470} = -5.983 \times 10^{-8} d/d$. Four of these had the same oscillation amplitude $\Delta_P = \Delta_{P,2470}$ as the best-fit estimate of CEP-2470, \ria{with} frequencies $f_P = a \, f_{P,2470}$ with  $a \in \{0.25, 0.5, 0.75, 1\}$. The fifth simulation used $\Delta_P = 4 \Delta_{P,2470}$ and $f_P =  f_{P,2470}$. For all, we kept all other parameters at the values of the stable reference model of CEP-2470 (no variations in the amplitude were added).  
 
Figure~\ref{fig:trendplusfluctsim} illustrates the results of this investigation. As expected, slow period fluctuations that are on the order of or longer than the sliding window duration are well-estimated by the local kernel method (two top panels).
%% though the size of the fluctuations appears systematically slightly underestimated when $f_P =  0.5 \,f_{P,2470}$ \riacomm{Use a $\lesssim$ instead of $=$?}. 
%% The reason for this is that this frequency corresponds to a period near to the full window length (shown as a thin horizontal line in the lower left corner of the top panel). 
The size of faster period fluctuations ($f_P \geq 0.75 \,f_{P,2470}$) is increasingly under-estimated (middle panel), and the kernel procedure eventually fails to detect fluctuations on timescales that approach the time interval within which our weights are non-negligible. The kernel procedure does, however, allow to recover the correct fluctuation frequency if the intensity of such fast fluctuations is increased ($\Delta_P = 4 \Delta_{P,2470}$, bottom panel), even though the fluctuation intensity is then underestimated by about a factor of four. Thus, we conclude that long-timescale fluctuations (relative to sliding window size) are well-estimated, whereas the detectability of shorter timescale fluctuations decreases for shorter fluctuation timescales and depends on the intensity of the fluctuations. The estimate of the characteristic size of the fluctuations is underestimated. This bias is discussed in detail in Appendix~\ref{app:bias}.

%% The results are shown in Figure~\ref{fig:trendplusfluctsim}. Slow fluctuations of the period are well estimated by the local kernel method, as can be seen in the top two panels, though the size of the fluctuations appears systematically slightly underestimated when $f_P =  0.5 \,f_{P,2470}$ \riacomm{Use a $\lesssim$ instead of $=$?}. The reason for this is that this frequency corresponds to a period near to the full window length (shown as a thin horizontal line in the lower left corner of the top panel). The size of an even shorter-period fluctuation ($f_P =  0.75 \,f_{P,2470}$) , although the fluctuation is on average still noticeable, is clearly under-estimated (middle panel). If the period of a Cepheid fluctuates with an amplitude and a frequency equal to those estimated for CEP-2470, the kernel procedure is no longer able to detect it (fourth panel from top in Figure~\ref{fig:trendplusfluctsim}), because the period of the fluctuation is practically equal to the interval where the Gaussian weights are non-negligible. A much stronger signal with $\Delta_P = 4 \Delta_{P,2470}$ and with the same frequency does appear in the estimates (bottom panel of Figure~\ref{fig:trendplusfluctsim}, dotted thick red line), though with an estimated fluctuation amplitude of $\Delta_{P,2470}$, that is, approximately one-quarter of the true one (thin black line). We can draw the conclusion that the method traces well fluctuations with longer periods than the window size, but signals with shorter timescales are estimated with a bias which increases with shortening periods.

The harmonic amplitudes $A_1$, which were kept constant in these simulations, were correctly reproduced by all five simulations and are not shown here for brevity.

\subsection{Adding amplitude variations} \label{subsec:combperampsim}

\begin{figure}
%\sidecaption[t]
\begin{center}
\includegraphics[scale=.65]{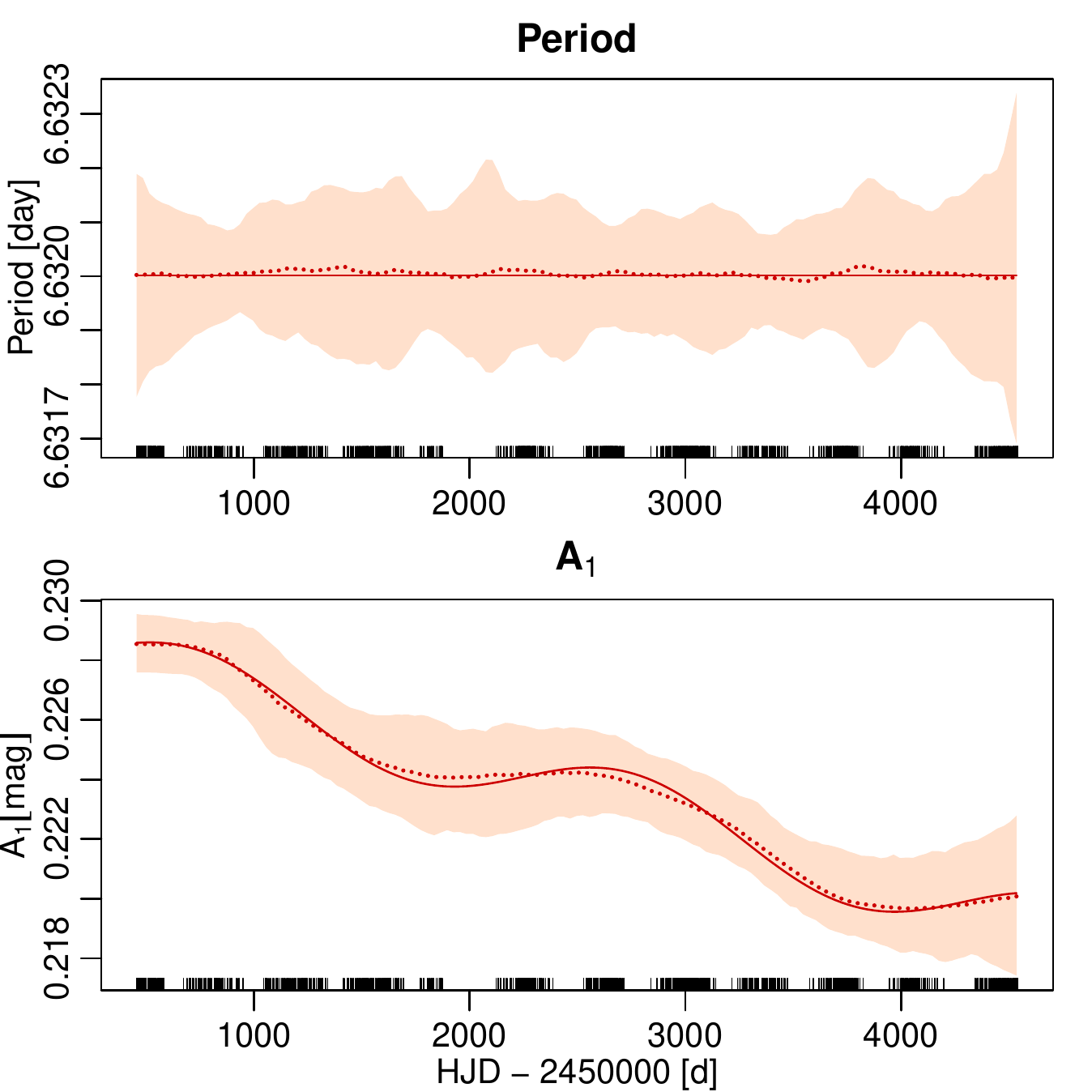}
\caption{Local kernel estimates of the modulation parameters of a Cepheid simulated using the estimated parameters of CEP-1748. The notation and the symbols are the same as in Figure~\ref{fig:trendplusfluctsim}.}
\label{fig:sim1748}       % Give a unique label
\end{center}
\end{figure}
\begin{figure}
%\sidecaption[t]
\begin{center}
\includegraphics[scale=.65]{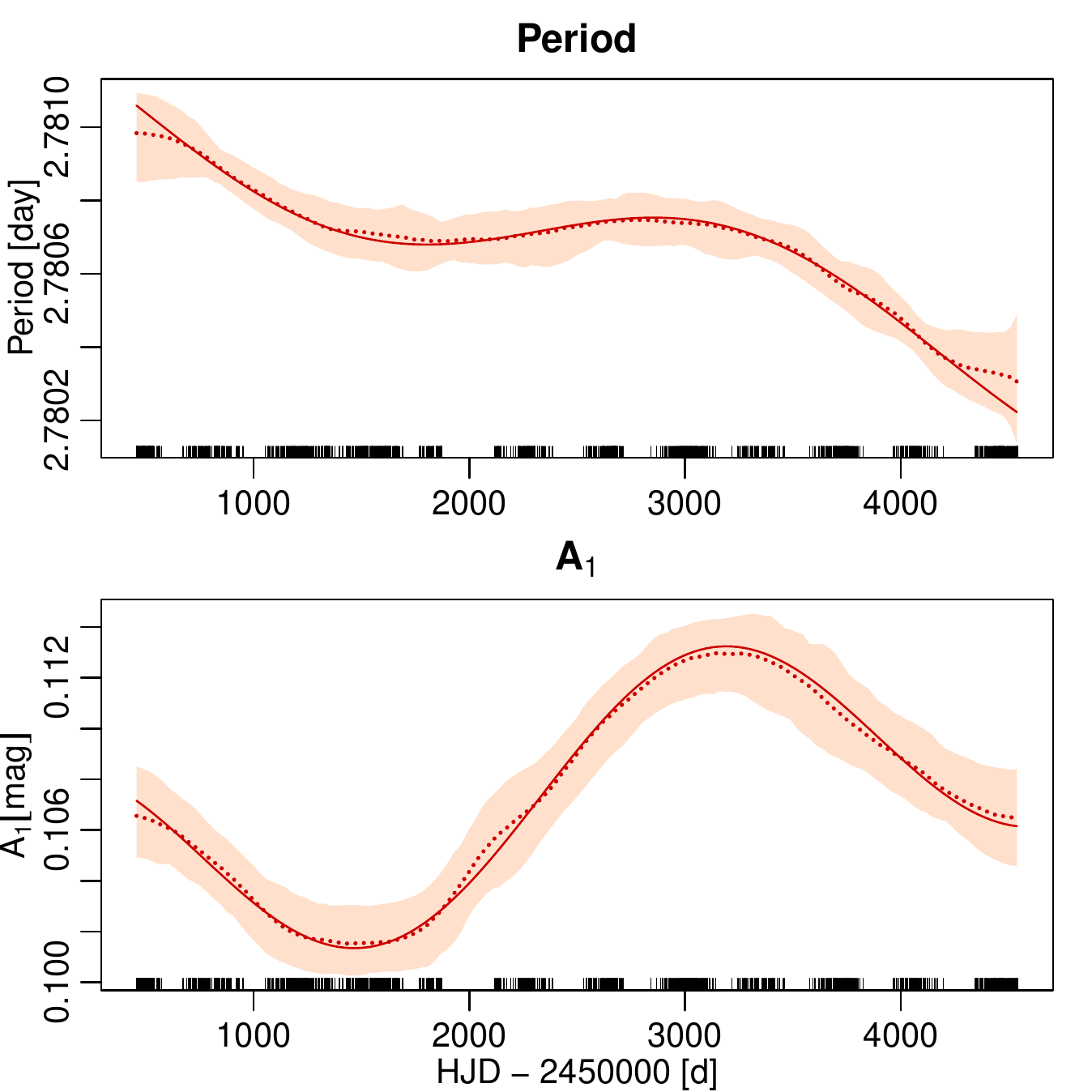}
\caption{Local kernel estimates of the modulation parameters of a Cepheid simulated using the estimated parameters of CEP-1405. The notation and the symbols are the same as in Figure~\ref{fig:trendplusfluctsim}.}
\label{fig:sim1405}       % Give a unique label
\end{center}
\end{figure}

We have verified the ability of the local kernel method to correctly detect amplitude modulations in the case of stable pulsation periods. To this end, we simulated variations in $A_1$ using CEP-1748's estimated $A_1$ fluctuations, setting both $\beta_{P,1748} = 0$ and $\Delta_{P,1748} = 0$. As Figure~\ref{fig:sim1748} shows, amplitude modulations on timescales longer than the window length are well-estimated (bottom panel). However, based on the above discussion of bias in detection fluctuations in pulsation period, we expect similar detection bias regarding fluctuations of $A_1$, cf. App.\,\ref{app:bias}. The assumed constant pulsation period is recovered as such (top panel).

Combining amplitude and period fluctuations, we have simulated a case in analogy to CEP-1405, cf. Figure~\ref{fig:sim1405}. We here adopted a fluctuation timescale that would be readily detectable using a 3-year sliding window (cf. App.\,\ref{subsec:trendpersim}). As Fig.~\ref{fig:sim1405} shows, the kernel method provides unbiased estimates of both the variations in $A_1$ and $P$, and reveals no issues related to the separation of the two phenomena. 

\section{Reliability and bias of the modulation parameter estimates} \label{app:bias}

%\begin{figure}
%\includegraphics[scale=.95]{per_vs_t_vs_modfr_sim.pdf}
%\caption{The kernel-estimated temporal variation of the period as the modulation frequency increases (from top to bottom). The solid black line is the simulated true period modulation. The thick blue line is the median estimate of the sliding window, the thin blue lines indicate the 0.025 and 0.975 quantiles.}
%\label{fig:per_vs_t_vs_modfr_sim}       % Give a unique label
%\end{figure}

\begin{figure}
%\sidecaption[t]
\begin{center}
\includegraphics[scale=.55]{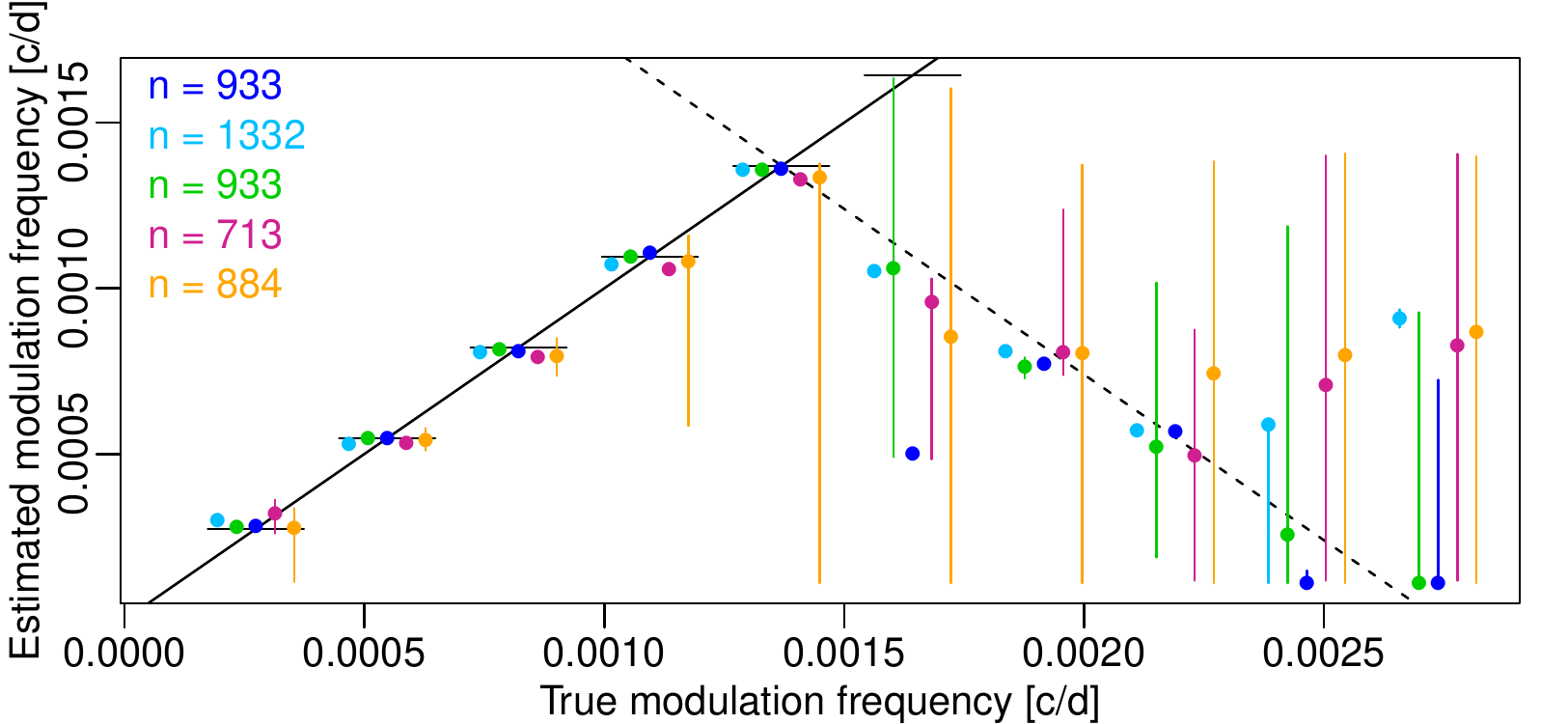}
%\caption{The modulation frequency estimated by model \eqref{eq:armodel} applied to local sliding window estimates of simulated period modulations. The differently coloured dots show the median of the estimates based on 200 different Gaussian noise realizations around each of five simulated modulated light curves, the segments indicate the span between the 0.025 and 0.975 quantile of the estimates. Cyan: simulated CEP-1527, green: simulated CEP-2217, blue: simulated CEP-2191, dark red: simulated CEP-1140, orange: simulated CEP-1833. All simulations used the same sequence of true modulation frequencies, but for visibility, the dots based on different Cepheids are offset from the true value. The solid black line is the $x=y$ line, where a perfect estimate should fall. The short horizontal segments are visual aids where the offset dots should fall in case of a good estimate. The dashed line indicates the yearly alias of the $x=y$ line.}
%\label{fig:modfreqest_sim}       % Give a unique label
%\end{center}
%\end{figure}
%\begin{figure}
%\begin{center}
\includegraphics[scale=.55]{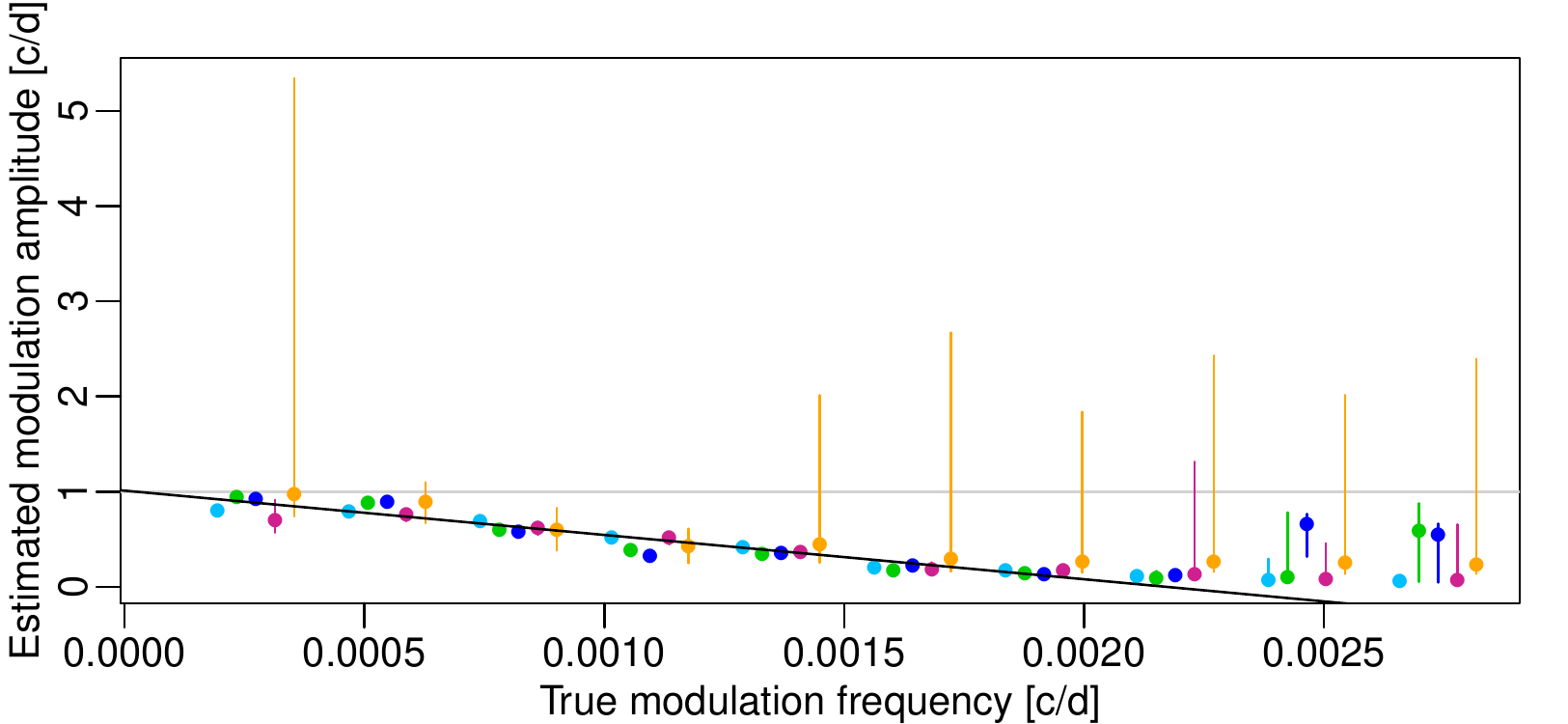}
\caption{The modulation frequency (top panel) and the ratio between the modulation amplitude and the true amplitude (bottom panel) estimated by model \eqref{eq:armodel} applied to local sliding window estimates of simulated period modulations. Coloured dots show the median of the estimates based on 200 different Gaussian noise realizations added to each of five simulated modulated light curves, and vertical segments indicate the range between the 0.025 and 0.975 quantiles. Cyan: using parameters of CEP-1527, green: CEP-2217, blue: CEP-2191, dark red: CEP-1140, orange: CEP-1833. All simulations used the same sequence of true modulation frequencies. For visibility, the dots based on different Cepheids are horizontally offset from the true value. In the top panel, the solid black line indicates $x=y$, i.e., the location of accurate estimates, with short horizontal segments aiding the eye as to where horizontally offset points  should lie. The dashed line indicates the yearly alias of the $x=y$ line. In the bottom panel, the horizontal grey line indicates $y = 1$, the location of the ratios in case of unbiased estimates. The solid black line is the best linear fit using the five medians at each modulation frequency corresponding to the seven lowest true modulation frequencies.}

\label{fig:modestbias_sim}       % Give a unique label
\end{center}
\end{figure}

Appendices~\ref{subsec:combpersim} and \ref{subsec:combperampsim} used simulated amplitude and period fluctuations to show that both types of fluctuations (separate or combined) are accurately recovered if the timescale of the fluctuations is at least half of the sliding window's timespan. However, they also showed that the sliding window introduces bias in the estimated intensity of period and/or amplitude fluctuations, which implies that estimates obtained via a heuristic model \eqref{eq:armodel} are also potentially biased or unreliable. Here, we investigate how the limitations of the kernel method influence our results, specifically regarding the frequency and amplitude of any detected fluctuations in $A_1$ and $P$.

%%%% original start of this App.: As the previous section shows, the kernel method, depending on the timescale of the modulation and on the choice of the window length, can be biased. This implies that the estimates gained from the heuristic model \eqref{eq:armodel} are also biased or unreliable. We need to investigate how the limitations of the kernel method influence these estimates, most importantly that of the frequency and the amplitude of the fluctuation. 

We selected five Cepheids spanning a range of different 
catalog pulsation periods (CEP-1527, $P_\mathrm{cat} = 1.49 d$, CEP-2217, $P_\mathrm{cat} = 2.31 d$, CEP-2191, $P_\mathrm{cat} = 4.21 d$, CEP-1140, $P_\mathrm{cat} = 8.19 d$, and  CEP-1833, $P_\mathrm{cat} = 19.16 d$) and different time sampling. 
We then simulated ten different periodic period modulations for each Cepheid that cover the frequency range detectable by our sliding window using modulation frequencies $f_i = i/3652.5, \; i = 1, 2, \ldots, 10$. The basic model in each case was the catalog pulsation period and the harmonic coefficients determined using the stable reference model. We here assumed high modulation amplitudes ($0.002 c/d$) in order to clearly illustrate the relationship of the detection bias as a function of fluctuation frequency (the expected ratio between true and estimated modulation amplitude does not depend on the signal-to-noise ratio). We estimated each of the simulated light curves (with 200 different added Gaussian white noise sequence for each) with our sliding window method, and fitted model \eqref{eq:armodel} to each of the estimated time series of periods.  

Figure \ref{fig:modestbias_sim} shows the results of this bias estimation. Its top panel shows the distribution of fitted modulation frequencies as a function of the true  modulation frequency. For modulation frequencies  $\lesssim 0.0015 c/d$, i.e., a 2-year modulation period, our method achieves essentially unbiased frequency estimates: in most cases the estimates for all stars are very close to the true modulation frequency. Specifically, we do not find a dependence of the fluctuation estimates on the catalogue pulsation period. The simulations involving the longest-period Cepheid (CEP-1833) however exhibit significantly higher variance, in particular at the lowest frequencies. This can be due to its long period: fewer full light curve cycles are observed in one window, and thus the change in period cannot be estimated as precisely as for a shorter-period Cepheid. Above the limit modulation frequency of $0.0015 c/d$, the estimated frequencies for all five simulated Cepheids are much more scattered, and interestingly, their median often falls not on the true value, but on its yearly alias.

To assess the bias of the modulation amplitude estimates from model \eqref{eq:armodel}, we computed the ratios between the estimated and the true amplitude, as shown the bottom panel of Fig.~\ref{fig:modestbias_sim}. As hinted at by Fig.~\ref{fig:trendplusfluctsim}, the kernel method's estimation of the modulation amplitude is biased.  Fig.~\ref{fig:modestbias_sim} suggests that this bias is nearly linear as a function of the true modulation frequency, very closely in the frequency range that can be reliably detected ($f_P < 0.0015 c/d$). Fitting this relationship, we obtain the empirical bias-correction formula
\begin{equation}
\rho \approx 1.012 - 465.372 f_P,
\end{equation}
which we use to estimate the true underlying modulation in Table \ref{tab:periodresults} and in Figures \ref{fig:frvar_per_distr}, \ref{fig:A1var_distr} and \ref{fig:pervarpars_vs_catalogP}. %\ria{In the future, we shall investigate this bias more closely using a greater number of stars, a wider range of pulsation periods and light curve shapes, and additional temporal sampling of the observations or modulation patterns.}
%%% However, we note that although we tried to cover the range of Cepheid primary periods in our sample, we selected differing time sampling sequences, and objects with widely different harmonic components, the fit is based on the simulations of only 5 examples, and with a single modulation amplitude.

\begin{figure}
%\sidecaption[t]
\begin{center}
\includegraphics[scale=.65]{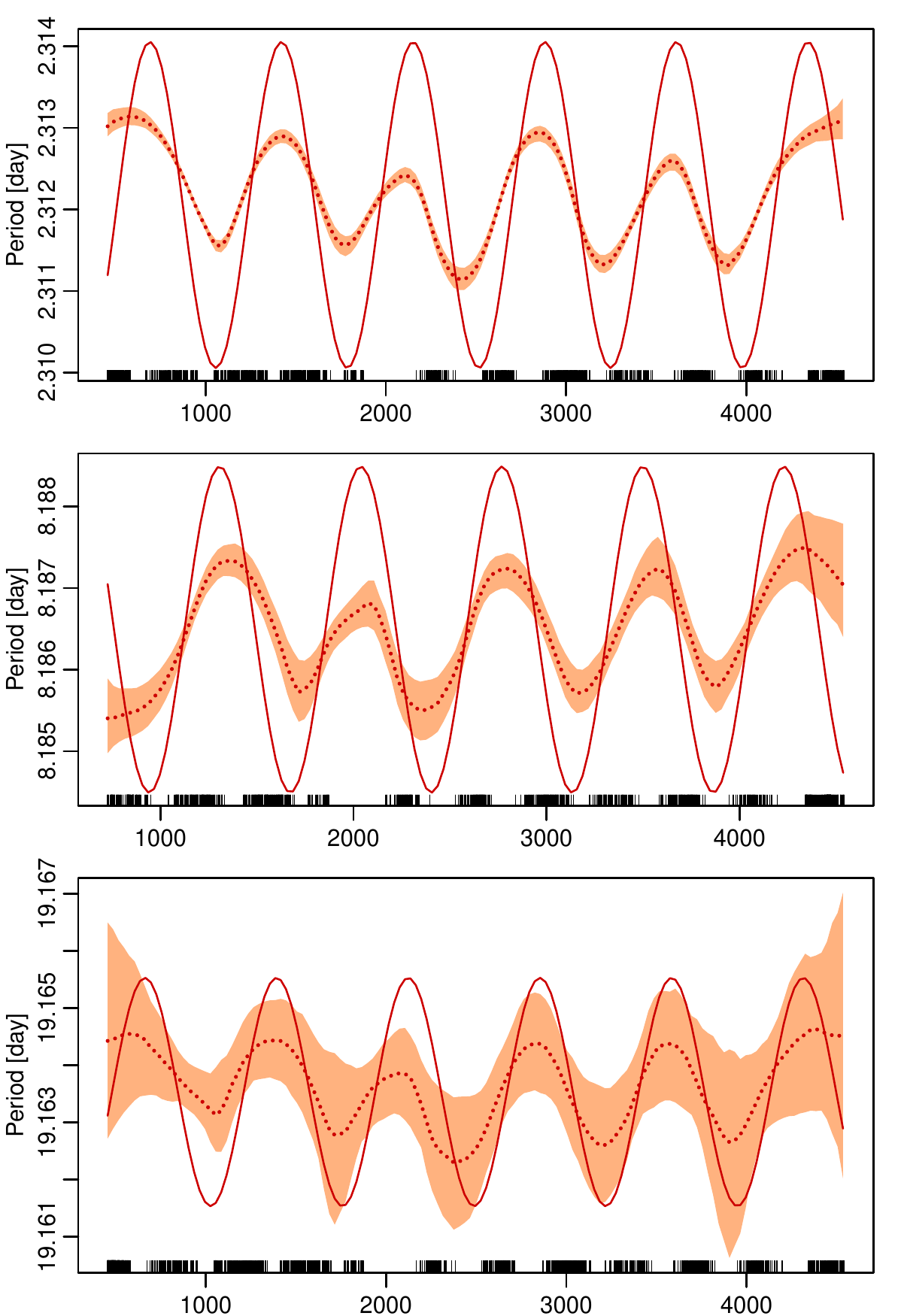}
\caption{Local kernel estimates of the period modulation in simulations of a modulation of $f_P = 0.00137 c/d$, using the stable parameters of CEP-2217 (top panel),  CEP-1140 (middle panel), and CEP-1833 (bottom panel). Solid dark red line: the true signal, dotted dark red line: median of the local kernel estimates on 200 random realizations, orange band: the stripe between the pointwise 0.025 and 0.975 quantiles of the estimates.}
\label{fig:shapedist_sim}       % Give a unique label
\end{center}
\end{figure}
Furthermore, the above simulations can be used to form an idea how much the particular time sampling influences locally the shape of the estimated pattern. Figure \ref{fig:shapedist_sim} presents the estimated modulations of the period in a case where the frequency of the modulation was fixed around the upper detection limit of our 3-year window (corresponding to a period of 2 years). The simulation in the uppermost panel uses the time sampling, catalog period and stable harmonic parameters of CEP-2217 (FO, harmonic order $H = 3$, $P_\mathrm{cat} = 2.31 d$), the middle one, that of  CEP-1140 (FU, harmonic order $H = 12$, $P_\mathrm{cat} = 8.19 d$), and the bottom panel, that of CEP-1833 (FU, harmonic order $H = 12$, $P_\mathrm{cat} = 19.16 d$). The systematic distortions caused by the different time samplings are noticeable, as well as the increasing statistical uncertainty due partly to the need of estimating more parameters from a similar amount of data. The shape of the estimated signal for the characteristics of CEP-2217 and 1140 is remarkably stable, the signal, even at this relatively high frequency is doubtless present, albeit under-estimated, and the deformations depend predominantly on the time sampling, but only little or very little on the noise. In our other two simulations, using CEP-1527 and 2191, the estimated signal pattern is even less variable than using CEP-2217. The larger variance of the estimates in the case of CEP-1833 implies the higher uncertainty of the estimates of the frequency and the amplitude.

\section{Figures of the temporal behaviour of a few variability parameters}\label{app:figtemporal}

\subsection{Pulsation period}\label{subsec:figperiods}

The figures show the kernel-estimated pulsation period (in days) versus Julian Date (in days). This function is plotted as a solid thick black line, together with its bootstrapped pointwise CI (thin black lines; see Section \ref{sec:slidingwindows}). The heavy orange line denotes the catalog period, which was used as known (non-optimized) value in the fitted stable reference model for the Cepheid. The orange band indicates the nonparametric bootstrap CI around this period, obtained from the procedure described in Section \ref{sec:slidingwindows}. The dotted black vertical lines indicate years after HJD$-2450000$. The grey background highlights time intervals where the deviation from the stable reference model was found significant by the multiple testing procedure of \citet{benjaminiyekutieli01}.  The times of the observations are  indicated with a rugplot at the bottom of the panels.

%\begin{figure*}
%%\sidecaption[t]
%\begin{center}
%\includegraphics[scale=.75]{ampch_all_period.pdf}
%\caption{}
%\label{fig:ampch_all_period}       % Give a unique label
%\end{center}
%\end{figure*}

\subsection{Amplitude of first harmonic term $A_1$}\label{app:figa1}

The figures show the kernel-estimated amplitude of the leading harmonic term (in magnitudes; for the definition, see in Section \ref{subsubsec:ampch}) versus Julian Date (in days). This function is plotted as a solid thick black line, together with its bootstrapped pointwise CI (thin black lines; see  Section \ref{sec:slidingwindows}). The heavy orange line denotes the best-fit amplitude from the stable reference model. The orange band indicates the nonparametric bootstrap CI obtained from the procedure described in Section \ref{sec:slidingwindows}. The dotted black horizontal and vertical lines are aids to the eye to estimate the extent and time interval of the changes. The grey background highlights time intervals where the deviation from the stable reference model was found significant by the multiple testing procedure of \citet{benjaminiyekutieli01}. The times of the observations are indicated with a rugplot at the bottom of the panels.

%\begin{figure*}
%%\sidecaption[t]
%\begin{center}
%\includegraphics[scale=.75]{ampch_all_period.pdf}
%\caption{}
%\label{fig:ampch_all_period}       % Give a unique label
%\end{center}
%\end{figure*}

\subsection{Light curve shape}\label{app:lcshape}

The figures show the kernel-estimated light curve shapes in two little- or non-overlapping windows, in which the peak-to-peak amplitude or the amplitude of the leading harmonic term was very different. The two curves are indicated by the different colours, and the observations in the different windows, by the different colours and shapes of the symbols. The times of the window centres are given in the legend. The observations close to the window centre and therefore more influential in the fit have darker red or blue colours, whereas those in the wings are shown in lighter shades. Approximate confidence intervals for the curves are also given as light red and light blue stripes around the estimated lines. They were computed based on the bootstrap repetitions of the time-varying window estimates, as described in  Section \ref{sec:slidingwindows}  and Appendix \ref{app:detailedmethods}. In both windows, the kernel-estimated coefficients on the repetitions in the window were used to reconstruct 250 bootstrap light curves. The stripes indicate the total span of these light curves. 

\begin{figure*}[h]
%\sidecaption[t]
\begin{center}
\includegraphics[scale=.78]{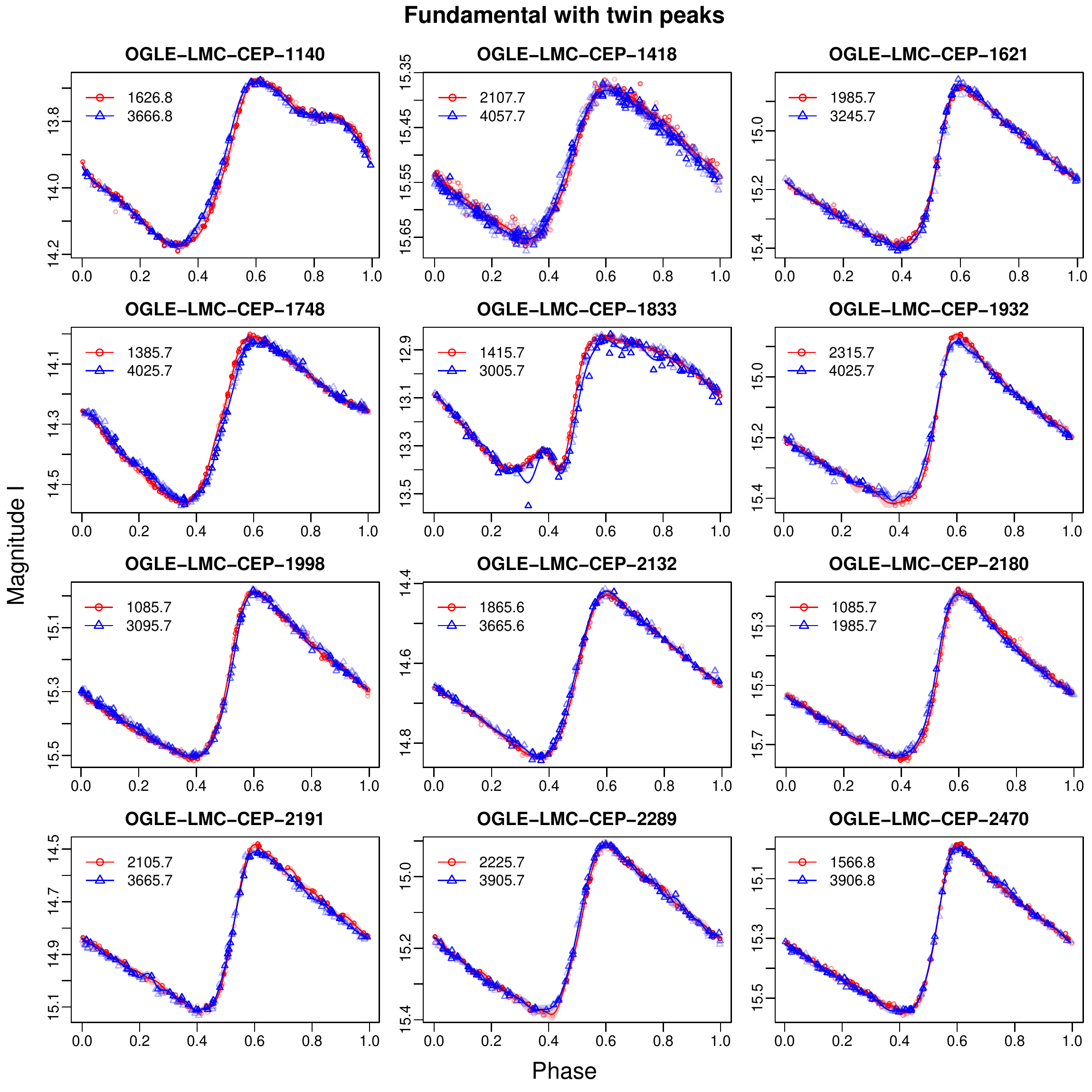}
\caption{Light curve shapes in two different windows for twin-peak fundamental Cepheids. The notation is given at the beginning of the section.}
\label{fig:twins_fu_lc}       % Give a unique label
\end{center}
\end{figure*}

\begin{figure*}[h]
%\sidecaption[t]
\begin{center}
\includegraphics[scale=.78]{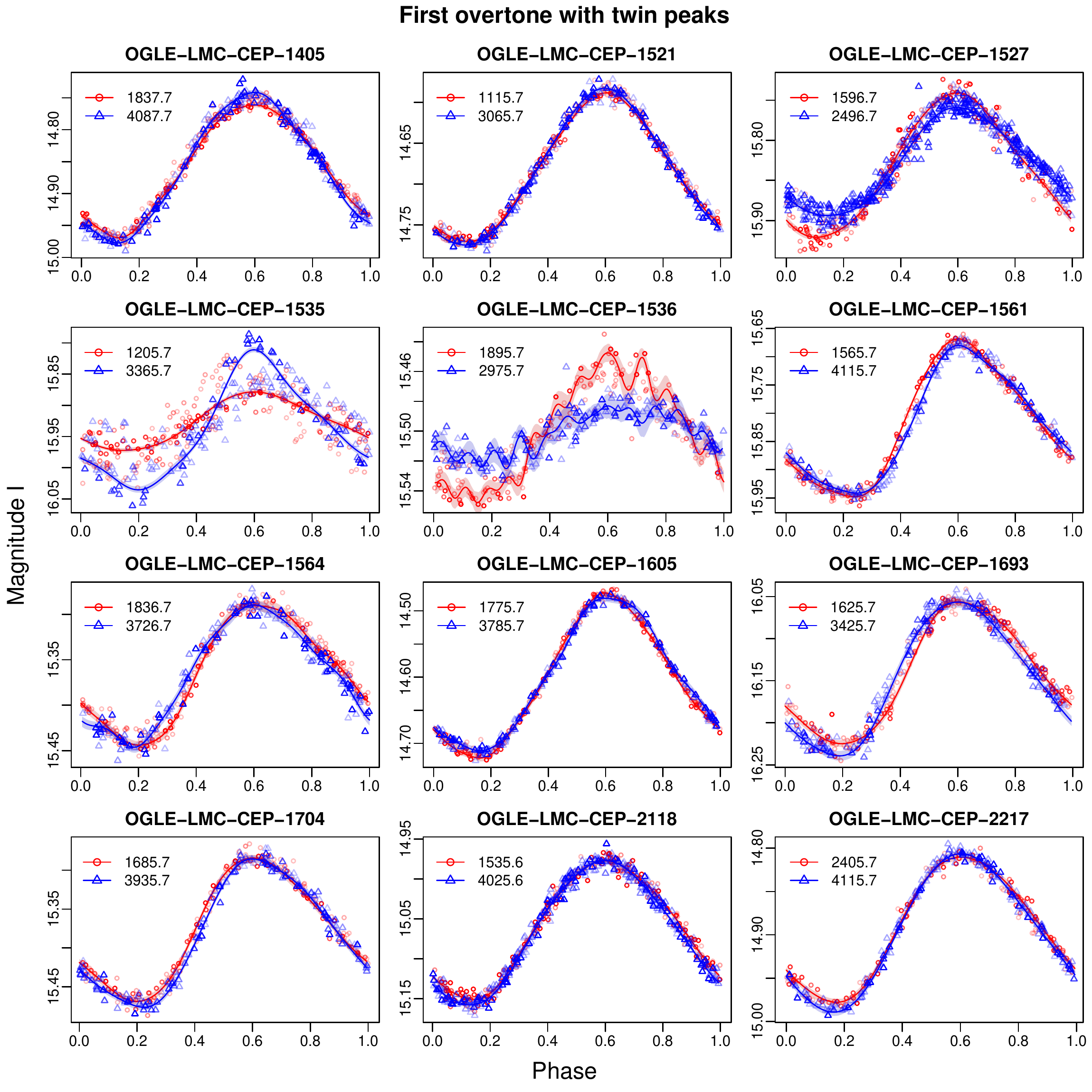}
\caption{{Light curve shapes in two different windows for twin-peak overtone Cepheids. The notation is given at the beginning of the section.}}
\label{fig:twins_fo_lc}       % Give a unique label
\end{center}
\end{figure*}

\begin{figure*}[h]
%\sidecaption[t]
\begin{center}
\includegraphics[scale=.78]{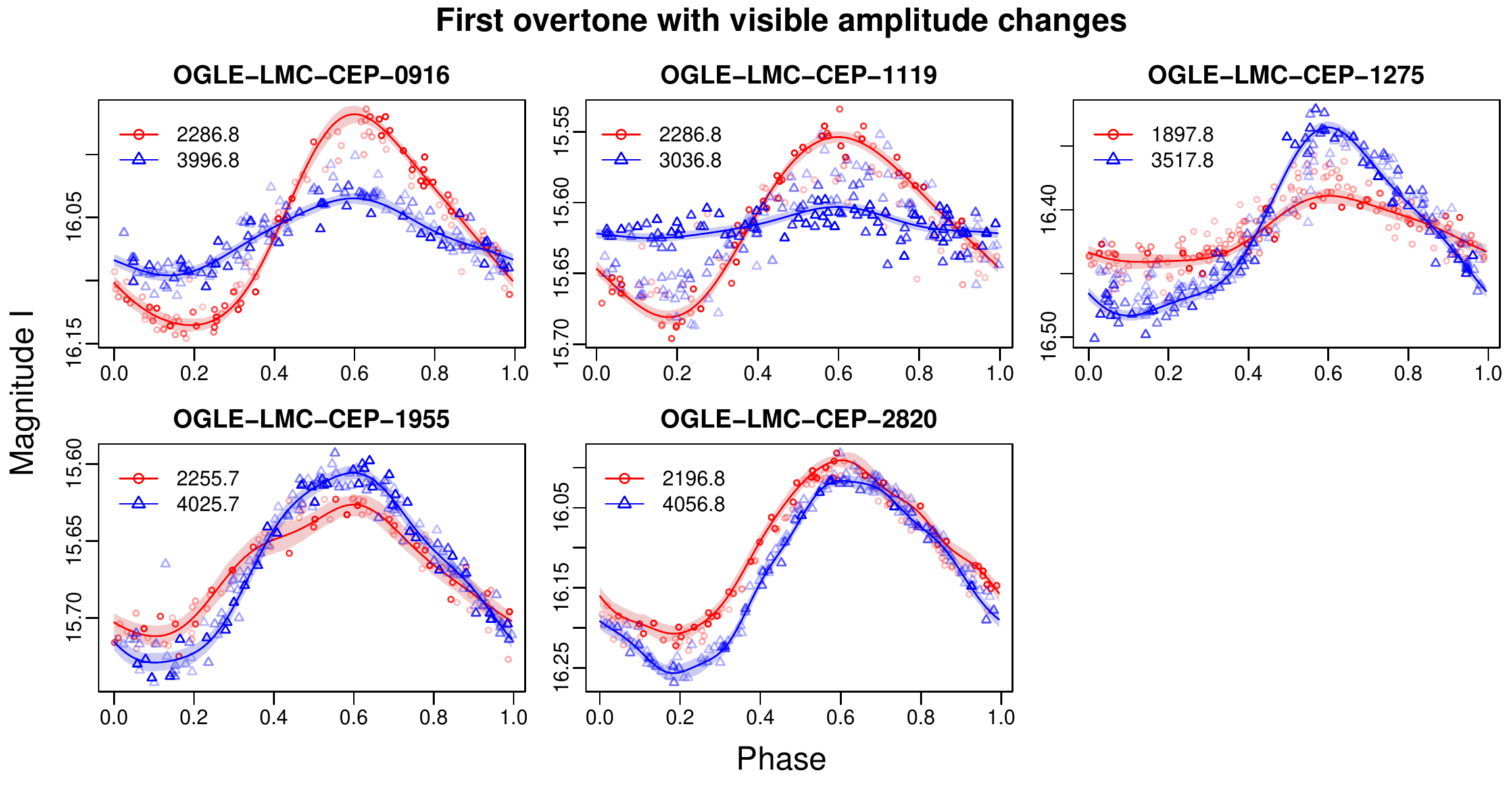}
\caption{Light curve shapes in two different windows for overtone Cepheids showing strong amplitude changes. The notation is given at the beginning of the section.}
\label{fig:ampch_lc}       % Give a unique label
\end{center}
\end{figure*}

\begin{figure*}[h]
%\sidecaption[t]
\begin{center}
\includegraphics[scale=.78]{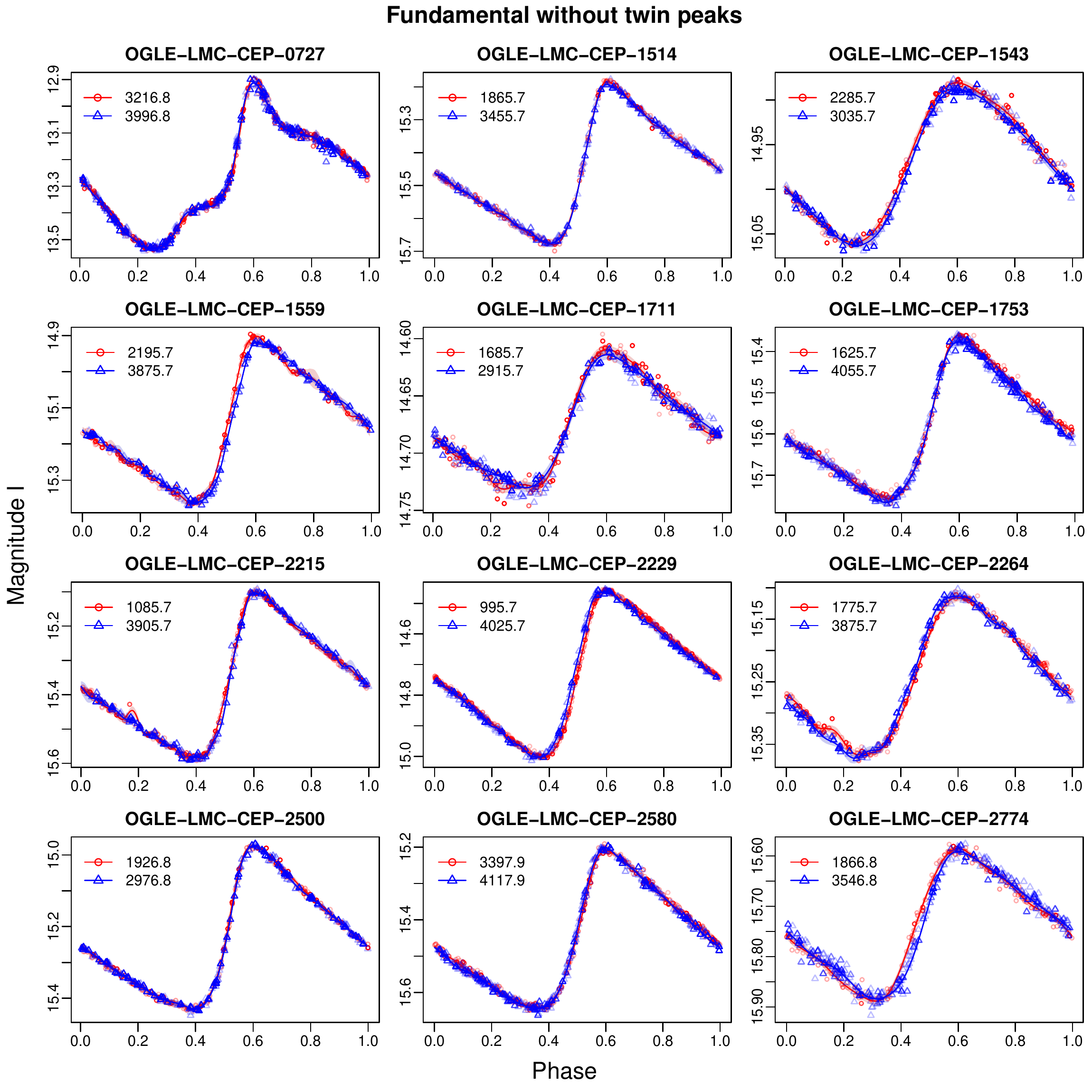}
\caption{Light curve shapes in two different windows for control fundamental Cepheids. The notation is given at the beginning of the section.}
\label{fig:control_fu_lc}       % Give a unique label
\end{center}
\end{figure*}

\begin{figure*}[h]
%\sidecaption[t]
\begin{center}
\includegraphics[scale=.78]{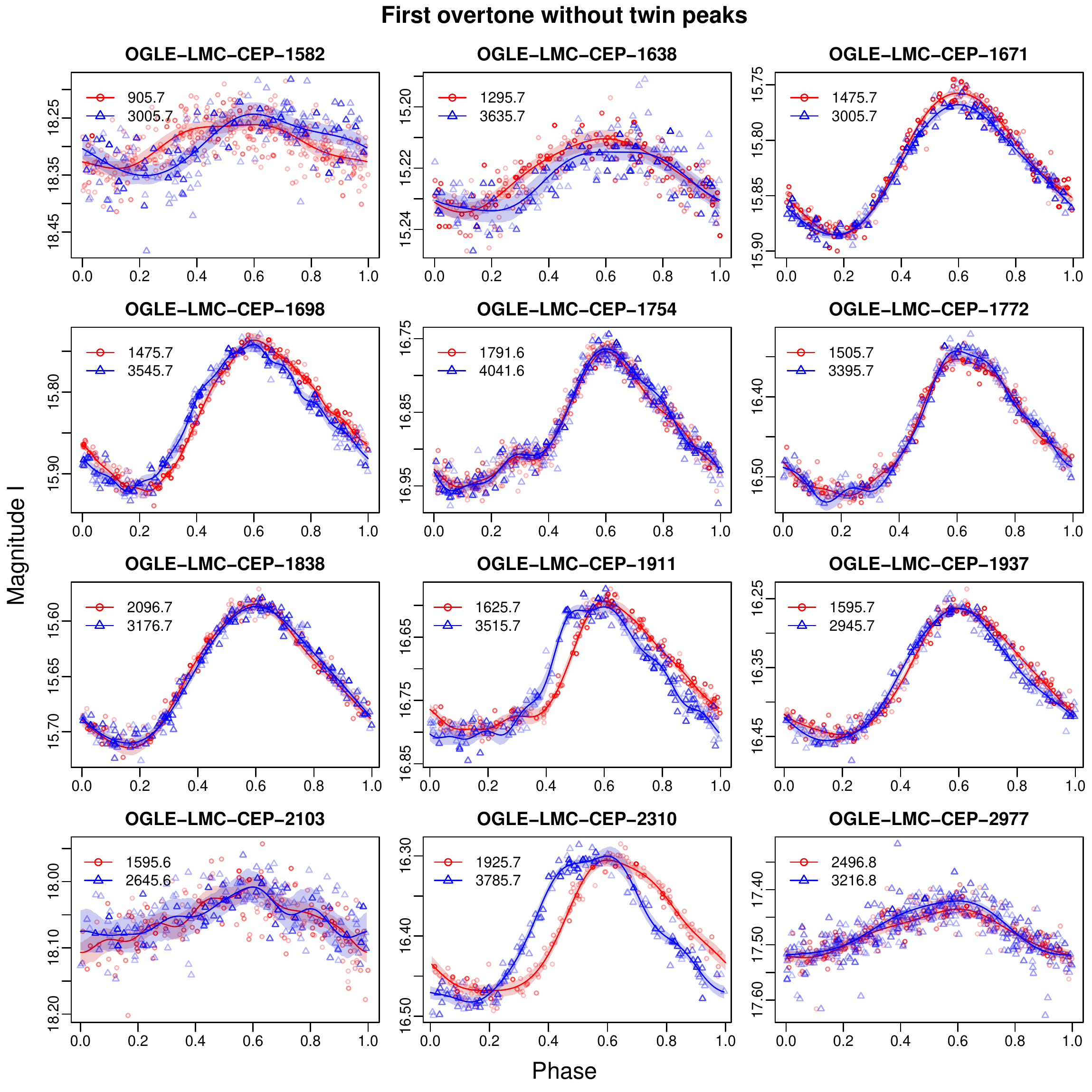}
\caption{Light curve shapes in two different windows for control overtone Cepheids. The notation is given at the beginning of the section.}
\label{fig:control_fo_lc}       % Give a unique label
\end{center}
\end{figure*}

\section{Table of trends and fluctuations in the pulsation period and the first harmonic amplitude}\label{app:tables}

\begin{table*}[!h]
\centering
\begin{tabular}{lrrrrrrr}
  \hline
CEP & $P$ & $\beta_P$ & $\sigma_{\beta_P}$ & $f_P$ & $\Delta P$ & $\sigma_{\Delta P}$ & $\Delta P_c$ \\  
 & [$d$] & [$10^{-7}d/d$] & [$10^{-7}d/d$] & [$1/d$] & [$10^{-4}d$] & [$10^{-4}d$] & [$10^{-4}d$] \\ 
\hline
\multicolumn{8}{c}{\smallskip{\bf FU Cepheids with twin peaks}} \\
  1140 & 8.1864915 & -0.5779 & 0.2094 & 0.00115 & 1.1214 & 0.0900 & 2.3633\\ 
  1418 & 2.8957105 & -0.3781 & 0.0525 & 0.00091 & 0.3207 & 0.0272 & 0.5437 \\ 
  1621 & 3.3588290 & -0.5340 & 0.1528 & 0.00066 & 1.3521 & 0.1350 & 1.9206\\ 
  1748 & 6.6320016 & -1.0912 & 0.1293 & 0.00118 & 0.8824 & 0.0611 & 1.9000 \\ 
  1833 & 19.1635310 & -21.7161 & 1.5753 & 0.00113 & 7.3639 & 1.0115 & 15.1125 \\ 
  1932 & 2.9467014 & -0.8024 & 0.0971 & 0.00027 & 2.0726 & 0.1049 & 2.3380 \\ 
  1998 & 3.0863817 & -0.4780 & 0.0361 & 0.00093 & 0.1383 & 0.0208 & 0.2390 \\ 
  2132 & 4.6823873 & 4.3943 & 0.1194 & 0.00179 & 0.3518 & 0.0424 & 1.9623 \\ 
  2180 & 2.5877856 & 0.2258 & 0.0751 & 0.00047 & 0.6269 & 0.0784 & 0.7883 \\ 
  2191 & 4.2070310 & -0.7288 & 0.1555 & 0.00061 & 1.6074 & 0.1469 & 2.2116 \\ 
  2289 & 3.7126964 & -0.3231 & 0.0532 & 0.00105 & 0.3232 & 0.0312 & 0.6197\\ 
  2470 & 3.2211085 & -0.5983 & 0.0510 & 0.00134 & 0.1886 & 0.0220 & 0.4847 \\ 
%\smallskip\\
\multicolumn{8}{c}{\smallskip{\bf FO Cepheids with twin peaks}}  \\
  1405 & 2.7807164 & -1.7996 & 0.1021 & 0.00027 & 1.6874 & 0.1284 & 1.9035 \\ 
  1521 & 3.2147769 & 4.0352 & 0.1497 & 0.00054 & 1.2717 & 0.1323 & 1.6710 \\ 
  1527 & 1.4923973 & -2.8863 & 0.0742 & 0.00039 & 1.8818 & 0.0858 & 2.2687 \\ 
  1535 & 1.2509786 & -0.6089 & 0.1356 & 0.00074 & 2.1608 & 0.1326 & 3.2261 \\ 
  1536 & 1.6112568 & -0.4181 & 0.3225 & 0.00039 & 3.3492 & 0.4349 & 4.0378 \\ 
  1561 & 1.5299801 & 0.0068 & 0.0478 & 0.00042 & 0.4888 & 0.0487 & 0.5975 \\ 
  1564 & 2.0629824 & 0.1446 & 0.0572 & 0.00088 & 0.5573 & 0.0397 & 0.9268 \\ 
  1605 & 3.0881609 & -0.0018 & 0.0984 & 0.00059 & 1.3976 & 0.0836 & 1.8933 \\ 
  1693 & 1.6639538 & -0.2544 & 0.0449 & 0.00130 & 0.2006 & 0.0173 & 0.4923 \\ 
  1704 & 1.8820681 & 2.4560 & 0.1133 & 0.00025 & 1.0959 & 0.1380 & 1.2205 \\ 
  2118 & 2.4461347 & -0.7408 & 0.0499 & 0.00037 & 1.0945 & 0.0641 & 1.3017 \\ 
  2217 & 2.3120543 & -7.7622 & 0.0872 & 0.00029 & 3.9691 & 0.1196 & 4.5357 \\ 
%\smallskip\\
\multicolumn{8}{c}{\smallskip{\bf FO Cepheids with visible changing amplitudes}} \\
  0916 & 1.3695732 & 0.9550 & 0.1352 & 0.00039 & 2.0659 & 0.0972 & 2.4928 \\ 
  1119 & 1.7431985 & -3.5252 & 0.5765 & 0.00079 & 4.2670 & 0.4105 & 6.6101 \\ 
  1275 & 0.9474017 & -0.6110 & 0.0696 & 0.00039 & 2.2950 & 0.0997 & 2.7668 \\ 
  1955 & 1.8879224 & -1.0391 & 0.1135 & 0.00108 & 0.7389 & 0.0670 & 1.4485 \\ 
  2820 & 2.1127656 & -0.3327 & 0.0585 & 0.00031 & 1.5344 & 0.0673 & 1.7731 \\ 
%\smallskip\\
\multicolumn{7}{c}{\smallskip{\bf FU Cepheids without twin peaks}}  \\
  0727 & 14.4891397 & 2.8043 & 0.5326 & 0.00100 & 2.5372 & 0.3123 & 4.6316 \\ 
  1514 & 3.0459183 & -0.0255 & 0.0379 & 0.00108 & 0.2073 & 0.0210 & 0.4063 \\ 
  1543 & 4.3781838 & -0.3817 & 0.1886 & 0.00074 & 1.4589 & 0.1237 & 2.1782 \\ 
  1559 & 3.6469543 & -0.1060 & 0.0510 & 0.00100 & 0.2689 & 0.0283 & 0.4941 \\ 
  1711 & 2.9189534 & -0.3011 & 0.1125 & 0.00162 & 0.3383 & 0.0391 & 1.3053 \\ 
  1753 & 2.5745626 & -0.0868 & 0.0290 & 0.00132 & 0.1232 & 0.0142 & 0.3111 \\ 
  2215 & 3.2408617 & 0.0805 & 0.0524 & 0.00074 & 0.3104 & 0.0435 & 0.4635 \\ 
  2229 & 4.8745455 & 0.0386 & 0.0377 & 0.00098 & 0.1970 & 0.0239 & 0.3545 \\ 
  2264 & 3.8808875 & -0.1143 & 0.0630 & 0.00083 & 0.6172 & 0.0447 & 0.9889 \\ 
  2500 & 3.2002470 & 0.1430 & 0.0321 & 0.00126 & 0.2428 & 0.0149 & 0.5703 \\ 
  2580 & 2.9569833 & 0.2428 & 0.0490 & 0.00084 & 0.2566 & 0.0196 & 0.4143 \\ 
  2774 & 3.8131712 & 0.6247 & 0.1062 & 0.00144 & 0.2607 & 0.0393 & 0.7664 \\ 
 %\smallskip\\
\multicolumn{7}{c}{\smallskip{\bf FO Cepheids without twin peaks}} \\
  1582 & 0.3187867 & -0.0053 & 0.0109 & 0.00061 & 0.1101 & 0.0104 & 0.1516 \\ 
  1638 & 2.3630077 & -0.0488 & 0.5678 & 0.00140 & 1.7129 & 0.2142 & 4.7344 \\ 
  1671 & 1.3642921 & -0.0081 & 0.0420 & 0.00147 & 0.2010 & 0.0172 & 0.6135 \\ 
  1698 & 1.5955712 & 0.0812 & 0.0537 & 0.00113 & 0.1722 & 0.0248 & 0.3533 \\ 
  1754 & 0.7113208 & -0.0100 & 0.0118 & 0.00125 & 0.0423 & 0.0044 & 0.0979 \\ 
  1772 & 0.9471192 & -0.0767 & 0.0105 & 0.00123 & 0.0737 & 0.0056 & 0.1670 \\ 
  1838 & 1.6431880 & -0.1005 & 0.0306 & 0.00127 & 0.1370 & 0.0152 & 0.3270 \\ 
  1911 & 0.7485379 & 0.0311 & 0.0126 & 0.00130 & 0.0578 & 0.0052 & 0.1418 \\ 
  1937 & 1.0593254 & 0.0085 & 0.0163 & 0.00142 & 0.0708 & 0.0061 & 0.2020 \\ 
  2103 & 0.3913544 & 0.3939 & 0.0880 & 0.00189 & 0.1347 & 0.0336 & 1.0074 \\ 
  2310 & 1.2713288 & 0.0863 & 0.0370 & 0.00044 & 0.3485 & 0.0387 & 0.4321 \\ 
  2977 & 0.4956663 & 0.0098 & 0.0161 & 0.00102 & 0.1089 & 0.0079 & 0.2033 \\ 
  \hline
\end{tabular}
{\vskip 2mm}
\caption{Summary of the results of fitting model \eqref{eq:armodel} to the modulations of the pulsation period $P$. $P$: Pulsation period (in days); $\beta_P$, $f_P$ and $\Delta P$: trend, oscillation frequency and oscillation amplitude of the pulsation period (see Section \ref{sec:trends+fluctuations}); $\sigma_{\beta_P}$ and $\sigma_{\Delta P}$: standard errors of $\beta_P$ and $\Delta_P$; $\Delta P_c$: the bias-corrected estimate of the amplitude of the period modulation.}
\label{tab:periodresults}
\end{table*}

% latex table generated in R 3.2.2 by xtable 1.7-4 package
% Mon Dec 21 17:11:10 2015
\begin{table*}[!h]
\centering
\begin{tabular}{lrrrrrrr}
  \hline
CEP & $A_1$ & $\beta_{A_1}$ & $\sigma_{\beta_{A_1}}$ & $f_{A_1}$ & $\Delta A_1$ & $\sigma_{\Delta A_1}$ & $\Delta A_{1c}$ \\ 
 & [mag] & [$10^{-6}$mag$/d$] & [$10^{-6}$mag$/d$] & [$1/d$] & [mmag] & [mmag] & [mmag] \\ 
\hline
\multicolumn{8}{c}{\smallskip{\bf FU Cepheids with twin peaks}} \\
  1140 & 0.2202 & -1.255 & 0.153 & 0.00037 & 1.92 & 0.17 & 2.28 \\ 
  1418 & 0.1207 & 0.065 & 0.155 & 0.00088 & 0.63 & 0.07 & 1.05 \\ 
  1621 & 0.2183 & 0.335 & 0.120 & 0.00037 & 2.91 & 0.14 & 3.46 \\ 
  1748 & 0.2241 & -2.057 & 0.207 & 0.00049 & 1.17 & 0.16 & 1.50 \\ 
  1833 & 0.2679 & -0.607 & 0.335 & 0.00150 & 0.65 & 0.12 & 2.05 \\ 
  1932 & 0.2083 & -1.598 & 0.100 & 0.00027 & 2.27 & 0.21 & 2.56 \\ 
  1998 & 0.2083 & -0.309 & 0.165 & 0.00081 & 0.78 & 0.09 & 1.23 \\ 
  2132 & 0.1638 & -0.612 & 0.101 & 0.00039 & 1.03 & 0.12 & 1.24 \\ 
  2180 & 0.2174 & -1.150 & 0.172 & 0.00078 & 0.92 & 0.10 & 1.42 \\ 
  2191 & 0.2353 & 0.944 & 0.212 & 0.00096 & 1.07 & 0.13 & 1.89 \\ 
  2289 & 0.1916 & -0.533 & 0.145 & 0.00076 & 1.35 & 0.10 & 2.05 \\ 
  2470 & 0.2118 & 0.648 & 0.199 & 0.00073 & 1.34 & 0.11 & 2.00 \\ 
\multicolumn{8}{c}{\smallskip{\bf FO Cepheids with twin peaks}}  \\
  1405 & 0.1073 & 1.529 & 0.277 & 0.00032 & 4.69 & 0.34 & 5.43 \\ 
  1521 & 0.0897 & -0.804 & 0.213 & 0.00074 & 1.57 & 0.13 & 2.35 \\ 
  1527 & 0.0647 & -2.419 & 0.409 & 0.00051 & 10.54 & 0.42 & 13.64 \\ 
  1535 & 0.0637 & -1.056 & 1.777 & 0.00074 & 15.42 & 0.98 & 23.03 \\ 
  1536 & 0.0251 & -2.358 & 0.591 & 0.00042 & 10.93 & 0.53 & 13.36 \\ 
  1561 & 0.1282 & -0.834 & 0.165 & 0.00054 & 2.39 & 0.15 & 3.15 \\ 
  1564 & 0.0746 & -0.459 & 0.119 & 0.00181 & 0.21 & 0.03 & 1.22 \\ 
  1605 & 0.1145 & 0.136 & 0.170 & 0.00061 & 1.13 & 0.12 & 1.55 \\ 
  1693 & 0.0853 & 1.085 & 0.179 & 0.00078 & 0.82 & 0.08 & 1.27 \\ 
  1704 & 0.0892 & 0.282 & 0.189 & 0.00118 & 0.59 & 0.07 & 1.28 \\ 
  2118 & 0.0870 & 0.199 & 0.172 & 0.00096 & 0.73 & 0.06 & 1.28 \\ 
  2217 & 0.0634 & 0.668 & 0.159 & 0.00025 & 2.50 & 0.25 & 2.79 \\ 
\multicolumn{8}{c}{\smallskip{\bf FO Cepheids with visible changing amplitudes}} \\
  0916 & 0.0425 & -9.139 & 1.425 & 0.00042 & 18.35 & 1.09 & 22.47 \\ 
  1119 & 0.0357 & 2.908 & 1.946 & 0.00076 & 14.92 & 0.84 & 22.68 \\ 
  1275 & 0.0439 & 1.149 & 0.998 & 0.00039 & 23.16 & 0.90 & 27.92 \\ 
  1955 & 0.0530 & 0.486 & 0.574 & 0.00025 & 9.20 & 0.66 & 10.24 \\ 
  2820 & 0.1109 & 1.236 & 0.306 & 0.00026 & 5.73 & 0.32 & 6.44 \\ 
\multicolumn{7}{c}{\smallskip{\bf FU Cepheids without twin peaks}} \\
  0727 & 0.2403 & -0.858 & 0.149 & 0.00068 & 0.71 & 0.09 & 1.02 \\ 
  1514 & 0.1870 & -0.068 & 0.100 & 0.00029 & 1.35 & 0.14 & 1.55 \\ 
  1543 & 0.0832 & -0.046 & 0.090 & 0.00066 & 0.63 & 0.07 & 0.90 \\ 
  1559 & 0.1811 & -0.109 & 0.121 & 0.00066 & 0.91 & 0.08 & 1.30 \\ 
  1711 & 0.0518 & -1.498 & 0.137 & 0.00047 & 0.87 & 0.10 & 1.10 \\ 
  1753 & 0.1584 & -0.099 & 0.146 & 0.00071 & 0.90 & 0.11 & 1.32 \\ 
  2215 & 0.1937 & -0.263 & 0.144 & 0.00025 & 1.29 & 0.22 & 1.44 \\ 
  2229 & 0.2218 & -0.128 & 0.122 & 0.00154 & 0.35 & 0.04 & 1.21 \\ 
  2264 & 0.1168 & -0.208 & 0.212 & 0.00076 & 0.77 & 0.11 & 1.17 \\ 
  2500 & 0.1852 & 0.250 & 0.123 & 0.00031 & 1.44 & 0.12 & 1.67 \\ 
  2580 & 0.1833 & -0.682 & 0.157 & 0.00051 & 1.10 & 0.10 & 1.42 \\ 
  2774 & 0.1275 & -0.357 & 0.226 & 0.00121 & 0.61 & 0.07 & 1.36 \\ 
\multicolumn{7}{c}{\smallskip{\bf FO Cepheids without twin peaks}} \\
  1582 & 0.0410 & 1.055 & 0.453 & 0.00034 & 4.38 & 0.56 & 5.14 \\ 
  1638 & 0.0105 & 0.842 & 0.142 & 0.00132 & 0.42 & 0.04 & 1.06 \\ 
  1671 & 0.0597 & -0.240 & 0.201 & 0.00088 & 1.05 & 0.09 & 1.75 \\ 
  1698 & 0.0838 & 0.157 & 0.152 & 0.00078 & 1.12 & 0.09 & 1.72 \\ 
  1754 & 0.0842 & 0.309 & 0.397 & 0.00103 & 0.88 & 0.13 & 1.66 \\ 
  1772 & 0.0851 & 0.047 & 0.166 & 0.00032 & 1.85 & 0.17 & 2.15 \\ 
  1838 & 0.0612 & 0.091 & 0.104 & 0.00130 & 0.26 & 0.03 & 0.63 \\ 
  1911 & 0.0968 & 0.733 & 0.262 & 0.00086 & 1.38 & 0.17 & 2.24 \\ 
  1937 & 0.0906 & -0.807 & 0.116 & 0.00100 & 0.48 & 0.06 & 0.88 \\ 
  2103 & 0.0365 & -2.420 & 0.540 & 0.00047 & 4.09 & 0.73 & 5.14 \\ 
  2310 & 0.0862 & 0.922 & 0.135 & 0.00034 & 2.79 & 0.20 & 3.27 \\ 
  2977 & 0.0461 & 1.426 & 0.374 & 0.00060 & 3.33 & 0.21 & 4.55 \\ 
   \hline
\end{tabular}
{\vskip 2mm}
\caption{Summary of the results of fitting model \eqref{eq:armodel} to the modulations of the amplitude $A_1$ of the first harmonic. $\beta_{A_1}$, $f_{A_1}$ and $\Delta A_1$: the trend, the oscillation frequency and the oscillation period of $A_1$ (see Section \ref{sec:trends+fluctuations}); $\sigma_{\beta_{A_1}}$ and $\sigma_{\Delta A_1}$: standard error of $\beta_{A_1}$ and $\Delta A_1$; $\Delta A_{1c}$: the bias-corrected estimate of the amplitude of the modulation. }
\label{tab:a1results}

\end{table*}

\end{document}